\documentclass[12pt,twocolumn]{aastex631}
\usepackage{graphics}
\usepackage{color}
\usepackage{topcapt}
\usepackage{amsmath}
\usepackage{empheq} 
\newcommand{\ltsimeq}{\la}
\newcommand{\gtsimeq}{\ga}

\newcommand{\msun}{M$_{\odot}$}
\newcommand{\mbary}{$M_{bary}$}
\newcommand{\mdyn}{$M_{dyn}$}
\newcommand{\mhalo}{$M_{200}$}
\newcommand{\mstar}{$M_*$}
\newcommand{\mhi}{$M_{HI}$}

\newcommand{\fb}{$f_{bary}$}
\newcommand{\vpv}{$V_{PV}$}

\newcommand{\vrot}{$V_{rot}$}
\newcommand{\vcor}{$V_{cor}$}
\newcommand{\vmax}{$V_{max}$}

\newcommand{\dpv}{$D_{PV}$}
\newcommand{\rmax}{$R_{max}$}
\newcommand{\rc}{$r_c$}

\newcommand{\kms}{km~s$^{-1}$}
\newcommand{\hi}{H{\sc i}}

\shortauthors{McQuinn et al.}
\shorttitle{The Turn-Down in the BTFR at Low Galaxy Masses}

\begin{document}
\title{THE TURN-DOWN OF THE BARYONIC TULLY-FISHER RELATION AND CHANGING BARYON FRACTION AT LOW GALAXY MASSES}

\author{Kristen.~B.~W. McQuinn}
\affiliation{Rutgers University, Department of Physics and Astronomy, 136 Frelinghuysen Road, Piscataway, NJ 08854, USA}
\email{kristen.mcquinn@rutgers.edu}

\author{Elizabeth~A.~K. Adams}
\affiliation{ASTRON, The Netherlands Institute for Radio Astronomy, Oude Hoogeveensedijk 4, 7991 PD, Dwingeloo, The Netherlands}
\affiliation{Kapteyn Astronomical Institute, University of Groningen, Postbus 800, 9700 AV Groningen, The Netherlands}

\author{John M. Cannon}
\affiliation{Department of Physics and Astronomy, Macalester College, Saint Paul, MN 55105, USA}

\author{Jackson Fuson}
\affiliation{Department of Physics and Astronomy, Macalester College, Saint Paul, MN 55105, USA}

\author{Evan D. Skillman}
\affiliation{University of Minnesota, Minnesota Institute for Astrophysics, School of Physics and Astronomy, 116 Church Street, S.E., Minneapolis, MN 55455, USA} 

\author{Alyson Brooks}
\affiliation{Rutgers University, Department of Physics and Astronomy, 136 Frelinghuysen Road, Piscataway, NJ 08854, USA}
\affiliation{Center for Computational Astrophysics, Flatiron Institute, 162 Fifth Avenue, New York, NY 10010, USA}

\author{Katherine L. Rhode}
\affiliation{Department of Astronomy, Indiana University, 727 East Third Street, Bloomington, IN 47405, USA}

\author{Martha P. Haynes}
\affiliation{Center for Astrophysics and Planetary Science, Space Sciences Building, Cornell University, Ithaca, NY 14853, USA}

\author{John L. Inoue}
\affiliation{Department of Physics and Astronomy, Macalester College, Saint Paul, MN 55105, USA}

\author{Joshua Marine}
\affiliation{Department of Physics and Astronomy, Macalester College, Saint Paul, MN 55105, USA}

\author{John.~J.~Salzer}
\affiliation{Department of Astronomy, Indiana University, 727 East Third Street, Bloomington, IN 47405, USA}

\author{Anjana K. Talluri}
\affiliation{University of Minnesota, Minnesota Institute for Astrophysics, School of Physics and Astronomy, 116 Church Street, S.E., Minneapolis, MN 55455, USA} 

\begin{abstract}
The ratio of baryonic-to-dark matter in present-day galaxies constrains galaxy formation theories and can be determined empirically via the baryonic Tully-Fisher relation (BTFR), which compares a galaxy’s baryonic mass (\mbary) to its maximum rotation velocity (\vmax). The BTFR is well-determined at \mbary $>10^8$ \msun, but poorly constrained at lower masses due to small samples and the challenges of measuring rotation velocities in this regime. For 25 galaxies with high-quality data and \mbary $\ltsimeq10^8$ \msun, we estimate \mbary\ from infrared and HI observations and \vmax\ from the \hi\ gas rotation. Many of the \vmax\ values are lower limits because the velocities are still rising at the edge of the detected \hi\ disks (\rmax); consequently, most of our sample has lower velocities than expected from extrapolations of the BTFR at higher masses. To estimate \vmax, we map each galaxy to a dark matter halo assuming density profiles with and without cores. In contrast to non-cored profiles, we find the cored profile rotation curves are still rising at \rmax\ values, similar to the data. When we compare the \vmax\ values derived from the cored density profiles to our \mbary\ measurements, we find a turndown of the BTFR at low masses that is consistent with $\Lambda$CDM predictions and implying baryon fractions of 1-10\% of the cosmic value. Although we are limited by the sample size and assumptions inherent in mapping measured rotational velocities to theoretical rotation curves, our results suggest that galaxy formation efficiency drops at masses below \mbary$\sim10^8$ \msun, corresponding to \mhalo$\sim10^{10}$ \msun.
\end{abstract} 


\section{Introduction}\label{sec:intro}
The baryonic Tully-Fisher relation (BTFR), which builds on the fundamental Tully-Fisher scaling relation \citep{Tully1977}, shows a strong correlation between the baryonic mass (\mbary) and the maximum rotational velocity (\vmax) of the gas in galaxies \citep[e.g., ][]{McGaugh2000, Bell2001, Gurovich2004, Geha2006, Begum2008a, Stark2009, McGaugh2012, Hall2012, Zaritsky2014, Lelli2016}. As \vmax\ traces the gravitational potential in a system and therefore provides a measure of the total halo mass assuming a dark matter (i.e., standard gravity)  paradigm, the BTFR  connects the baryon content of galaxies to their halo mass (\mhalo) and is thought to arise naturally from galaxy formation scenarios \citep[e.g.,][]{Dalcanton1997, Mo1998}.

The BTFR has a constant slope over the mass range \mbary\ $\sim10^8 - 10^{12}$ \msun\ corresponding to the velocity range \vmax\ $\sim50-300$ km s$^{-1}$ \citep[e.g.,][]{McGaugh2012, Lelli2016}. The constant power-law scaling implies the baryon content and rotation speed in galaxies scale with the total halo mass and virial velocity. This is somewhat unexpected given the variations in how baryons inhabit their halos, which depends on the complex, energetic, and stochastic processes involved in galaxy evolution, and the variations in halos properties, including the mass concentration and angular momentum distribution. 

What happens to the BTFR below \mbary\ $\sim10^8$ \msun\ is unclear, but of particular interest. In such low-mass galaxies, the energetics of baryon physics can increasingly overcome the shallow potential wells of the systems, with direct consequences for the fraction of baryons that are retained in a halo. The baryonic physics driving these changes can be internal to the galaxy, such as stellar feedback-driven winds \citep[e.g.,][]{Dekel1986, MacLow1999, Martin1999, McQuinn2019a}, or external, such as the metagalactic UV radiation field during the epoch of reionization that is expected to heat a galaxy's gas content and suppress additional gas accretion to a system \citep[e.g.,][]{Babul1992, Efstathiou1992, Hoeft2006, Okamoto2008, Faerman2013}. The combined impact is predicted to reduce the baryonic content at lower galaxy masses, which can also be thought of as a decrease in the efficiency of galaxy formation. 

If the baryon fraction drops as predicted at lower galaxy masses, this means the baryon-to-dark matter ratio is decreasing and  the slope of the BTFR should steepen. Empirically constraining whether this change occurs, the corresponding characteristic mass scale, and how sharply the BTFR slope changes can put limits on galaxy formation physics and resolve the discrepancy between the observed and predicted galaxy mass functions \citep[e.g.,][]{Sawala2015}.

It is important to note that some of the same processes that reduce the baryon content in galaxies can have other effects that complicate interpretation. Stellar feedback not only removes baryons, it can re-distribute baryons and dark matter within a host system, changing the shape of a galaxy's inner density profile. Briefly, feedback can drive gas out of a galaxy's center and this outward redistribution of gas can be dynamically important enough to alter the dark matter distribution \citep{Navarro1996}. This is seen in some hydrodynamical simulations where feedback alters the overall gravitational potential and transforms a steeply rising ``cuspy'' inner density profile into a flatter ``cored'' profile. Such changes may be temporary, with cusps reforming as gas cools after star formation activity decreases, and flows back into the center of a system, although predictions vary \citep[e.g.,][]{Governato2010, Chan2015, Pontzen2012, DiCintio2014a, DiCintio2014b,  Onorbe2015, Tollet2016, Benitez-Llambay2019}. The possible range in inner density profiles due to baryonic physics introduces challenges in interpreting the placement of individual low-mass galaxies on the BTFR; this can be at least partially overcome if the BTFR is populated with a large enough sample of dwarf galaxies and with measurements from high quality data.

Extending the BTFR below \mbary\ $\sim10^8$ \msun\ (or below \vmax\ $\sim50$ \kms) has proven challenging because of the relatively small sample size of galaxies available for study. Such very low-mass galaxies are difficult to detect due to their low optical luminosities and \hi\ fluxes, and not all detected low-mass galaxies have gas kinematics that are suitable tracers of the gravitational potential (i.e., gaseous disks with disordered motion, warped disks, etc.\ yield uncertain measurements of the bulk rotational motion). As a result, the low-mass end of the BTFR remains sparsely populated \citep[e.g.,][]{Lelli2019}. 

Here, we study the low-mass end of the BTFR using a sample of 25 isolated, gas-rich galaxies, the majority of which have \mbary\ between $10^7 - 10^8$ \msun. We build this large sample by combining low-mass galaxies recently catalogued from the Arecibo Legacy Fast ALFA survey \citep[ALFALFA;][]{Giovanelli2005, Haynes2018} with well-studied low-mass galaxies in the nearby universe. We use measurements of \mbary\ and gas rotational velocities to explore the placement of this larger sample on the BTFR. We then map the galaxies to dark matter halos to estimate \vmax, explore changes wrought by feedback on the measured gas rotational velocities, and determine how this impacts the low-mass end of the BTFR. 

An important distinction in this work is that our sample consists of galaxies that are gas-rich and isolated, which largely removes the impact of environment. There has been considerable attention in constraining the stellar mass-halo mass (SMHM) relation at low masses using gas-poor satellite galaxies around the Milky Way (MW), M~31, and MW analogs \citep[e.g.,][]{Jethwa2018, Drlica-Wagner2020, Nadler2020, Carlsten2020, Mao2020}. The SMHM relation provides constraints on how dark matter halos may be populated at low masses and, in many ways, is analogous to the BTFR. A notable difference is that the SMHM is derived from satellite galaxies. Thus, interpreting the SMHM is dependent on environmental effects, which introduce a set of challenges. For example, low mass galaxies evolving within the virial radius of a galaxy or galaxy group can lose a significant fraction of their mass \citep[e.g.,][]{Penarrubia2008, Putman2021}. High-resolution hydrodynamical solutions of low-mass galaxies have shown that the present day \mhalo\ values can be as much as two orders of magnitude lower than the peak halo mass for galaxies, with comparable reductions in stellar mass. The combined mass loss effects can add significant scatter and uncertainty in the low-mass end of the SMHM relation \citep{Munshi2021}. Yet, despite adding in complexities due to environment, the SMHM relation can be populated down to lower masses given the proximity of the MW and M~31 satellites than is possible for the BTFR that uses more distant gas-rich galaxies. Note that studies of satellites around MW analogs are impacted similarly by detection limits as the gas-rich field galaxies, but they provide a needed check on whether the MW and M~31 satellite populations are representative. Thus, the low-mass end of the BTFR and the SHMH relation complement each other in providing constraints on galaxy formation models. 

This manuscript is organized as follows. In Section~\ref{sec:samples}, we describe the sample selection and baryonic mass measurements in detail. In Section~\ref{sec:velocities}, we discuss the challenges inherent in measuring and interpreting rotational velocities in very low-mass galaxies, report \hi\ rotational velocities measured from position-velocity slices based on a new technique presented in \citet{McQuinn2021} with some modifications, and apply asymmetric drift corrections to the velocities. In Section~\ref{sec:btfr_empirical}, we place our empirical measurements on the BTFR. In Section~\ref{sec:dm} we estimate the maximum rotational velocities by mapping each galaxy to a dark matter halo in both a cuspy Einasto density profile and a cored Einasto density profile. In Section~\ref{sec:results}, we use these new values to re-examine the galaxy sample on the BTFR and the implications for the baryon fractions in low-mass galaxies. We also explore the SMHM relation for our sample and, as our sample is comprised of gas-rich galaxies, the {\em baryonic mass - halo mass} (BMHM) relation. In Section~\ref{sec:discuss}, we discuss our results and present a summary of our conclusions. Finally, in the Appendix, we provide an atlas of the velocity fields with our rotational velocity measurements, a comparison with previous rotational velocity measurements from the literature, and the subsequent fitting of our data to families of theoretical rotation curves. 

\section{Galaxy Sample and Properties}\label{sec:samples}
Our goal is to populate the low-mass end of the BTFR with a statistically-significant sample of dwarf galaxies. To that end, we have assembled a sample of galaxies with \mbary\ $\ltsimeq10^8$ \msun\ in the local universe with well-measured galaxy properties. We also include a small number of galaxies with \mbary\ $\gtsimeq10^8$ \msun\ to connect our sample with the range in masses included in previously published work on the BTFR. 

The galaxies were selected from surveys of low-mass systems in the nearby universe including the Survey of \hi\ in Extremely Low-mass Dwarfs \citep[SHIELD;][]{Cannon2011}, the VLA-ACS Nearby Galaxy Survey Treasury \citep[VLA-ANGST;][]{Ott2012}, and the Local Irregulars That Trace Luminosity Extremes, The \hi\ Nearby Galaxy Survey \citep[LITTLE THINGS;][]{Hunter2012}. SHIELD is an off-shoot of the ALFALFA \hi\ survey and includes a volume-complete sample of galaxies with \mhi\ $\ltsimeq 10^{7.2}$ \msun\ in the local universe. The overall aim of SHIELD is to characterize galaxy properties at the low-mass end of the \hi\ mass function. The VLA-ANGST program is a volume-complete survey of low-mass galaxies that lie outside the Local Group but within 4 Mpc. The LITTLE THINGS program includes an assortment of low-mass galaxies within the Local Volume with a variety of properties. Two galaxies, Ho~II (DDO~50) and DDO~154, were observed as part of The \hi\ Nearby Galaxy Survey \citep[THINGS;][]{Walter2008}, but are included here as part of the LITTLE THINGS sample.

\subsection{Selection Critiera}\label{sec:criteria}
Table~\ref{tab:properties} lists the sample of galaxies used in our analysis. From the SHIELD, VLA-ANGST, and LITTLE THINGS programs, we selected galaxies that have (i) accurate distances measured using the tip of the red giant branch (TRGB) method from HST optical imaging, (ii) \hi\ velocity fields that are conducive to measuring the bulk rotation of the gas based on interferometric observations of the 21-cm spectral line, and (iii) high inclination angles (i.e., $i>30^{\circ}$) which avoids uncertain inclination corrections from dominating the error budget on the \hi\ rotational velocity. The distance and inclination angle critieria are quantitative while the assessment of the velocity fields is subjective. Galaxies were rejected if adopting a single inclination correction was problematic, the \hi\ disks showed signs of interactions (which makes fitting the velocity fields problematic), or the \hi\ disks displayed strongly disordered motion. 

Our strict selection criteria limit the number of galaxies that can be used for our analysis, but is intended to yield a more robust result and a more straightforward interpretation. Rejecting galaxies based on these criteria does not constitute a bias for the current study, as these criteria are independent of the internal kinematics of the galaxies. The total number of galaxies selected from the combined surveys for our study is 25, including seven from SHIELD, twelve from VLA-ANGST, and six from LITTLE THINGS.

\begin{table*}
\begin{center}
\caption{Galaxy Sample and Measured Properties}
\label{tab:properties}
\end{center}
\setlength{\tabcolsep}{3.9pt}                          
\begin{center}
\vspace{-15pt}
\begin{tabular}{l | cc | lcc | lcc | c}
\hline 
\hline 
Galaxy          	& D    				& Ref. 	& $F_{3.6~\micron}$	& Ref.	& log            		& $S_{HI}$		& Ref.	& log            			& log           		 \\
                	& (Mpc)			&	&  (mJy)		& 	& (M$_*$/\msun)  	& (Jy \kms)		&	& (M$_{HI}$/\msun) 		& (\mbary/\msun)	\\
\hline                          
\multicolumn{10}{c}{SHIELD Galaxies}  \\                      
\hline
AGC110482 &  7.82$\pm$0.21 		& 1	& 1.14$\pm$0.30	& This work 	& 7.39$\pm0.12$ 	& $1.33\pm0.04$	& 8	&  7.28$\pm$0.03 		&  7.70$\pm0.06$	 \\
AGC111164 &  5.11$\pm$0.07 		& 1	& 0.41$\pm$0.19	& This work	& 6.57$\pm0.20$ 	& $0.65\pm0.04$	& 8	&  6.60$\pm$0.03 		&  6.96$\pm0.08$ \\
AGC229053 &  12.50$^{+0.26}_{-0.17}$ 	& 2	& 0.15$\pm$0.04	& This work	& 6.91$\pm0.12$ 	& $0.77\pm0.04$	& 8	&  7.45$\pm$0.03	 	&  7.66$\pm0.03$	  \\
AGC731921 &  11.51$\pm$0.29 		& 2	& 1.09$\pm$0.14	& This work	& 7.71$\pm0.12$ 	& $1.26\pm0.04$	& 8	&  7.59$\pm$0.03 		&  8.02$\pm0.06$	  \\
AGC739005 &  8.63$^{+0.18}_{-0.22}$ 	& 2	& 0.57$\pm$0.18	& This work	& 7.17$\pm0.14$ 	& $1.16\pm0.05$	& 8	&  7.31$\pm$0.03	 	&  7.62$\pm0.05$	   \\
AGC742601 &  7.00$\pm$0.18 		& 2	& 0.25$\pm$0.04	& This work	& 6.63$\pm0.12$ 	& $0.88\pm0.06$	& 8	&  7.01$\pm$0.04 		&  7.25$\pm0.04$	   \\
AGC749237 &  11.62$^{+0.20}_{-0.16}$ 	& 1	& 0.88$\pm$0.06	& This work	& 7.62$\pm0.12$ 	& $1.80\pm0.05$	& 8	&  7.76$\pm$0.02	 	&  8.07$\pm0.04$	 \\
\hline                          
\multicolumn{10}{c}{VLA-ANGST Galaxies}  \\
\hline
DDO99 	& 2.63$\pm$0.10 		& 3	& 8.27$\pm$1.12 	& 5	& 7.30$\pm$0.12	& 33.0$\pm$3.3 	& 9	& 7.73 $\pm$ 0.05 		& 7.96 $\pm$ 0.05	 \\
DDO125 	& 2.60$\pm$0.07 		& 3	& 24.1$\pm$3.3	& 5	& 7.76$\pm$0.12	&  21.7$\pm$2.2 	& 9	& 7.54 $\pm$ 0.05 		& 8.01 $\pm$ 0.07	 \\
DDO181 	& 3.12$\pm$0.06 		& 3	& 5.93$\pm$0.8 	& 5	& 7.31$\pm$0.12	& 12.2$\pm$1.2 	& 9	& 7.45 $\pm$ 0.05 		& 7.76 $\pm$ 0.05	\\
DDO183 	& 3.28$\pm$0.08 		& 3	& 6.09$\pm$0.83 	& 5	& 7.36$\pm$0.12	& 10.5$\pm$1.1 	& 9	& 7.42 $\pm$ 0.05 		& 7.77 $\pm$ 0.06	\\
NGC3109 	& 1.34$\pm$0.05 		& 3	& 302$\pm$41.0 	& 5	& 8.28$\pm$0.12	& 1110$\pm$90	& 10	& 8.67 $\pm$ 0.05 		& 8.91 $\pm$ 0.05	\\
NGC3741 	& 3.23$\pm$0.12 		& 3	& 4.77$\pm$0.65 	& 5	& 7.24$\pm$0.12	& 74.7$\pm$7.5 	& 9	& 8.26 $\pm$ 0.05 		& 8.42 $\pm$ 0.05	\\
SextansA 	& 1.44$\pm$0.06 		& 3	& 37.0$\pm$5.0 	& 5	& 7.43$\pm$0.12	& 190	 		& 11	& 7.97 $\pm$ 0.06 		& 8.18 $\pm$ 0.05	\\
SextansB 	& 1.43$\pm$0.02 		& 3	& 49.7$\pm$6.7 	& 6	& 7.55$\pm$0.12	& 112			& 11	& 7.73 $\pm$ 0.05 		& 8.03 $\pm$ 0.05	\\
UGC04483 	& 3.53$\pm$0.13 		& 3	& 1.91$\pm$0.26 	& 5	& 6.92$\pm$0.12	& 13.6		 	& 12	& 7.60 $\pm$ 0.03		& 7.79 $\pm$ 0.05	 \\
UGC08508 	& 2.64$\pm$0.10 		& 3	& 8.11$\pm$1.1 	& 5	& 7.30$\pm$0.12	& 18.3$\pm$1.8 	& 9	& 7.48 $\pm$ 0.05 		& 7.78 $\pm$ 0.05	 \\
UGC08833 	& 3.20$\pm$0.12 		& 3	& 2.45$\pm$0.33 	& 5	& 6.95$\pm$0.12	& 6.3$\pm$0.6 	& 9	& 7.18 $\pm$ 0.05 		& 7.46 $\pm$ 0.05	\\
UGCA292 	& 3.77$\pm$0.14 		& 3	& 1.43$\pm$0.2 	& 5	& 6.85$\pm$0.12	& 14.3		 	& 12	& 7.68 $\pm$ 0.05 		& 7.85 $\pm$ 0.05	\\
\hline                          
\multicolumn{10}{c}{LITTLE THINGS Galaxies}  \\ 
\hline
DDO53 	& 3.68$\pm$0.17	 	& 3	& 5.9		 	& 6	& 7.45$\pm$0.12	& 21.5$\pm$2.2 	& 9	& 7.84 $\pm$ 0.06 		& 8.08 $\pm$ 0.05	\\
DDO126 	& 4.97$\pm$0.23 		& 3	& 7.35$\pm$1.0 	& 6	& 7.81$\pm$0.12	& 28.5		 	& 11	& 8.22 $\pm$ 0.06 		& 8.45 $\pm$ 0.05	\\
DDO154 	& 4.04$\pm$0.07 		& 3	& 4.6			& 7	& 7.42$\pm$0.12	& 106		 	& 13	& 8.50 $\pm$ 0.02 		& 8.65 $\pm$ 0.02	\\
F564v3	& 8.83$\pm$0.41 		& 3	& 0.6		 	& 7	& 7.22$\pm$0.12	& 1.89		 	& 14	& 7.54 $\pm$ 0.06 		& 7.80 $\pm$ 0.05	\\
HoII		& 3.47$\pm$0.05 		& 3	& 77.5$\pm$9.8 	& 6	& 8.45$\pm$0.12	& 219.3	 	& 15	& 8.79$\pm$ 0.05 		& 9.06 $\pm$ 0.05	 \\
WLM 		& 0.97$\pm$0.03 		& 4	& 88.1$\pm$11.9	& 6	& 7.46$\pm$0.12	&  292$\pm$74 	& 16	& 7.81 $\pm$ 0.11 		& 8.06 $\pm$ 0.09	\\
\hline              
\end{tabular}
\end{center}
\tablecomments{Distance References: 1: \citet{McQuinn2014}; 2: \citet{McQuinn2021}; 3: \citet{Tully2013}; 4: \citet{McQuinn2017}. $F_{3.6\micron}$ References: 5: \citet{Ott2012}; 6: \citet{Dale2009}; 7: \citet{Lelli2016}.  $S_{HI}$ References: 8: \citet{Haynes2018}; 9: \citet{Begum2008b}; 10: \citet{Barnes2001}; 11: \citet{Huchtmeier1989}; 12: \citet{Huchtmeier2003}; 13: \citet{Hoffman2019}; 14: \citet{Honey2018}; 15: \citet{Walter2007}; 16: \citet{Barnes2004}. We adopt an uncertainty of 0.12 on all values of log(M$_*$/\msun) and a 10\% uncertainty on $S_{HI}$ measurements without reported uncertainties. \mbary $=$\mstar $+ 1.33 \times$\mhi.}
\end{table*}

\subsection{Previously Measured Galaxy Properties}\label{sec:data_previous}
For the majority of the galaxy properties, we use existing measurements in the literature in our analysis. The exceptions are stellar masses for the SHIELD galaxies, and the rotational velocities of the neutral hydrogen gas and the radius at which this velocity is measured for the full sample. The stellar masses for the SHIELD galaxies are determined using the same approach applied to the VLA-ANGST and LITTLE THINGS systems, namely by measuring the infrared fluxes from {\it Spitzer Space Telescope} imaging and assuming a mass-to-light ratio. The \hi\ rotational velocities and radii are determined using a new method, as discussed in Section~\ref{sec:pv}. In the following subsections, we provide a detailed description of measured galaxy properties, with values and appropriate references for the measurements listed in Tables~\ref{tab:properties} \& \ref{tab:velocities}. 

\subsubsection{Distances}
Distances ($D$) were determined from $HST$ imaging of the resolved stars using the TRGB method. Using high-quality distances is critical to accurately interpreting the BTFR as the \mbary\ depends on the square of the distance. This is particularly important for analysis at the low-mass end of the BTFR as these low-mass galaxies are preferentially detected at closer distances where velocity-based distance estimates are prone to larger uncertainties from higher relative peculiar velocities. The TRGB distances are from \citet{McQuinn2014, McQuinn2017, McQuinn2021} and from the CosmicFlows-3 program \citep{Tully2013}. 

\subsubsection{Stellar Masses}
Stellar masses (\mstar) were determined using 3.6$\micron$ fluxes ($F_{3.6~\micron}$) from the {\em Spitzer Space Telescope} imaging, assuming a mass-to-light ratio, and adopting the secure TRGB distances. We use integrated 3.6$\micron$ fluxes as a tracer of the stellar mass as older stellar populations dominate both the galaxy infrared luminosity and the stellar masses, thereby reducing the dependence of the stellar mass calculation on the star formation history of the galaxy \citep[e.g.,][]{Charlot1996, Madau1998, Bell2001, Li2007, Zhu2010}. Stellar masses based on mass-to-infrared light ratios are commonly used in calculating the baryonic masses of galaxies for BTFR analysis, including for the more massive galaxies from \citet{Lelli2019} that we present in our comparison analysis. For the mass-to-light ratio, we adopt a ratio of 0.5 in solar units, to be consistent with the value used in \cite{Lelli2019}. 

For the VLA-ANGST and LITTLE THINGS galaxies, we use existing measurements of 3.6$\micron$ fluxes, listed in Table~\ref{tab:properties} with the original source of the data.  For the SHIELD galaxies, we measured the 3.6$\micron$ fluxes from Spizter imaging obtained as part of the larger SHIELD program (PID 14040) in a multi-step process. First, we identified foreground stars and background galaxies around the SHIELD galaxies using the higher resolution $HST$ optical imaging. Second, we excised these sources from the images and interpolated over the regions that were removed. Third, we measured the flux within concentric annuli to create a curve of growth for each galaxy. We used the geometry (i.e., semi-major axis, ellipticity, and position angle) determined from the resolved stars \citep{McQuinn2014, McQuinn2021} as a starting point. We then fit isophotes to the outer regions of each galaxy, adjusting the geometry based on the best-fitted isophote. Using this final geometry, concentric ellipses were generated and the 3.6$\micron$ fluxes were measured within each annulus. Fourth, the median background flux was determined in an area off-source and used to background-subtract the fluxes in each annulus after adjusting for area. Finally, we determined the final extent of each galaxy based on the annulus where the surface brightness reached the median background surface brightness and calculated aperture corrections following the methodology in \citet{Dale2009}. We then applied a mass-to-light ratio of 0.5 to the aperture corrected flux contained inside this final outer ellipse. 

For the stellar mass measurements, we adopt an uncertainty of 0.12 dex in log space. This uncertainty is based on the range in mass-to-infrared light ratios at 3.6$\micron$ of 0.45$-$0.6 (i.e., a range of 0.15 which translates to an uncertainty of 33\% in stellar mass or 0.12 dex); it also corresponds to the upper end of the uncertainties quoted for various mass-to-light ratio calibrations at 3.6$\micron$ \citep[e.g.,][]{Zhu2010, Meidt2014, McGaugh2014, Schombert2019}, including the uncertainty on the mean value of the mass-to-light ratio determined from the BTFR \citep{Schombert2014}. An uncertainty in log stellar mass of 0.12 dex is larger than the range in stellar masses calculated from fluxes measured from the original 3.6$\micron$ images and the cleaned 3.6$\micron$ images for the SHIELD galaxies, providing a check that our adopted uncertainty is larger than the uncertainty that the somewhat subjective cleaning process has on the accuracy of the stellar masses. Two exceptions to this are AGC111164 and AGC739005 where the uncertainties in the  measured 3.6$\micron$ fluxes resulted in larger uncertainties on the calculated \mstar. 

\subsubsection{Gas Masses}
Gas masses ($M_{gas}$) are based on \hi\ fluxes ($S_{HI}$) measured from single-dish radio telescopes reported in a variety of sources, adopting the secure TRGB distances, and assuming an additional factor of 1.33 to account for the mass of helium. We do not include any additional corrections for molecular gas as observational constraints on the molecular gas fractions in the galaxy mass regime of interest are scarce. We expect the molecular gas to make a much smaller contribution to the total baryonic mass than either the stellar or atomic gas components. We also do not make any adjustments for the ionized gas component, which may be an important contributor to the overall baryon mass budget if the ionized gas in the circumgalactic medium of the galaxies is taken into account \citep{Gnedin2012, Wright2019}. 

\begin{table*}
\begin{center}
\caption{Gas Kinematics}
\label{tab:velocities}
\end{center}
\vspace{-15pt}
\begin{center}
\begin{tabular}{l | crrc |  cc | ccc | c}
\hline 
\hline 
		& Rising or	& Width 	&		&		&			&			&		&		&		&	   \\
Galaxy          	& Plateau	& of PV	& PA		& $i$		& \dpv			& \rmax		& \vpv		& \vrot		& \vcor	& log  \\
                	& (R/P)		& (\arcsec)	& ($^{\circ}$)	& ($^{\circ}$)	& (\arcsec)  		& (kpc)    		& (\kms)	& (\kms)	& (\kms)	& (\mdyn/\msun) \\
\hline           
\multicolumn{11}{c}{SHIELD Galaxies}  \\                      
\hline
AGC110482 & R		& 50		& 84		& $55\pm8$	& $32\pm5$		& $0.6\pm0.1$	& $16\pm1$	& $10\pm1$	& $13\pm2$	& 7.4 $\pm$ 0.1  \\
AGC111164 & R		& 50		& 326		& $50\pm8$	& $28\pm8$		& $0.4\pm0.1$	& $17\pm1$	& $11\pm2$	& $15\pm2$	& 7.2 $\pm$ 0.1 \\ 
AGC229053 & R		& 50		& 190		& $50\pm8$	& $60\pm11$		& $1.8\pm0.3$	& $31\pm1$	& $20\pm3$	& $22\pm3$	& 8.3 $\pm$ 0.1 \\
AGC731921 & R		& 50		& 110		& $40\pm8$	& $64\pm5$		& $1.8\pm0.2$	& $48\pm1$	& $ 37\pm6$	& $38\pm6$	& 8.8 $\pm$ 0.1 \\
AGC739005 & R		& 50		& 308		& $55\pm8$	& $36\pm9$		& $0.8\pm0.2$	& $33\pm1$	& $20\pm2$	& $22\pm2$	& 7.9 $\pm$ 0.1 \\
AGC742601 & R		& 50		& 266		& $45\pm8$	& $28\pm4$		& $0.5\pm0.1$	& $16\pm1$	& $11\pm2$	& $15\pm2$	& 7.4 $\pm$ 0.1 \\
AGC749237 & R		& 50		& 254		& $54\pm8$	& $35\pm3$		& $1.0\pm0.1$	& $39\pm1$	& $24\pm3$	& $26\pm3$	& 8.2 $\pm$ 0.1  \\
\hline              
\multicolumn{11}{c}{VLA-ANGST Galaxies}  \\                      
\hline                          
DDO99 	& R		& 210		& 7		& 90$\pm8$	& 339$\pm$2		& 2.2$\pm$0.1 	& 40$\pm$3	& 20$\pm$1 	& 22$\pm$1	& 8.4 $\pm$ 0.1 \\
DDO125 	& R		& 240		& $-55$	& 66$\pm8$	& 266$\pm$2		& 1.7$\pm$0.1 	& 24$\pm$1 	& 13$\pm$1 	& 16$\pm$2	& 8.0 $\pm$ 0.1  \\
DDO181 	& R		& 240		& 80		& 65$\pm8$	& 233$\pm$3		& 1.8$\pm$0.1 	& 36$\pm$4 	& 20$\pm$3 	& 22$\pm$3	& 8.3 $\pm$ 0.1 \\
DDO183 	& R		& 180		& 35		& 90$\pm8$	& 173$\pm$2		& 1.4$\pm$0.1 	& 25$\pm$4 	& 13$\pm$2 	& 16$\pm$2	& 7.9 $\pm$ 0.1  \\
NGC3109 	& P		& 390		& 183		& 83$\pm8$	& 1793$\pm$3	& 5.8$\pm$0.2 	&112$\pm$2	& 57$\pm$1 	& 58$\pm$1	& 9.7 $\pm$ 0.1  \\
NGC3741 	& P		& 150		& 106		& 67$\pm8$	& 410$\pm$2 	& 3.2$\pm$0.1 	& 53$\pm$2	& 29$\pm$2 	& 31$\pm$2	& 8.8 $\pm$ 0.1 \\
SextansA 	& P		& 240		& 55		& 38$\pm8$	& 366$\pm$3		& 1.3$\pm$0.1 	& 38$\pm$1 	& 31$\pm$6 	& 32$\pm$6	& 8.5 $\pm$ 0.1 \\
SextansB 	& R		& 378		& 45		& 53$\pm8$	& 503$\pm$4 	& 1.7$\pm$0.1 	& 53$\pm$1	& 33$\pm$4 	& 34$\pm$4	& 8.7 $\pm$ 0.1 \\
UGC04483 	& R		& 60		& 184		& 64$\pm8$	& 77$\pm$2		& 0.7$\pm$0.1 	& 27$\pm$1 	& 15$\pm$1 	& 18$\pm$2	& 7.7 $\pm$ 0.1 \\
UGC08508 	& R		& 210		& 288 		& 63$\pm8$	& 182$\pm$3		& 1.2$\pm$0.1 	& 40$\pm$1 	& 22$\pm$2 	& 24$\pm$2	& 8.2 $\pm$ 0.1 \\
UGC08833 	& R		& 90		& 236		& 30$\pm8$	& 101$\pm$4		& 0.8$\pm$0.1 	& 20$\pm$1 	& 20$\pm$5 	& 22$\pm$5	& 7.9 $\pm$ 0.1  \\
UGCA292 	& R		& 138		& 175		& 52$\pm8$	& 108$\pm$3		& 1.0$\pm$0.1 	& 17$\pm$1	& 11$\pm$1 	& 15$\pm$2	& 7.7 $\pm$ 0.1  \\
\hline
\multicolumn{11}{c}{LITTLE THINGS Galaxies}  \\                      
\hline                          
DDO53 	& R		& 138		& 131		& 33$\pm8$	& 111$\pm$2 	& 1.0$\pm$0.1 	& 20$\pm$4	& 19$\pm$5 	& 21$\pm$5	& 8.0 $\pm$ 0.1  \\
DDO126 	& R		& 98		& 120		& 59$\pm8$	& 255$\pm$2 	& 3.1$\pm$0.1 	& 59$\pm$4	& 35$\pm$4	& 36$\pm$4	& 9.0 $\pm$ 0.1 \\
DDO154 	& P		& 134		& 220		& 49$\pm8$	& 572$\pm$4 	& 5.6$\pm$0.1 	& 81$\pm$2	& 54$\pm$7 	& 55$\pm$7	& 9.6 $\pm$ 0.1 \\
F564v3	& R		& 57		& 30		& 51$\pm8$	& 75$\pm$4	 	& 1.6$\pm$0.1 	& 39$\pm$2	& 25$\pm$3 	& 27$\pm$3	& 8.4 $\pm$ 0.1   \\
HoII		& P		& 146		& 175		& 34$\pm8$	& 528$\pm$2 	& 4.4$\pm$0.1 	& 48$\pm$1	& 42$\pm$9 	& 43$\pm$9	& 9.3 $\pm$ 0.1 \\
WLM 		& P		& 480		& 175		& 90$\pm8$	& 965$\pm$4 	& 2.3$\pm$0.1 	& 55$\pm$6	& 27$\pm$3 	& 29$\pm$3	& 8.6 $\pm$ 0.1 \\
\hline              
\end{tabular}
\end{center}
\tablecomments{The width and position angle (PA) are of the PV slices used to measure the bulk rotational motion of the \hi\ gas. \dpv\ and \vpv\ provide measures of the spatial extent and range in velocity from the PV slices; these values are converted into \rmax\ and \vrot\ after taking into account the distance to a galaxy and its inclination angle. Inclination angles with uncertainties for the SHIELD galaxies are from \citet{McNichols2016} and \citet{McQuinn2021}. Inclination angles for the VLA-ANGST and LITTLE THINGS galaxies are from \citet{Karachentsev2013} with adopted uncertainties of $8^{\circ}$. All inclination angles were derived from optical imaging; see text for details. \vcor\ represents the asymmetric drift corrected rotational velocity. \mdyn\ is based on \vcor\ and \rmax. Also noted is whether the velocity field appears to plateau, suggesting the data is reaching the flat part of the rotation curve, or is still rising.}
\end{table*}

\subsubsection{Inclination Angles}
Inclination angles ($i$) are based on measurements of the major and minor axes measured from optical imaging of the stars and correcting for the expected aspect ratio of dwarf irregular galaxies \citep[e.g.,][]{Staveley-Smith1992, Johnson2017}.  While a few galaxies have inclination angles determined from the \hi\ disks available in the literature, many do not. Thus, for uniformity, we use the optically-based inclination angles. However, as a measure of the inclination angle uncertainty, we consider the difference between the optically-based inclination angles and the inclination angles reported from rotation curve fitting of the \hi\ data for LITTLE THINGS galaxies from \citet{Iorio2017}. We find the mean difference in the inclination angles to be $8^{\circ}$, which is larger than the $5^{\circ}$ in uncertainty reported on the optically-based inclination angles for the SHIELD galaxies \citep{McNichols2016, McQuinn2021}\footnote{Uncertainties on the optically-based inclination angles of the VLA-ANGST and LITTLE THINGS galaxies were not reported \citep{Karachentsev2013}}. Thus, while uncertainties on inclination angles are generally quoted to be much smaller, we conservatively adopt a $8^{\circ}$ uncertainty and propagate this to the uncertainties in the corrected velocities. We apply this value to the full sample to ensure the uncertainties will be comparable, but note that this may not be representative of all galaxies. As stated above, we mitigate uncertainties associated with projection effects by limiting our sample to galaxies with $i\geq30^{\circ}$. Our minimum value of $i\geq30^{\circ}$ is what is also applied to the comparison sample of massive galaxies we use from \citet{Lelli2019}, but slightly lower than the value of $i\geq40^{\circ}$ adopted in the study by \citet{Read2017}. 
 
\section{Gas Rotational Velocities}\label{sec:velocities}
\subsection{Background}
The BTFR uses empirical measurements of the gas rotational velocity to connect the baryon content in galaxies to the full gravitational potential. Thus, ideally what is needed are well-determined, maximum rotational velocities in galaxies where rotation dominates the bulk motion of the gas (i.e., turbulent motions make a smaller contribution). Such velocities can be derived via kinematic modelling from \hi\ observations with multiple resolution elements across a galaxy whose \hi\ velocity fields show clear signatures of ordered rotation out to radii that are past the flat part of the rotation curve. For low-mass galaxies, achieving these conditions is the exception rather than the rule. 

The inherent difficulties in connecting observed gas velocity fields to the gravitational potential in low-mass galaxies are endemic to studies of the low-mass end of the BTFR and have introduced pervasive uncertainties in the literature. For example, eight galaxies sample the mass regime \mbary\ $<10^8$ \msun\ from the Faint Irregular Galaxy GMRT Survey \citep[FIGGS;][]{Begum2008a, Begum2008b}, but the velocity fields of many of the galaxies have never been published nor have the reduced observations been made available. It is difficult to assess whether the galaxies have ordered gas rotation or if the reported velocities for these systems reach the flat part of the rotation curve, but it is probably not justifiable to assume this is the case, given how common disordered motions are in dwarf galaxies and the small galaxy sizes. There are five galaxies with \mbary\ $<10^8$ \msun\ from LITTLE THINGS, but four systems have rotational velocities measured on the rising part of the rotation curve \citep{Iorio2017}. An earlier study of the gas kinematics in SHIELD galaxies placed twelve galaxies, including three systems in the present analysis, on the BTFR \citep{McNichols2016}, but caution is advised in the interpretation of these points as most of the measured velocities are lower limits for maximum rotational velocity. Similarly, the very low-mass galaxy Leo~P \citep[\mstar $=6\times10^5$ \msun;][]{McQuinn2015b}, also discovered in the ALFALFA survey \citep{Giovanelli2013}, has the requisite data on the gas kinematics, but the rotational velocity for Leo~P is known to be a lower limit for \vmax\ of the galaxy \citep{Bernstein-Cooper2014}. 

Despite the uncertainties in the velocity measurements, combinations of these galaxies have been previously placed on the BTFR. Collectively, they appear to agree with an extrapolation from more massive galaxies and have been used as evidence of a continuation of the BTFR slope at low masses \citep[e.g.,][]{McGaugh2012, Papastergis2015, Bernstein-Cooper2014, Iorio2017}. The challenges are often acknowledged, particularly when velocities measured on the rising part of the rotation curve are used, and sometimes the galaxies are presented in analysis with a separate notation, but the systems are still included in the interpretation of the lower mass end of the BTFR. Thus, while existing measurements of low-mass galaxies are often used as evidence for a continuation of the BTFR below $10^8$ \msun\ without any change in slope, we argue that they provide far more limited constraints than such a conclusion demands.

In this section, we discuss the steps we take to mitigate the uncertainties related to using rotational velocities as tracers of mass in low-mass galaxies, which has implications for our downstream analysis. We also discuss the shortcomings of the analysis. We then present measurements of the gas rotational velocities in our sample, and, finally, we apply asymmetric drift corrections to our measured velocities to account for the contribution of turbulent, random motion of the gas to the gravitational support of the galaxies.

\begin{figure*}
\begin{center}
\includegraphics[width=0.48\textwidth]{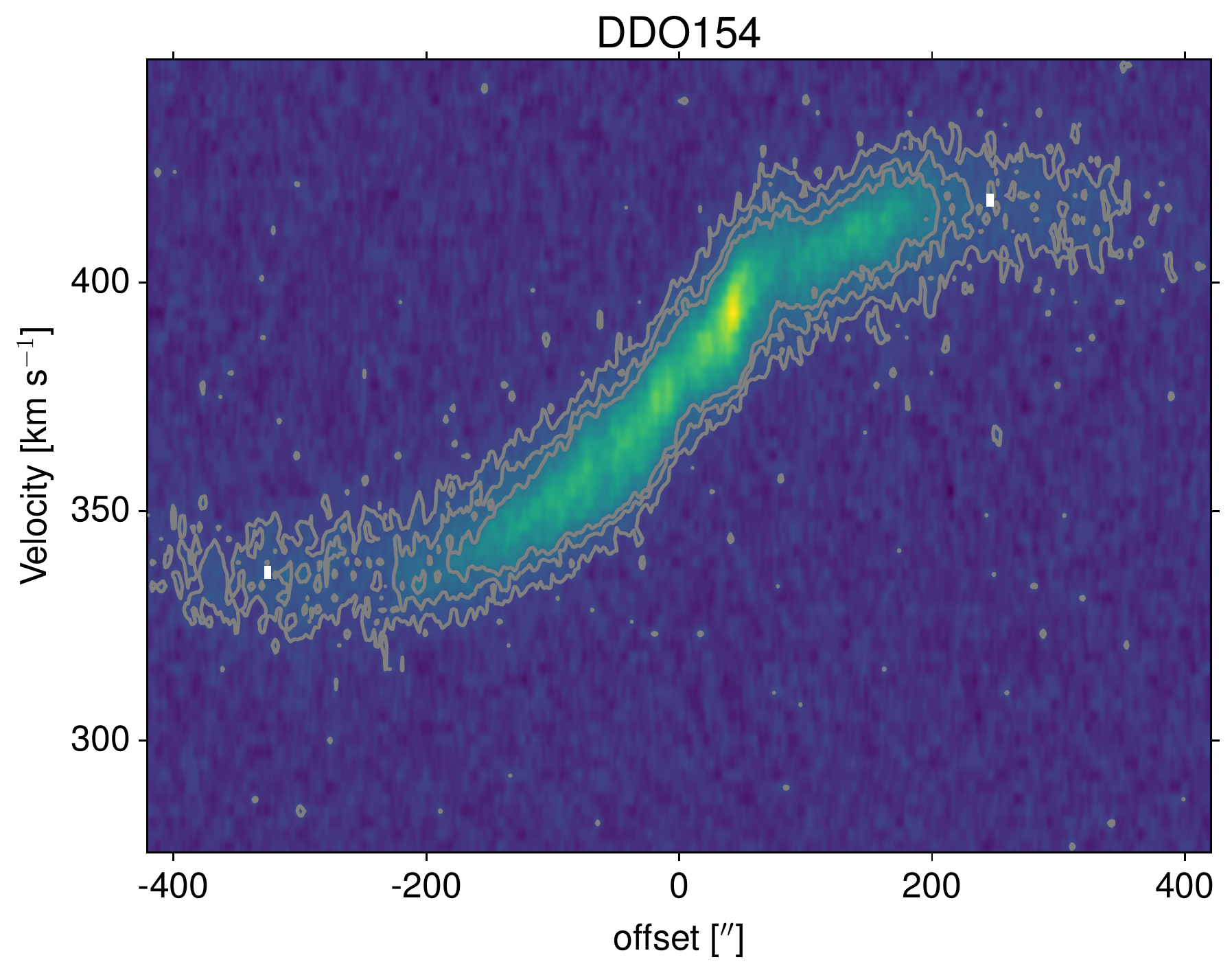}
\includegraphics[width=0.48\textwidth]{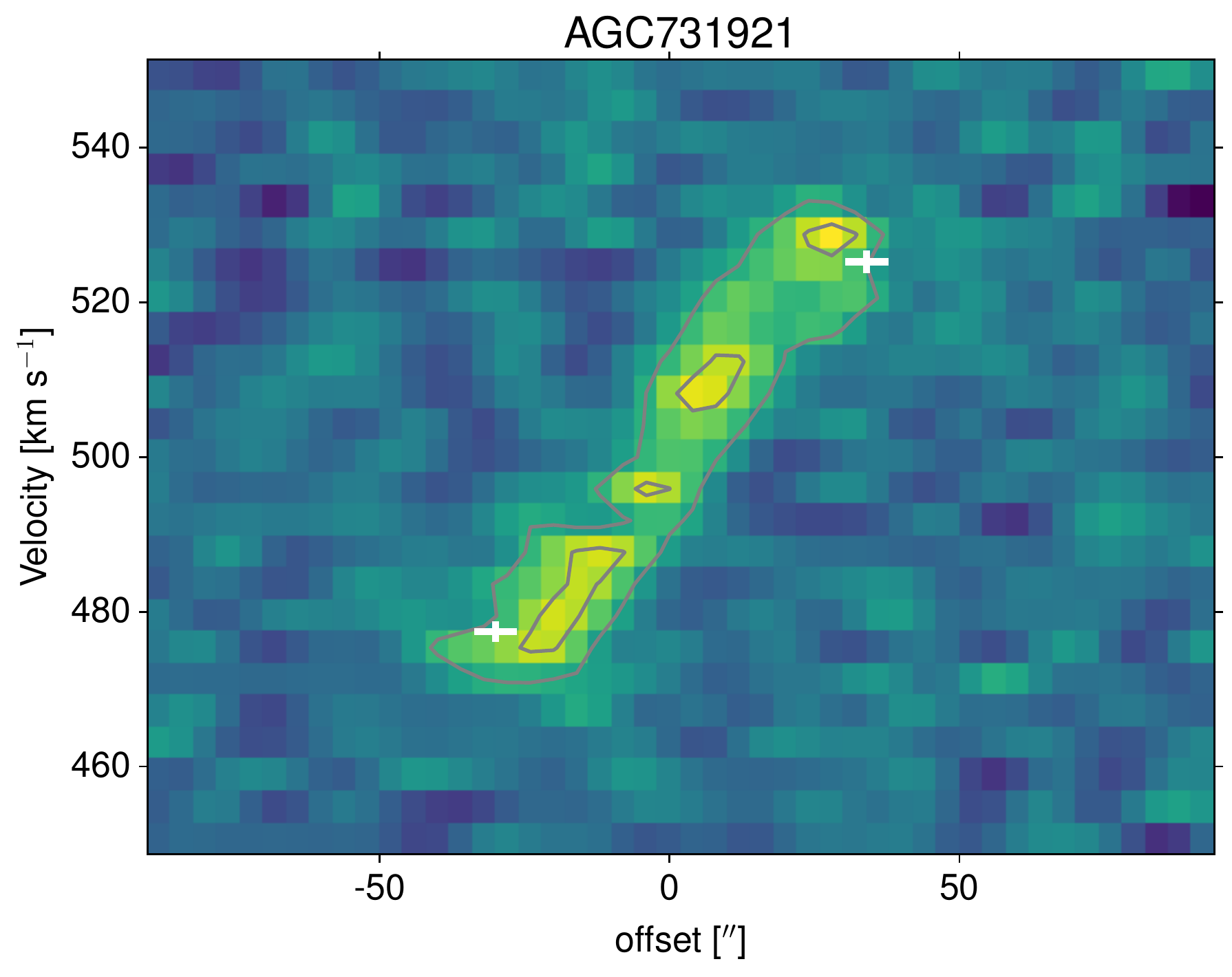}
\end{center}
\caption{Examples PV diagrams for a galaxy with a velocity field that plateaus (left; DDO154) and one that is still rising (right; AGC731921). \hi\ contours are plotted at 3, 6, and, for DDO154, 9$\times$ the rms value. The location of the best-fitting \vpv\ and \dpv\ values and their uncertainties are shown in white. An atlas of PV diagrams for the rest of the sample is presented in Appendix~\ref{app:pv}.}
\label{fig:pv_examples}
\end{figure*}

\subsection{Mitigating the Uncertainties in Measuring Rotational Velocities in Low-mass Galaxies}\label{sec:mitigate}
We briefly describe known challenges associated with using the measured gas kinematics in low-mass galaxies as a tracer of the total galaxy mass, and how we address these issues. First, the velocity fields of star-forming galaxies contain disordered and turbulent motions and, in low-mass galaxies, the magnitudes of these can be comparable to the amplitude of the rotation. This is mitigated by restricting our analysis to galaxies that show more ordered rotation and avoiding systems with disordered velocity fields or signs of interaction (see Section \ref{sec:samples}). Combining new results from the low-mass galaxies in the SHIELD program with existing data on VLA-ANGST and LITTLE THINGS galaxies, we have now achieved a statistically-significant sample of low-mass galaxies with velocity fields that show predominantly ordered motion. 

Second, the detected \hi\ extent in very low-mass galaxies can be limited and may be covered by only a few resolution elements in the observations. Thus, the \hi\ velocity fields are often unsuitable for the kinematic modelling needed to create resolved rotation curves. For example, in a previous study of twelve SHIELD galaxies, rotation curve analysis was undertaken on only five systems and with limited success \citep[][Figure~13]{McNichols2016}. As a substitute, position-velocity (PV) diagrams are used to estimate \vrot\ in low-mass galaxies, where half the maximum range of velocities associated with the galaxy is taken as a measure of \vrot\ after correcting for inclination. 

Previous analysis of PV diagrams define the ``maximum range of velocities'' by eye \citep{McNichols2016}. Recently, a new approach to measuring the velocity range from PV diagrams was introduced for the SHIELD galaxies which entails a more robust, re-produceable method for measuring the velocity and spatial extent of \hi,  including uncertainties \citep{McQuinn2021}. We adopt a modification of this approach for measuring the gas rotational velocities in our sample and discuss this in detail in Section~\ref{sec:pv} with an atlas of the PV diagrams presented in Appendix~\ref{app:pv}.

Third, the spatial extent of the detected \hi\ is modest because the background UV radiation field truncates low column density \hi\ \citep[e.g.,][]{Maloney1993}. As a result, the velocities measured at the outermost \hi\ points lie in the inner regions of the galaxy halos and may still be on the rising part of the rotation curves. Whether or not \vrot\ is approaching a maximum can be determined by examining the velocity fields and PV diagrams from spatially resolved \hi\ kinematics. If the velocity reaches a plateau in the outer regions, this suggests the measurements have reached the flat part of the rotation curve with the measured velocity approaching \vmax. For example, the left panel in Figure~\ref{fig:pv_examples} presents the PV diagram for DDO~154 where the velocity appears to plateau or flatten. In contrast, the right panel presents the PV diagram for AGC~731921 where the velocity is still increasing at the last detected point. As expected, DDO~154 is more massive (log(\mbary/\msun) of 8.65$\pm$0.02 vs.\ 7.99$\pm$0.05), closer (4.04$\pm$0.07 vs.\ 11.51$\pm$0.29 Mpc), and the detected \hi\ disk reaches significantly farther (5.6$\pm$0.1 vs.\ 1.8$\pm$0.2 kpc) than AGC~731921, which is typical of the galaxies with measured rotational velocities that plateau. 

Based on inspection of the PV diagrams, we find the velocities are still rising in 19 and plateau in 6 galaxies in our sample; we make note of this in Table~\ref{tab:velocities}. We use these empirical measurements to place the galaxies on the BTFR, recognizing that the majority of the velocities are lower limits. 

We further address this issue by mapping the measured velocities and radial extent of the detected \hi\ disks to theoretical rotation curves generated using different dark matter density profiles (see Section~\ref{sec:dm}). In this way, we are able to estimate \vmax\ under different assumptions and explore where these estimates place the galaxies on the BTFR. 

The modest extent of the detected \hi\ disks has a secondary impact on our analysis as the inner rotation curves of low-mass galaxies are non-uniform \citep[e.g.,][]{Oman2015, Santos-Santos2018}. While likely not the only factor driving the diversity observed in inner rotation curves, the dynamic, and often stochastic, star formation processes in low-mass galaxies can result in fluctuations in the gas velocities that are not easily quantified. For galaxies with extended rotation curves, the differences in the inner regions of galaxies do not impact the interpretation as strongly, as \vmax\ should be relatively stable. This issue is not readily mitigated when studying an individual galaxy, but the impact of the non-uniformity of velocities at smaller radii might be minimized if a large enough sample is studied. The present work includes 25 galaxies which improves statistics in this regard. Future work with a larger sample, including the addition of galaxies meeting our selection criteria from the full SHIELD sample with a sub-sample suitable for rotation curve fitting, may help to define the mean relation between \vmax\ and \mbary\ and improve interpretations. 

Despite these caveats, without the ability to measure gas velocities in the flat part of rotation curves in low-mass galaxies, mapping empirical data to different models with plausible star formation and feedback prescriptions offers a way forward to examine the low-mass end of the BTFR in a $\Lambda$CDM context.

\subsection{Rotational Velocities Measured from PV Diagrams}\label{sec:pv}
As mentioned above, the \hi\ velocity fields in very low-mass galaxies are often not suitable for kinematic modelling due to the limited spatial extent of their detected \hi\ disks, the fact that their rotational velocity can be comparable to the velocity dispersion of the \hi, and/or their somewhat irregular gas motions. However, the bulk rotational motion of the gas can be estimated using position-velocity (PV) slices from the datacubes. \citet{McQuinn2021} presented a new method to robustly measure the maximum velocity range of the \hi\ and its spatial extent from PV slices, with uncertainties, for the SHIELD galaxies. Here, we modify this approach slightly and apply it uniformly to VLA \hi\ datacubes for the VLA-ANGST and LITTLE THINGS galaxies to determine the \hi\ rotational velocity and related radius for our full sample.

PV slices through the datacubes were created using position angles (PA) empirically determined for optimal extraction of velocity information. Where available, we started with previously reported PA values. We then iterated to determine the angle which best represented the velocity gradient in each galaxy. For the SHIELD galaxies, the widths of the PV slices were chosen to be greater than the major axis beam size, ensuring that the slices are representative of the bulk projected motion of the gas. Specifically, a PV width of 50\arcsec\ was uniformly used, which is wide enough for the PV slice to include all detected flux from the SHIELD galaxies. The VLA-ANGST and LITTLE THINGS galaxies typically have higher linear resolution due to their closer distances than the SHIELD galaxies. Thus, for consistency, for the VLA-ANGST and LITTLE THINGS galaxies we adopted a width that is approximately equal to the physical scale of 50\arcsec\ at a distance of 10 Mpc (i.e., 2.4 kpc), corresponding to one of the farthest distances for the SHIELD sample. The PAs and PV widths are listed in Table~\ref{tab:velocities}.

\begin{figure}
\includegraphics[width=0.45\textwidth]{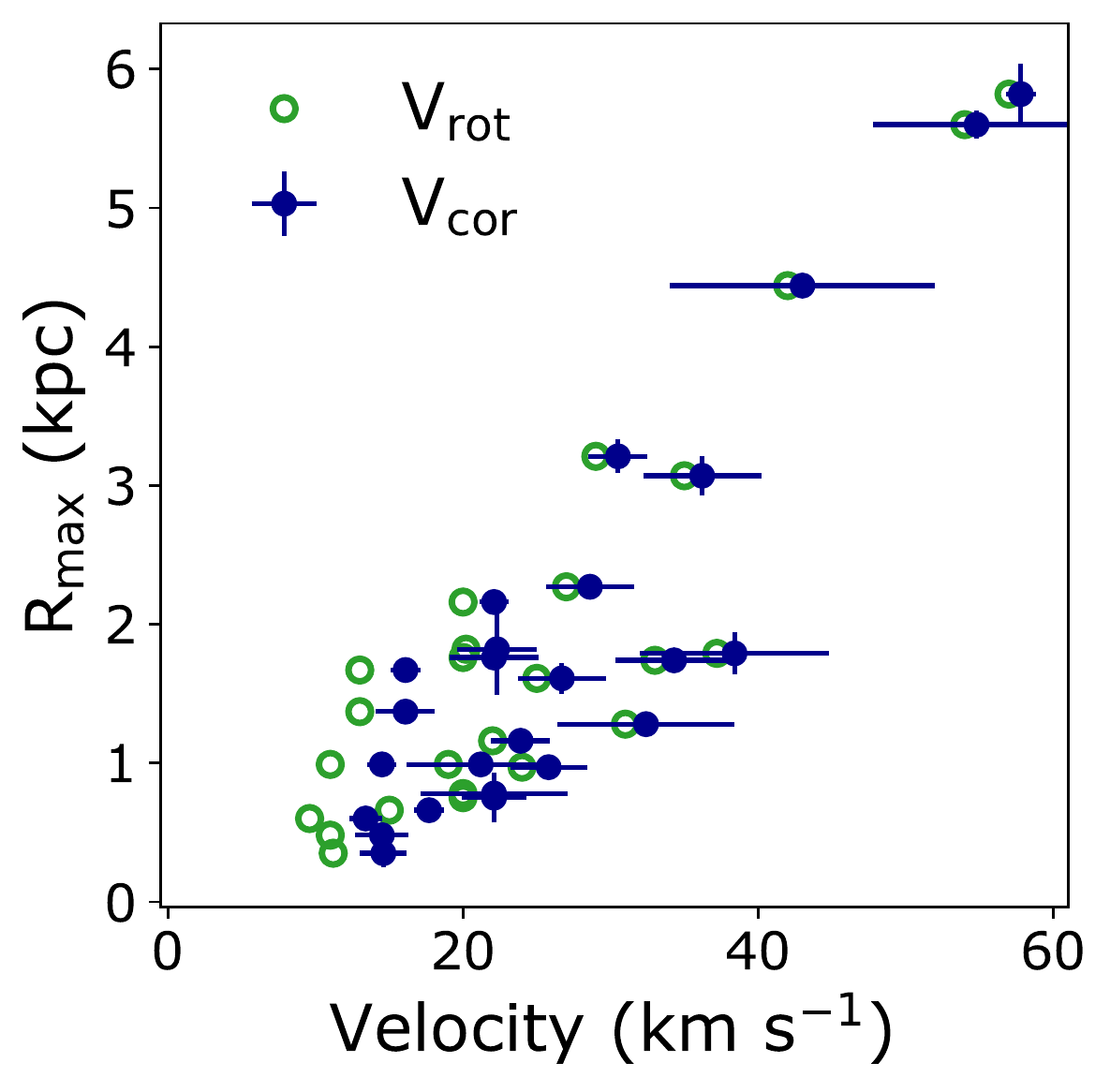}
\caption{Rotational velocities (\vrot; open green circles) and asymmetric drift corrected velocities (\vcor, filled blue circles) compared with the maximum physical radius of the \hi\ (\rmax) for the SHIELD, VLA-ANGST, and LITTLE THINGS galaxies. The asymmetric drift corrections have the largest relative impact on galaxies with the slowest measured rotational velocities and little impact on galaxies with \vrot\ $\gtsimeq 30$ \kms.}
\label{fig:vel_rmax}
\end{figure}

\citet{McQuinn2021} measured the velocity and radius of the \hi\ from the PV diagrams based on slicing the PV diagrams in orthogonal bins. The velocity at each spatial bin was found by fitting a Gaussian function to the velocity profile extracted from the spatial axis. The difference between the center velocities of the Gaussian functions with the minimum and maximum values (which did not always correspond to the maximum extent of the fitting) was then taken as the maximum range of velocities. Similarly, the spatial extent was determined by fitting Gaussian functions to the profile extracted along the velocity axis. The difference between the center offset values of the Gaussian profiles with the minimum and maximum offset values was taken as the diameter of the gas. For velocity fields that plateau, the fitting along the velocity axis will return a spatial offset that is not as meaningful as there are many spatial offsets for the same velocity. 

\begin{figure*}
\includegraphics[width=0.95\textwidth]{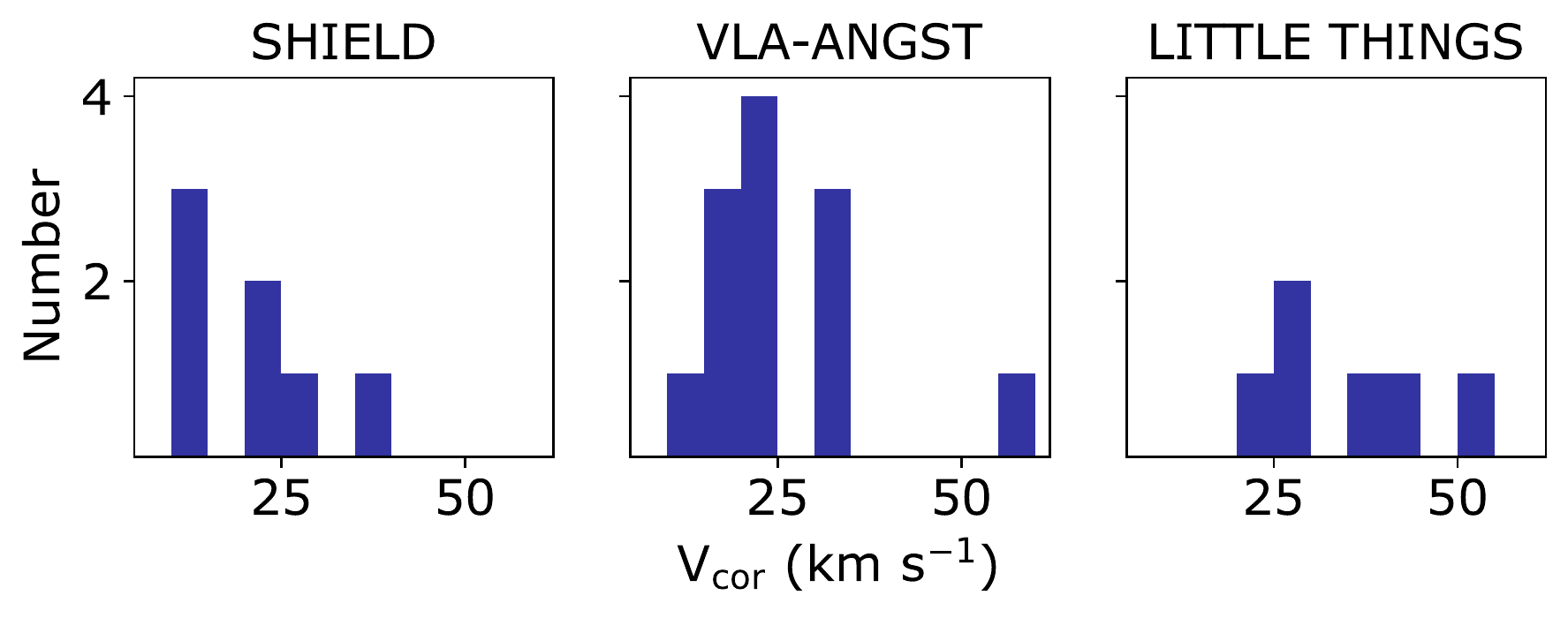}
\caption{Histograms of the asymmetric drift corrected velocities, \vcor, from each survey. SHIELD includes the galaxies with the slowest rotational measurements in our analysis while LITTLE THINGS includes galaxies that sample a slightly higher range in rotational velocities.}
\label{fig:vasd}
\end{figure*}

We have adapted this method for broader applicability to flat rotation curves\footnote{Naturally, galaxies with flat rotation curves are more amenable to kinematic modelling, but this approach is driven by the desire to treat all data consistently.} by determining the furthest spatial extent and then robustly fitting a Gaussian function to the velocity profiles in the corresponding spatial bin. In order to the determine the spatial extent, Gaussian functions were fit to the velocity profile of every spatial bin, and, in order to avoid bias based on the binning scheme, the starting location of the binning was iterated over to include all possible combinations. In order for a fit to the velocity profile to be accepted, the peak of the Gaussian profile had to reach a user-specified value above the rms. We fine-tuned the peak rms cut-off for each galaxy; while a non-standardized approach, this provided the most robust fits to the data which were obtained with different flux limits and spatial resolution. We designate the maximum range in velocities as \vpv, and the maximum spatial offset in arcsec as \dpv; the $\pm2\sigma$ error on the offsets was defined as the beam size or the bin size, whichever was larger. We also allowed for the possibility of a velocity profile composed of two Gaussian functions as it is possible for gas of two velocity components to be along the same line-of-sight (e.g., extraplanar gas). A double Gaussian profile was required to have a lower Bayes information criterion than a single Gaussian profile and at least one component had to have a peak above the user-specified rms limit. The final measured \dpv\ and \vpv\ values are listed in Table~\ref{tab:velocities}. We present a comparison of the rotational velocities estimated from PV diagrams with values derived from rotation curve fitting in Appendix~\ref{app:pv}.
 
The PV slices shown in Figure~\ref{fig:pv_examples} provide examples of \vpv\ and \dpv\ determinations for DDO~154 from the VLA-ANGST datacube and for AGC~731921 from the SHIELD sample. Appendix~\ref{app:pv} includes the PV slices for the remainder of the sample, marked with our measured maximum ranges in velocity, \vpv, and diameter, \dpv.

The de-projected rotational velocity, \vrot, is determined by taking half the value of \vpv\ and correcting for inclination (\vrot $=\frac{1}{2}$\vpv/$sin~i$); uncertainties in \vrot\ include uncertainties in the inclination angles. Finally, the physical radius at which the velocity is measured, \rmax, is calculated by taking half the angular diameter \dpv\ and adopting the distance to each galaxy in Table~\ref{tab:properties} (\rmax $=\frac{1}{2}$\dpv\ $\cdot~Distance$). All velocity and spatial measurements are listed in Table~\ref{tab:velocities}. The measured \vpv\ for the SHIELD galaxies are consistent with the values reported in \citet{McQuinn2021}, although the spatial offsets are slightly larger. Our modifications to the method are able to measure \vpv\ in the flat part of rotation curves, which is important for some of the VLA-ANGST and LITTLE THINGS galaxies; thus we use these updated values for SHIELD in our analysis to be consistent in our measurements across the sample.

\subsection{Asymmetric Drift Corrections to the Velocities}\label{sec:asd}
Gas rotational velocities are a measure of the centripetal support against gravity. Yet, the motion of gas in galaxies is also comprised of random motions that deviate from circular rotation and act as pressure support against gravity. In low-mass galaxies that have slower rotation speeds, the asymmetric deviations from circular motion can make an appreciable contribution to support the galaxy against gravity \citep[e.g.,][]{Tully1978}. Thus, both circular and turbulent motion must be accounted for when using motion as a tracer of the total gravitational potential of a galaxy.

The turbulent motion of the gas, which requires an ``asymmetric drift'' correction \citep{Oort1965}, is often quantified using the gas velocity dispersion after taking into account the gas surface density as a function of radius \citep[e.g.,][]{Skillman1987}. For our sample, particularly for the SHIELD galaxies, the \hi\ is detected in only a few channels and the measured dispersion may not be representative of the whole galaxy. Therefore, we assume a constant value for the gas velocity dispersion of $\sigma_{asd} = 8\pm2$ \kms, based on previous measurements of the \hi\ velocity dispersion in low-mass galaxies, including some of the VLA-ANGST galaxies used in our analysis \citep{deBlok2006, Swaters2009, Warren2012, Stilp2013}. Note that this asymmetric drift correction is still an approximation to account for pressure support; gradients in gas pressure may also contribute to supporting against gravity at small radii \citep{Valenzueula2007}.

The assumed  $\sigma_{asd}=8\pm2$ \kms\ is applied as an asymmetric drift correction to \vrot\ using a simple, consistent approach for all galaxies. Our new method for this correction is predicated on the assumption that both the circular and turbulent motion are well-represented by Gaussian profiles and therefore the associated dispersion or Gaussian standard deviation can be added in quadrature; once combined, this wider Gaussian profile then includes both circular and turbulent motion. This approach is valid as a Gaussian shape is a good approximation of the 21-cm spectral line profiles of the galaxies, particularly for the SHIELD systems. Specifically, we first assume \vrot\ is $\frac{1}{2}$ the full-width at half-maximum (FWHM) of a Gaussian profile whose standard deviation can be calculated using the relation:

\begin{equation}
\sigma_{rot} = \frac{2}{2.355} \cdot V_{rot}
\label{eq:asd1}
\end{equation}

\noindent We then assume $\sigma_{asd}$ is well-representative of a Gaussian standard deviation and add this dispersion in quadrature with $\sigma_{rot}$:

\begin{equation}
\sigma_{cor}^2 = \sigma_{rot}^2 + \sigma_{asd}^2 \\
\label{eq:asd2}
\end{equation}

\noindent where $\sigma_{cor}$ is the corrected term that now includes both bulk rotational and asymmetric motion. Finally, the asymmetric drift corrected velocity, \vcor, is calculated using the relationship between standard deviation and the FWHM, again assuming a Gaussian form:

\begin{equation}
V_{cor} = \frac{1}{2} \cdot 2.355 \cdot \sigma_{cor}
\label{eq:asd3}
\end{equation}

\noindent Values for \vcor\ are listed in Table~\ref{tab:velocities}; uncertainties include the measured uncertainties in \vrot\ from \vpv, the uncertainties in the inclination corrections, and the assumed uncertainty of 2 \kms\ for $\sigma_{asd}$. 

Figure~\ref{fig:vel_rmax} compares the measured rotational velocities, \vrot\ (light green open circles), and the velocities corrected for asymmetric drift, \vcor\ (dark blue filled circles), with the physical radius, \rmax, at which it was measured. For clarity, we only show uncertainties on \vcor; uncertainties on \vrot\ are of the same order. The shift in velocities due to the asymmetric drift correction is clearly seen at lower velocities but has little impact for velocities $\gtsimeq 30$ \kms. Figure~\ref{fig:vel_rmax} also demonstrates that larger velocities are measured at larger radial distances in galaxies. For the slowest rotating galaxies, predominantly made up of the SHIELD galaxies, a number of the velocities are measured within a 1 kpc radius of the center of the galaxy. 

Figure~\ref{fig:vasd} shows the distribution of \vcor\ for the galaxies from each survey. SHIELD includes the slowest rotating systems. VLA-ANGST and LITTLE THINGS include galaxies which overlap with the higher velocity range of the SHIELD systems. VLA-ANGST and LITTLE THINGS also include one galaxy each with a higher rotational velocity (i.e., \vcor\ $\sim50$ \kms) that overlaps with well-studied more massive galaxies.

\begin{figure}
\begin{center}
\includegraphics[width=0.49\textwidth]{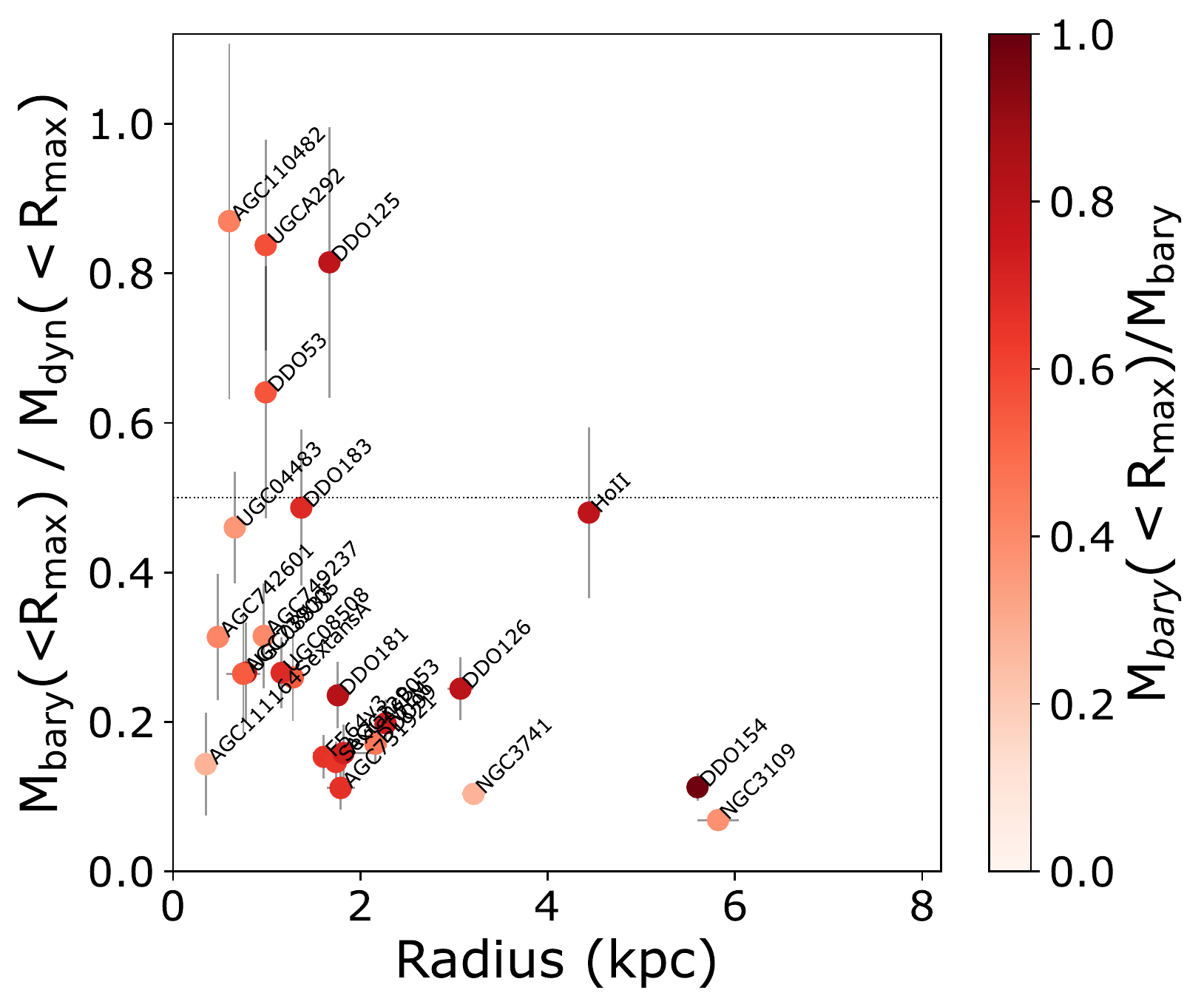}
\end{center}
\caption{Ratio of \mbary($<$\rmax) to \mdyn$<$\rmax) as a function of the radius at which the velocity measurements were made (\rmax). While only a rough surrogate for determining whether the galaxies are dark matter-dominated at \rmax, the comparison provides a useful check for our analysis mapping the measured kinematics of galaxies to theoretical rotation curves in Section~\ref{sec:rot_curves}. Points are color-coded by the fraction of baryonic mass within \rmax; the horizontal line marks where the baryonic mass is 50\% of the dynamical mass at \rmax.} 
\label{fig:Mbary_Mdyn}
\end{figure}

\subsection{Baryonic Masses at \rmax\ and Dynamical Mass Estimates}\label{sec:Mabary_Mdyn}
At small radii, a galaxy's mass distribution may or may not be dark matter dominated. If the bulk rotational motion of the gas includes important dynamical contributions from the baryons, then mapping a velocity to a theoretical rotation curve will have systematic uncertainties. To ensure this is not impacting our downstream analysis, we compare the baryonic mass to the dynamical mass based on the gas kinematics. 

We first estimate \mhi\ within \rmax\ by summing the \hi\ flux in the interferometric \hi\ datacubes within an ellipse defined by \dpv\ and position angles in Table~\ref{tab:velocities} and assuming the inclination angles listed in Table~\ref{tab:properties}. To account for the mass of helium, the resulting \mhi\ are scaled by a factor of 1.33.

For the SHIELD galaxies, we then estimated \mstar\ within \rmax\ using a similar approach. We summed the 3.6\micron\ flux in the cleaned IRAC images within an ellipse defined by \dpv\ and the adopted geometry described above. The fluxes were converted to stellar mass by assuming a mass-to-light ratio of 0.5.

For the VLA-ANGST and LITTLE THINGS galaxies where we do not have cleaned 3.6\micron\ imaging available, we use the full value of \mstar. Thus, the \mstar\ values will be biased towards high values, making our assessment of whether the galaxies are dark matter dominated at \rmax\ conservative. While this may overestimate \mstar\ within \rmax, it will not impact the answer significantly since \rmax\ is typically at larger distances in these galaxies and likely encloses the majority of \mstar. In addition, the VLA-ANGST and LITTLE THINGS galaxies are \hi\ dominated so any associated error on \mstar\ has a smaller impact on the total value of \mbary\ within \rmax. 

We estimate the dynamical mass of the sample using the approximation derived by applying the virial theorem to dwarf galaxies \citep[][and references therein]{Hoffman1996}. We use the asymmetric drift corrected velocities so as to include both rotational and pressure support in the calculation, and the radius at which the rotational velocities were measured:

\begin{equation}
M_{dyn} = \frac{V_{cor}^{2} \cdot R_{max}}{G} \\
\label{eq:Mdyn}
\end{equation}

\noindent which simplifies to:  
\begin{equation}
M_{dyn} = 2.325 \times 10^{5}M_{\odot}\left(\frac{V_{cor}^{2}}{\rm km^{2}~s^{-2}}\right)\left(\frac{R_{max}}{\rm kpc}\right)
\label{eq:Mdyn_simplify}
\end{equation}

\noindent where \mdyn\ is the dynamical mass in \msun, G is the universal constant of gravitation. These values are listed in Table~\ref{tab:velocities}.

Figure~\ref{fig:Mbary_Mdyn} shows the ratio of \mbary\ enclosed within \rmax\ to \mdyn\ determined at the same radius as a function of \rmax; the points are color-coded by the fraction of baryonic mass at \rmax\ for each galaxy. The ratio of \mbary($<$\rmax)/\mdyn($<$\rmax) provides an indication as to the whether the galaxies are dark matter-dominated at \rmax, although it is only a conservative surrogate as we are integrating the baryonic mass components out to \rmax. In addition, for the ANGST and LITTLE THINGS galaxies, \mbary($<$\rmax) is based on the total stellar mass values instead of \mstar($<$\rmax). As seen in Figure~\ref{fig:Mbary_Mdyn}, the majority of galaxies have \mbary($<$\rmax)/\mdyn\ $\ltsimeq50$\%; these approximate upper limits suggest that, as expected, the galaxies are dark matter dominated at \rmax. There are a few galaxies where the baryonic mass has a larger contribution. Without a detailed mass decomposition, it is less certain whether these galaxies are dark matter dominated at \rmax.  For completeness, we keep these galaxies in our full analysis, color-code all points by their ratio of \mbary($<$\rmax)/\mdyn($<$\rmax) in the BTFR analysis based on fitting to theoretical rotation curves (Figure~\ref{fig:btfr_fits}, and note that this is an additional caveat in the mapping of these individual galaxies to dark matter halos shown below.

\begin{figure}
\begin{center}
\includegraphics[width=0.49\textwidth]{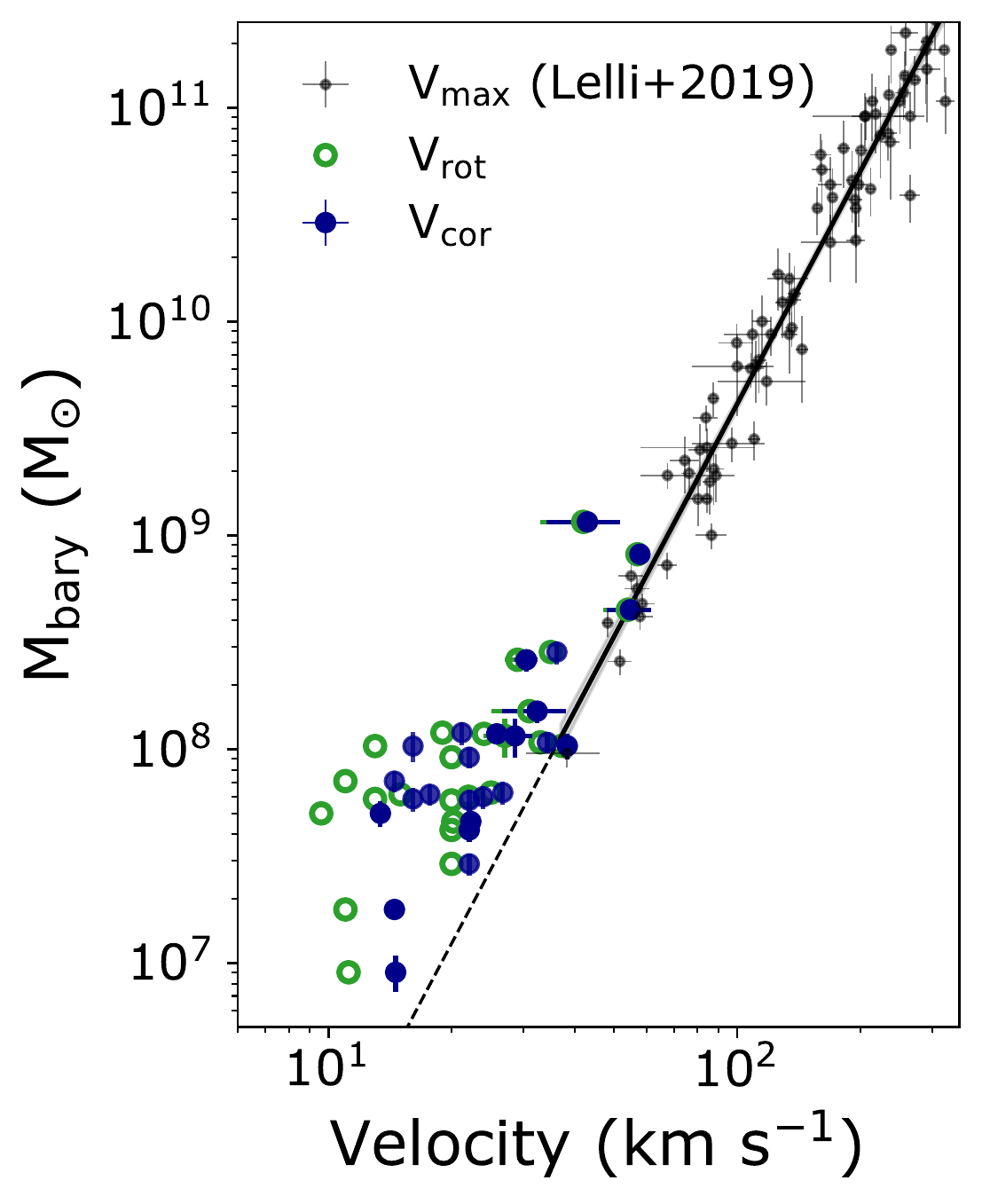}
\end{center}
\caption{The BTFR with our empirical measurements of baryonic mass and \hi\ rotational velocities. Inclination corrected rotational velocities (\vrot) are shown with open green circle; asymmetric drift corrected velocities (\vcor) are shown with solid blue circles. Galaxies with rotational velocities that are reaching the flat part of the rotation curve (i.e., where the velocity fields in the PV diagrams plateau) are plotted with uncertainties. All other velocities are measured on the rising part of the rotation curves and are lower limits for the maximum rotational velocities (\vmax) in the galaxies. \vmax\ measured for higher mass galaxies with well-resolved rotation curves are plotted as black circles \citep{Lelli2019}. The best-fitting line to those points is shown as a solid line, with the 1-$\sigma$ dispersion shaded in grey; an extrapolation of the line to lower masses is shown as a dashed line.}
\label{fig:btfr_empirical}
\end{figure}

\section{The BTFR with the Empirical Velocities}\label{sec:btfr_empirical}
The measured rotational velocities of the galaxies enable an initial exploration of the low-mass end of the BTFR. Figure~\ref{fig:btfr_empirical} shows the BTFR with \vrot, our rotational velocities measured from the PV slices corrected for inclination angle, for the SHIELD, VLA-ANGST, and LITTLE THINGS galaxies (green). For the galaxies with velocity fields that appear to flatten, we plot \vrot\ with uncertainties. For the remainder of the galaxies where the velocities were measured on the rising part of the rotation curve, representing lower limits for \vmax, we plot \vrot\ without uncertainties. We also show our asymmetric drift corrected velocities, \vcor\ (dark blue). As previously noted, the asymmetric drift correction has the largest impact on the galaxies with the lowest values of \vrot, which can now be seen to correspond to the lower mass galaxies.

We also include measurements for more massive, gas-rich irregular and spiral galaxies which populate the BTFR to higher galaxy masses (\mbary\ $\sim 10^9 - 10^{11}$ \msun; black points). For this comparison sample, we selected galaxies with high-quality rotation curves from \citet{Lelli2019} that met three criteria. First, the galaxies' rotation curves flattened, reducing uncertainties in interpreting the velocities used to place these galaxies on the BTFR. Second, we applied the same inclination angle restrictions used on our sample, namely $i>30^{\circ}$, to mitigate against uncertain projection effects. Third, we used galaxies with uncertainties on \mbary\ that were $\leq$ 50\% of \mbary. This latter criterion was applied to reject mass measurements based on distances with high uncertainties.\footnote{The distances to the galaxies were determined from a number of techniques with varying accuracy.   However, the methods used to measure the distances to the galaxies are not specified in \citet{Lelli2019}. Thus, we opted to filter the sample based on \mbary\ uncertainties.} We chose our limit of  50\% of \mbary\ as the distribution of uncertainties reported in \citet{Lelli2019} is bi-modal, with a clear separation at this value. 

Our final comparison sample includes 77 galaxies from the catalog of \citet{Lelli2019}. The stellar masses are based on 3.6$\micron$ fluxes and adopting the same constant mass-to-light ratio used to determine the stellar masses for the SHIELD, VLA-ANGST, and LITTLE THINGS galaxies. The velocities represent the circular velocities measured at the peak of the rotation curves, \vmax, the observational analog to the peak velocity in theoretical rotation curves.

Figure~\ref{fig:btfr_empirical} shows the best-fitting line (solid black; slope $= 3.61\pm0.10$; intercept $2.39\pm0.21$) and 1-$\sigma$ deviation (grey) determined from our selected subsample of galaxies from \citet{Lelli2019}, which is consistent within the uncertainties with their reported best-fitting line based on the full sample. Also shown is an extrapolation of this best-fitting line to the lower mass regime of our sample (dashed black). Even with the asymmetric drift corrections, nearly all of the galaxies below \mbary\ $\sim10^8$ \msun\ have lower velocities for a given baryonic mass compared with the relation extrapolated from more massive galaxies. On the one hand, we are reporting \vcor\ values determined at radii that are within the gas and stellar disk, whereas the \mbary\ includes the total gas and star content. Thus, our points may be expected to be offset slightly to the left of the extrapolation. On the other hand, if this were the only factor, then \mbary\ would have to be over-estimated by a factor of ten compared to the values they would have if they were located along the extrapolated BTFR fit (dashed line). This is significantly larger than the differences between \mbary($<$\rmax) estimated above compared to \mbary, which seems unlikely. Instead, the most probable explanation that the majority of the galaxies lie to the left of the extrapolated BTFR is because their rotational velocities were measured on the rising part of the rotation curves and, thus, are lower limits for \vmax.

\section{Estimating \vmax\ by Mapping Galaxies to Dark Matter Halos}\label{sec:dm}
Under the assumption that galaxies reside in dark matter halos, and that these halos have a specific set of properties, the gas rotational velocity measured at a given radius can be used to estimate the dark matter halo mass, \mhalo. We use our values of \vcor\ and \rmax\ to map each of the galaxies in our sample to a cold dark matter halo. Once mapped to \mhalo, we have estimates of the associated value of \vmax, and can calculate additional variables such as the baryon fraction (i.e., \mbary/\mhalo). We use these estimated values to re-assess the distribution of low-mass galaxies on the BTFR and compare with predictions from simulations. 

Our basic approach is to generate a family of rotation curves based on monotonically increasing halo masses using an assumed dark matter density profile and mass-concentration relation. The values of \vcor\ and \rmax\ for a galaxy are then overlaid on the rotation curves; the rotation curve with the most closely matched velocity at the value of \rmax\ is selected and the associated dark matter halo mass is assigned to the galaxy. The uncertainties in \vcor\ and \rmax\ are used to bracket the range of possible rotation curves, and therefore halo masses and \vmax\ values, for each galaxy. 

\begin{figure*}
\begin{center}
\includegraphics[width=0.98\textwidth]{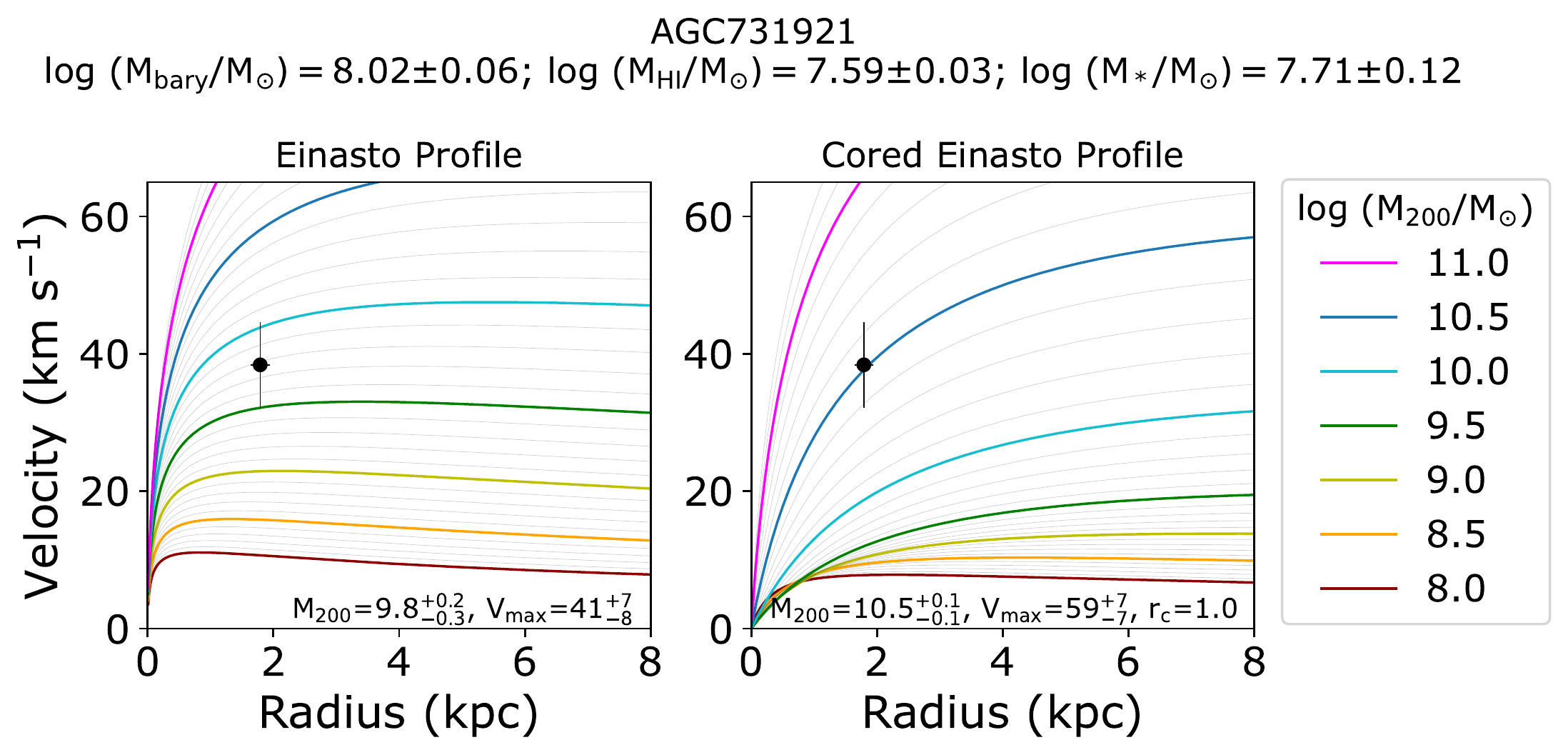} 
\end{center}
\caption{Families of rotation curves generated for log(\mhalo/\msun) $=10^8-10^{11.2}$, in increments of 0.1 dex (grey lines) and 0.5 dex (colored lines; see legend for details) based on a steeply rising Einasto density profile (left) and a more slowly rising cored Einasto density profile (right). Overplotted on both panels as a filled black circle are the \vcor\ and \rmax\ values for one galaxy, AGC~731921. \mbary, \mhi, and \mstar\ measurements are listed at the top below the galaxy name. The values of log(\mhalo/\msun) and \vmax\ (in units of \kms) based on the best-fitting rotation curves are listed at the bottom of each panel. The core radius \rc\ (in units of kpc), for the cored Einasto profile is shown at the bottom of the right panel. Similar figures for all other galaxies are presented in Appendix~\ref{app:rotcurves}. For uniformity, we plot rotation curves for the same range in halo mass, but note that not all curves are appropriate for a given galaxy (i.e., in the current figure, \mhalo $-10^8$ \msun is not appropriate given that \mbary $\sim 10^8$ \msun, resulting in a rotation curve that crosses the others at small radii).}
\label{fig:rot_curve_example}
\end{figure*}

We consider two dark matter density profiles, specifically a cuspy Einasto profile \citep[hereafter, simply referred to as an Einasto profile;][]{Einasto1965}, and a ``cored'' Einasto profile. The Einasto profile was chosen as a reference point for a cold dark matter profile where baryonic effects have not significantly altered the density profiles.\footnote{The Einasto profile has been shown to be a more accurate description of spherically averaged CDM density profiles than a Navarro--Frenk--White \citep[NFW;][]{Navarro1996} profile \citep[see, e.g.,][and references therein]{Dutton2014}.} The cored Einasto profile was motivated by its connection to high-resolution simulations that converge for low mass halos (log(\mhalo/\msun)$\sim9.5$) and where feedback prescriptions carve dark matter cores, allowing us to compare our empirical results with predictions in a self-consistent way. We adopt the parameters for the cored Einasto profile based on fits to low-mass galaxies from the FIRE simulations from \citet{Lazar2020}. We considered a third profile, the cored Dekel-Zhao profile \citep{Freundlich2020}, but found that it did not give satisfactory fits to the data. While similar to the cored Einasto profiles for the lower mass end of our sample, the rotation curves based on the mean cored Dekel-Zhao profile for the higher mass galaxies rose more gently. Thus, our values of \vmax\ and \rmax\ intersected with rotation curves that were still rising significantly and mapped these higher mass galaxies to dark matter halos that tended to be unrealistically massive. 

Note that we also considered an alternative approach for estimating \vmax\ using the relationship between the 21-cm spectral line width, $W_{50}$, and \vmax\ determined from studies of more massive galaxies \citep{Lelli2019} that is independent of assuming a dark matter halo. However, since the detected \hi\ velocity fields in the majority of our galaxies do not plateau, any conversion from $W_{50}$ to \vmax\ will still be a lower limit for the true maximum velocity in a system \citep[e.g.,][]{Brooks2017}.

\subsection{Assumed Density Profiles}\label{sec:profiles}
\subsubsection{The Einasto Profile}
The Einasto profile has the following form:

\begin{equation}
\rho(r) = \rho_s \exp \left( -\frac{2}{\alpha} \left[ \left( \frac{r}{r_s} \right)^{\alpha} - 1 \right] \right)
\label{eq:einasto}
\end{equation}

\noindent where $r_s$ is the scale radius, which we determine based on the concentration of a halo $c = R_{vir}/r_s$ (see Section \ref{sec:concentration} for details), $\rho_s$ is the density at $r_s$, and $\alpha$ describes how quickly the profile steepens (i.e., the radial dependence of the profile slope). The value of $\alpha$ is determined from the formula of \citet{Gao2008}:

\begin{equation}
\alpha = 0.155 + 0.0095~\nu_{\rm vir}^2
\label{eq:alpha}
\end{equation}

\noindent where $\nu_{\rm vir}$ is a dimensionless peak-height parameter. 

The rotation curves are then calculated for a set of halos ranging from log(\mhalo/\msun) $=8.5-11.2$ in steps of 0.1 dex based on:

\begin{equation}
V(r) = \sqrt{\frac{GM(<r)}{r}}
\label{eq:rotcurve_einasto}
\end{equation}

\noindent where \mhalo\ is the halo mass defined at 200$\times$ the critical density of the universe, $M(<r)$ is the mass enclosed by the density profiles at radius $r$, and G is the gravitational constant. \vmax\ for any given halo is then simply the maximum value of $V(r)$. 

\subsubsection{The Cored Einasto Profile}
The cored Einasto profile is a simple modification to Equation~\ref{eq:einasto} that adds a core radius, \rc, and has the following form:

\begin{equation}
\rho(r) = \rho_s \exp\left( -\frac{2}{\alpha} \left[ \left( \frac{r + r_c}{r_s} \right)^{\alpha} - 1 \right] \right)
\label{eq:cored_einasto}
\end{equation}

The formation of a cored profile and the size of a core depend not only on stellar feedback, but also the ratio of stellar to total halo mass. That is, in low-mass galaxies with the same stellar mass, star formation and feedback will have a larger impact on the re-distribution of matter in systems residing in a smaller halo. One caveat to this statement is that when the stellar mass makes up a substantial portion of the total mass in the galaxy center, the baryons contract the density profile instead of forming a core, and the deeper gravitational potential makes feedback less effective at producing outflows \citep{Governato2012, DiCintio2014b, Chan2015}. This results in a cuspier inner profile and a more steeply rising rotation curve. 

We use the formalism for \rc\ based on fits to results from the FIRE simulations which produce cores in galaxies with \mstar\ $>10^6$ \msun. Specifically, we use Equation~12 from \citet{Lazar2020} for the best-fitting relation of \rc\ based on \mstar/\mhalo, which we reproduce for clarity here with the appropriate values from their Table~1:

\begin{equation}
r_c = 10^{1.21} \left(0.71 + \frac{M_*/M_h}{7.2\times10^{-3}}\right)^{-2.31} \left(\frac{M_*/M_h}{0.011}\right)^{1.55}
\label{eq:rc_einasto}
\end{equation}

\noindent where \rc\ is in units of kpc. For each galaxy, we use the measured \mstar\ value and calculate \rc\ for each assumed \mhalo\ value over the same range of \mhalo\ values as the Einasto profile above (i.e., log(\mhalo/\msun) $=8.0-11.2$ in steps of 0.1 dex). The results yield a unique value of \rc\ for a given value of \mstar\ at each assumed value of \mhalo, with smaller \mstar/\mhalo\ ratios producing smaller cores.  Because the baryon retention at low galaxy masses is predicted to depend on a number of variables \citep[i.e., stellar feedback, environment, reionization; see Section~\ref{sec:intro} and, e.g.,][]{Munshi2021}, this approaches avoids assuming a constant core size for a given value of \mstar\ (which would imply a fixed \mstar/\mhalo\ ratio). Note that since we are using the equation for \rc\ based on the fits from \citet{Lazar2020}, our calculations also depend on the feedback prescriptions used in the FIRE simulations. Note, also, that \rc\ calculated from Equation~\ref{eq:rc_einasto} only represents the predicted mean core radius. 

Based on these cored Einasto profiles, we then generate a family of rotation curves using Equation~\ref{eq:rotcurve_einasto}, and again define \vmax\ as the maximum value of $V(r)$ for each halo.

\subsection{Concentration Parameter}\label{sec:concentration}
The concentration parameter, which is defined as the ratio of the virial radius of a halo to the scale radius ($c=R_{vir}/r_s$), quantifies the concentration of mass within a halo. The mean concentration varies as a function of halo mass and is often parameterized as a power law for a given redshift, based on a well-determined cosmology \citep[e.g.,][]{Planck2018}, with significant scatter about the mean \citep[e.g.,][]{Bullock2001, Dutton2014, Lazar2020}. 

In our analysis, we adopt the mean mass-concentration relation from \citet{Bullock2001} for both Einasto profiles. The FIRE simulations follow a mass-concentration relation from \citet{Wang2019}, with similar mean values in our mass range of interest to those from \citet{Bullock2001}. However, as noted above, the mass-concentration relations represent only the mean values. The considerable scatter about the mean, regardless of the relation used, contributes systematic uncertainties in our analysis and is a limitation in our comparisons below.

\subsection{Family of Rotation Curves}\label{sec:rot_curves}
Figure~\ref{fig:rot_curve_example} shows examples of the rotation curves generated from the Einasto profile (left) and the cored Einasto profile (right) for log(\mhalo/\msun) of $8.0-11.2$. In both panels, the family of rotation curves is plotted in log(\mhalo/\msun) steps of 0.1 dex in light gray with every 0.5 dex interval highlighted in a different color; see figure legend for details. For the cored profile, the radius of the core changes depending on \mhalo\ (i.e., \mstar\ is single valued for each galaxy but since each curve is associated with a different mass halo, the core size changes with larger cores created in lower halo masses and vice-versa; see Section \ref{sec:profiles} and Equation~\ref{eq:rc_einasto}). Overplotted as a filled black circle in each panel of Figure~\ref{fig:rot_curve_example} are the values of \vcor\ and \rmax\ for the galaxy, with uncertainties; we use a circle to demarcate that \vrot\ was measured on the rising part of the rotation curve. Similar plots for the rest of the sample are shown in Appendix~\ref{app:rotcurves}; galaxies where \vrot\ was measured on the flat part of the rotation curve are shown with a black star symbol instead of a circle.

\begin{table*}
\begin{center}
\caption{Estimated Properties Based on Fits to Theoretical Rotation Curves}
\label{tab:fit_values}
\end{center}
\vspace{-15pt}
\begin{center}
\begin{tabular}{l | rc | rcc}
\hline 
\hline 
		& \multicolumn{2}{c}{Einasto}			& \multicolumn{3}{c}{Cored Einasto}	\\
\hline
Galaxy          	& log(\mhalo/\msun)		& \vmax	& log(\mhalo/\msun)	& \vmax	& \rc	\\
                	&  				& (\kms)	&			& (\kms)	& (kpc)\\
\hline           
\multicolumn{6}{c}{SHIELD Galaxies}  \\                      
\hline
AGC110482 & 8.3 $^{+ 0.2 }_{- 0.2}$ & 14$^{+ 2}_{- 3}$ & 10.1 $^{+ 0.1 }_{- 0.1}$ & 41$^{+ 6}_{- 5}$ & 1.2 \\
AGC111164 & 8.5 $^{+ 0.3 }_{- 0.2}$ & 16$^{+ 4}_{- 3}$ & 9.7 $^{+ 0.1 }_{- 0.1}$ & 33$^{+ 5}_{- 4}$ & 0.4 \\
AGC229053 & 9.0 $^{+ 0.1 }_{- 0.2}$ & 23$^{+ 2}_{- 4}$ & 9.7 $^{+ 0.1 }_{- 0.1}$ & 30$^{+ 5}_{- 4}$ & 1.0 \\
AGC731921 & 9.8 $^{+ 0.2 }_{- 0.3}$ & 41$^{+ 7}_{- 8}$ & 10.5 $^{+ 0.1 }_{- 0.1}$ & 59$^{+ 7}_{- 7}$ & 1.0 \\
AGC739005 & 9.1 $^{+ 0.1 }_{- 0.2}$ & 25$^{+ 2}_{- 4}$ & 10.1 $^{+ 0.1 }_{- 0.1}$ & 44$^{+ 5}_{- 5}$ & 0.7 \\
AGC742601 & 8.5 $^{+ 0.2 }_{- 0.3}$ & 16$^{+ 3}_{- 4}$ & 9.7 $^{+ 0.1 }_{- 0.1}$ & 33$^{+ 4}_{- 4}$ & 0.5 \\
AGC749237 & 9.3 $^{+ 0.1 }_{- 0.2}$ & 29$^{+ 2}_{- 5}$ & 10.4 $^{+ 0.1 }_{- 0.1}$ & 54$^{+ 7}_{- 6}$ & 1.0 \\
\hline              
\multicolumn{6}{c}{VLA-ANGST Galaxies}  \\                      
\hline
DDO99 & 8.9 $^{+ 0.1 }_{- 0.1}$ & 21$^{+ 4}_{- 2}$ & 9.9 $^{+ 0.1 }_{- 0.1}$ & 33$^{+ 5}_{- 4}$ & 1.5  \\
DDO125 & 8.5 $^{+ 0.1 }_{- 0.1}$ & 16$^{+ 2}_{- 2}$ & 9.9 $^{+ 0.1 }_{- 0.1}$ & 29$^{+ 4}_{- 3}$ & 2.4  \\
DDO181 & 8.9 $^{+ 0.2 }_{- 0.1}$ & 21$^{+ 4}_{- 2}$ & 9.9 $^{+ 0.1 }_{- 0.1}$ & 33$^{+ 5}_{- 4}$ & 1.5  \\
DDO183 & 8.5 $^{+ 0.2 }_{- 0.2}$ & 16$^{+ 3}_{- 3}$ & 9.8 $^{+ 0.1 }_{- 0.1}$ & 29$^{+ 4}_{- 4}$ & 1.9  \\
NGC3109 & 10.3 $^{+ 0.1 }_{- 0.1}$ & 59$^{+ 5}_{- 5}$ & 10.8 $^{+ 0.1 }_{- 0.1}$ & 71$^{+ 9}_{- 7}$ & 1.7  \\
NGC3741 & 9.4 $^{+ 0.1 }_{- 0.1}$ & 31$^{+ 3}_{- 3}$ & 10.0 $^{+ 0.1 }_{- 0.1}$ & 38$^{+ 5}_{- 5}$ & 1.0  \\
SextansA & 9.6 $^{+ 0.3 }_{- 0.3}$ & 36$^{+ 9}_{- 8}$ & 10.3 $^{+ 0.2 }_{- 0.1}$ & 51$^{+ 12}_{- 6}$ & 0.8  \\
SextansB & 9.6 $^{+ 0.2 }_{- 0.2}$ & 36$^{+ 6}_{- 6}$ & 10.4 $^{+ 0.1 }_{- 0.1}$ & 55$^{+ 7}_{- 6}$ & 0.9  \\
UGC04483 & 8.7 $^{+ 0.1 }_{- 0.1}$ & 18$^{+ 2}_{- 1}$ & 9.9 $^{+ 0.1 }_{- 0.1}$ & 38$^{+ 5}_{- 5}$ & 0.6  \\
UGC08508 & 9.1 $^{+ 0.1 }_{- 0.1}$ & 25$^{+ 2}_{- 3}$ & 10.1 $^{+ 0.1 }_{- 0.1}$ & 42$^{+ 6}_{- 5}$ & 0.9  \\
UGC08833 & 9.0 $^{+ 0.4 }_{- 0.3}$ & 23$^{+ 8}_{- 5}$ & 10.0 $^{+ 0.1 }_{- 0.2}$ & 42$^{+ 5}_{- 9}$ & 0.5  \\
UGCA292 & 8.4 $^{+ 0.1 }_{- 0.2}$ & 15$^{+ 1}_{- 3}$ & 9.6 $^{+ 0.1 }_{- 0.1}$ & 27$^{+ 4}_{- 4}$ & 1.1  \\
\hline
\multicolumn{6}{c}{LITTLE THINGS Galaxies}  \\                      
\hline
DDO53 & 8.9 $^{+ 0.3 }_{- 0.3}$ & 21$^{+ 6}_{- 4}$ & 10.2 $^{+ 0.1 }_{- 0.2}$ & 45$^{+ 6}_{- 10}$ & 1.1  \\
DDO126 & 9.6 $^{+ 0.2 }_{- 0.1}$ & 36$^{+ 6}_{- 3}$ & 10.4 $^{+ 0.1 }_{- 0.1}$ & 51$^{+ 7}_{- 6}$ & 1.5  \\
DDO154 & 10.2 $^{+ 0.2 }_{- 0.2}$ & 55$^{+ 9}_{- 8}$ & 10.4 $^{+ 0.1 }_{- 0.1}$ & 57$^{+ 6}_{- 6}$ & 0.6  \\
F564v3 & 9.2 $^{+ 0.2 }_{- 0.1}$ & 27$^{+ 4}_{- 3}$ & 10.0 $^{+ 0.1 }_{- 0.1}$ & 38$^{+ 6}_{- 4}$ & 1.0  \\
HoII & 9.9 $^{+ 0.2 }_{- 0.3}$ & 44$^{+ 8}_{- 9}$ & 10.6 $^{+ 0.2 }_{- 0.3}$ & 56$^{+ 13}_{- 13}$ & 2.5  \\
WLM & 9.3 $^{+ 0.1 }_{- 0.1}$ & 29$^{+ 2}_{- 3}$ & 10.1 $^{+ 0.1 }_{- 0.1}$ & 40$^{+ 6}_{- 5}$ & 1.3  \\
\hline              
\end{tabular}
\end{center}
\tablecomments{Estimated halo masses (\mhalo) and maximum circular velocities (\vmax) based on fits to families of rotation curves from Einasto and cored Einasto profiles. For the fits to the cored Einasto profiles, we also list the radius of core, \rc.}
\end{table*}

As seen in Figure~\ref{fig:rot_curve_example}, the velocity profiles change as a function of mass, but the curves do not cross. An exception is seen for the cored Einasto rotation curve at the lowest mass halo, \mhalo\ $=10^8$ \msun. Given that \mbary\ $> 10^8$ \msun\ for the galaxy shown, this is an inappropriate \mhalo\ scale; however, our analysis includes galaxies that are lower-mass and therefore we include this rotation curve to be consistent with the plots of the full galaxy sample that appear in Appendix~\ref{app:rotcurves}. 

From Figure~\ref{fig:rot_curve_example}, the most closely matched rotation curve to the \vcor\ and \rmax\ pair of values is selected and the associated \mhalo\ is assigned to the galaxy; uncertainties on \mhalo\ are based on the range of rotation curves encompassed by the uncertainties on \vcor\ and \rmax. For the assigned \mhalo, we calculate the maximum velocity of the halo, \vmax, from Equation~\ref{eq:rotcurve_einasto}; uncertainties in \vmax\ are based on the range in halo masses encompassed by the fits of \vcor\ and \rmax\ to the rotation curves. The best-fitting log(\mhalo) values are listed at the bottom of each panel in units of \msun, as are the corresponding values of \vmax\ in units of \kms. For the cored Einasto profile, \rc\ is also shown, in units of kpc. For easy comparison to the baryonic content, \mbary, \mhi, and \mstar\ for the galaxy are listed at the top of the figure. Similar figures for all galaxies are presented in Appendix~\ref{app:rotcurves}; values from the fits for each galaxy are provided in Table~\ref{tab:fit_values}.

\begin{figure}
\begin{center}
\includegraphics[width=0.49\textwidth]{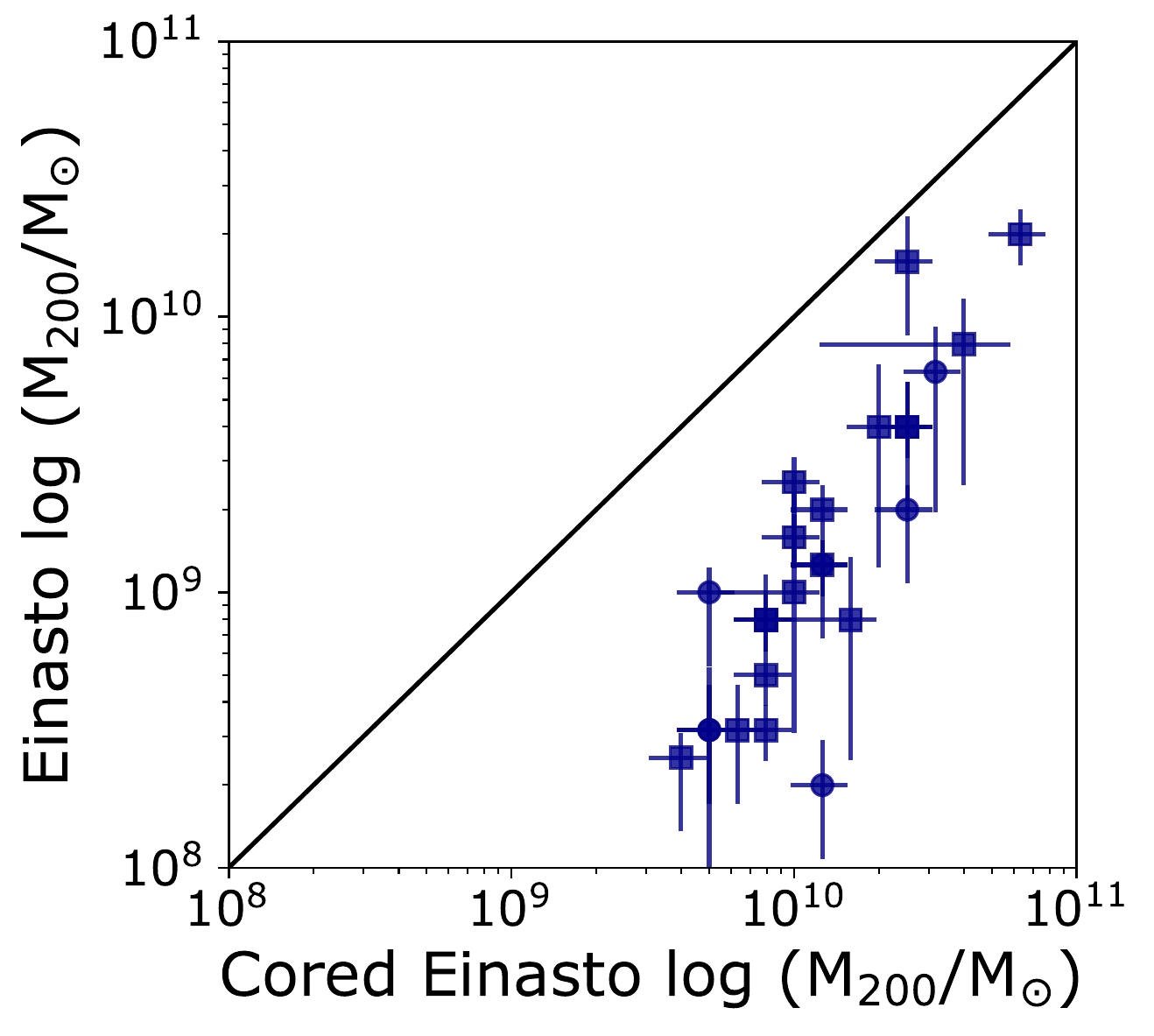}
\end{center}
\caption{Comparison of \mhalo\ for our galaxy sample estimated from fits to the Einasto and cored Einasto rotation curves. The steeply rising nature of the Einasto profile places the lowest mass galaxies in halos that are up to $\sim2$ orders of magnitude less massive than fits to the cored Einasto profiles. The solid black lines represent perfect agreement between the two sets of values.}
\label{fig:mstar_mhalo}
\end{figure}

\subsection{Comparison of the Data and the Theoretical Rotation Curves}\label{sec:compare_einasto}
Immediately apparent in Figure~\ref{fig:rot_curve_example} is the significant difference in the inner shape of the rotation curves between the cuspy Einasto profile (left panel) and the cored Einasto profiles (right panel), as expected. The  family of Einasto rotation curves flatten at much smaller radii than the cored profiles. As a result, the values of \vcor\ measured at \rmax\ for many of the galaxies in our sample fall well out onto the flat part of the Einasto rotation curves (see Appendix~\ref{app:rotcurves} for examples), in contrast to what is typically observed. This simply reflects the conclusions of many previous studies that the rotation curves of low mass galaxies are better represented with models that have cored profiles \citep[e.g.,][]{deBlok2001, Read2017}. Despite the fact that the Einasto density profiles do not represent the data well, we carry the results from these profiles forward in our analysis for completeness.

When \vcor\ and \rmax\ are matched to the more slowly-rising cored Einasto rotation curves, the galaxies are placed in more massive halos, resulting in more realistic values of \vmax\ in the sense that they are greater than \vcor. For a few that are plateauing, corresponding to the most massive (\mstar\ $\sim10^8$ \msun) galaxies where \vrot\ was measured at a larger radii, \vmax\ $\sim$ \vcor. However, there are some exceptions. For three of the six galaxies where the velocity fields plateau and \vcor\ is a good representation of the maximum velocity in the galaxy (i.e., NGC~3109, Sextans~A, WLM), the \vcor --\rmax\ values fall somewhat below the plateau of the cored Einasto rotation curves. This suggests that the cored Einasto profile as implemented here rises too slowly compared with some dwarf galaxies. A mis-match between some of the observations and models is unavoidable as observed inner rotation curves can have a range of slopes whereas simulated inner rotation curves based on a mean mass-concentration relation and a mean core radius relation are inherently uniform. Note, however, that the best-fitting value of \mhalo\ for WLM from the cored Einasto profiles agrees within the uncertainties with \mhalo\ derived from more detailed modeling of the stellar and gas kinematics in the galaxy \citep{Leung2021}.

Figure~\ref{fig:mstar_mhalo} compares the values of \mhalo\ based on fits to the Einasto and cored Einasto rotation curves, which are significantly different. The steeply rising nature of the Einasto rotation curves result in placing galaxies in lower mass halos. The differences in the halo masses increases for galaxies with smaller values of \vcor\ and \rmax\ as the shape of the cuspy versus cored rotation curves are greatest in the inner regions. The lowest mass halos derived from the Einasto rotation curves are close to $\sim$2 orders of magnitude smaller than those derived from the cored Einasto curves. In addition, the range in \mhalo\ occupied by the galaxies based on the cored Einasto rotation curve span a small range in values.

\begin{figure}
\begin{center}
\includegraphics[width=0.47\textwidth]{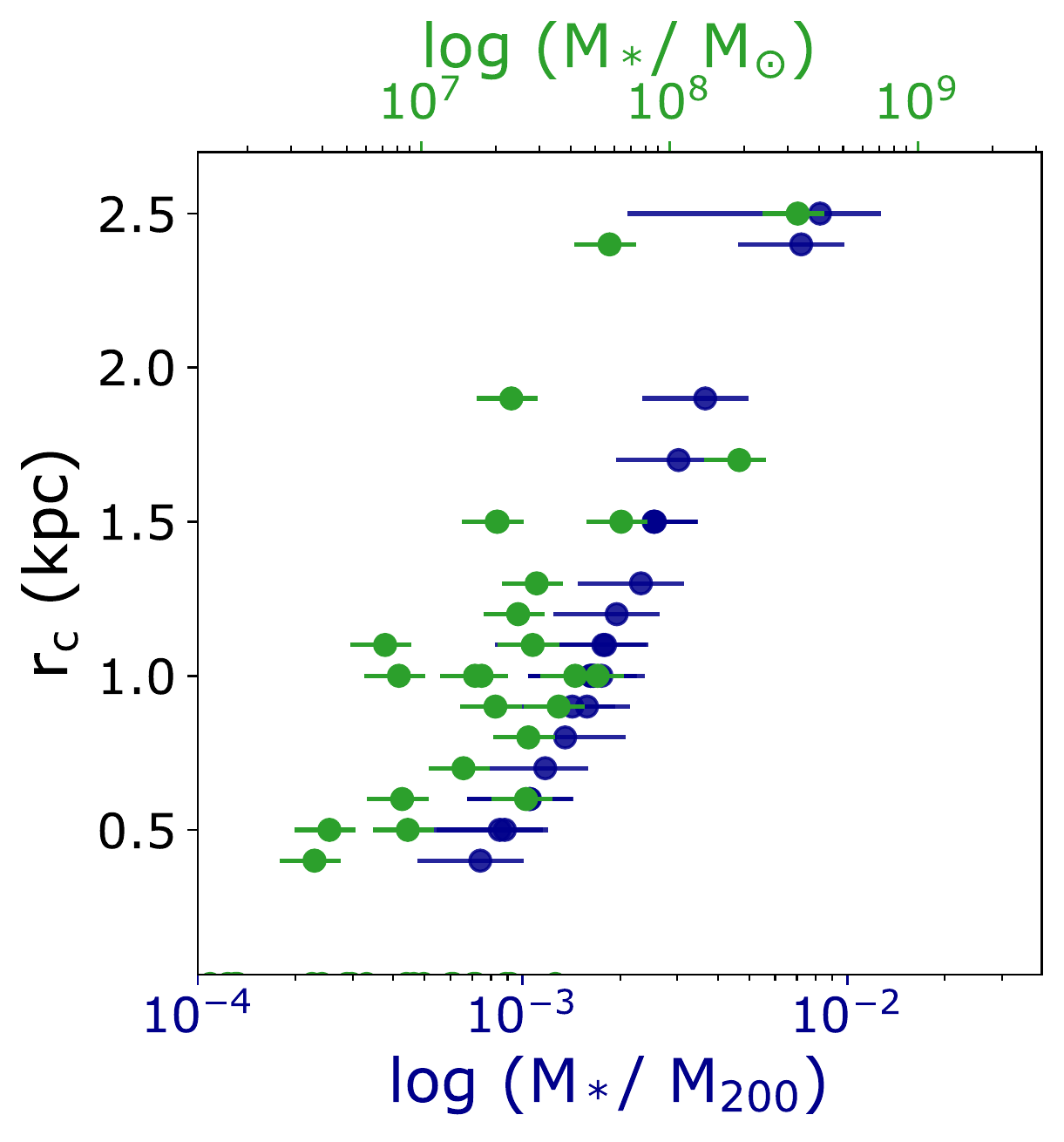}
\end{center}
\caption{Radius of the cores in the density profiles from the fits to theoretical rotation curves based on cored Einasto profiles as a function of \mstar\ (green points; top x-axis scale) and \mstar/\mhalo\ (blue points; bottom x-axis scale). Values in the top and bottom x-axis scales are independent of each other. As the cored profiles are calibrated to the FIRE simulations, the core radius is dependent on the FIRE feedback prescriptions which create cores down to \mstar\ $\sim10^6$ \msun. Larger cores are formed in galaxies with higher \mstar\ and higher ratios of \mstar/\mhalo, as expected in this mass regime.}
\label{fig:core_size}
\end{figure}

Figure~\ref{fig:core_size} presents the core radius, \rc, of the density profiles from the fits to theoretical rotation curves based on the cored Einasto profiles as as a function of \mstar\ (green points; top x-axis scale) and \mstar/\mhalo\ (blue points; bottom x-axis scale); top and bottom x-axis scales are independent of one another. The values of \rc\ are provided in Table~\ref{tab:fit_values} and range from 0.4 to 2.5 kpc.  The core radius depends on the calibrations from the FIRE simulations where cores are created down to \mstar\ $\sim10^6$ \msun, which is below the mass of the lowest-mass galaxy in our sample. The largest cores are formed at the higher mass end of our sample (\mstar\ $\gtsimeq10^8$ \msun) with the highest baryon fraction (i.e., larger \mstar/\mhalo\ ratios). 

\begin{figure*}
\begin{center}
\includegraphics[width=0.49\textwidth]{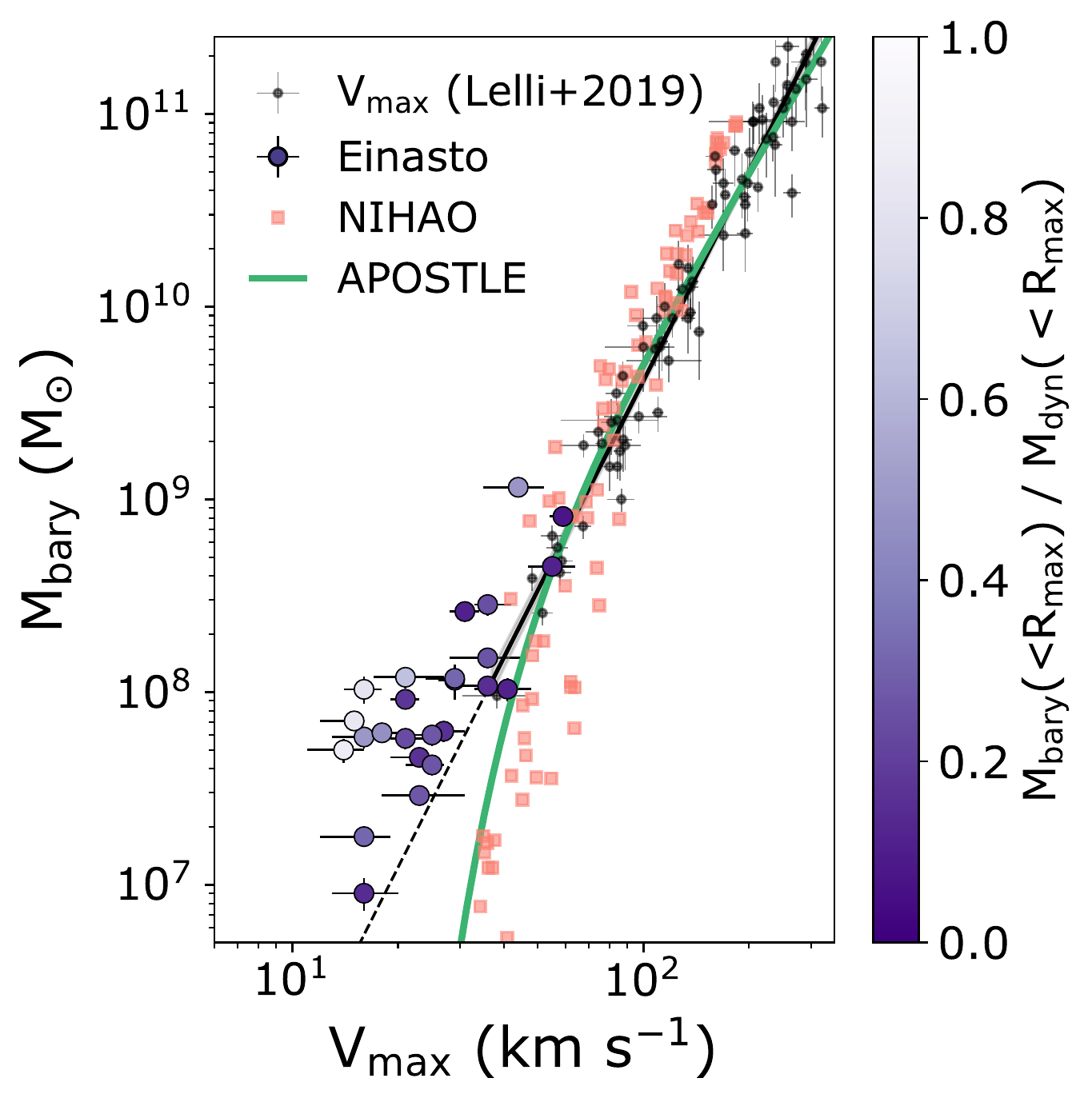}
\includegraphics[width=0.49\textwidth]{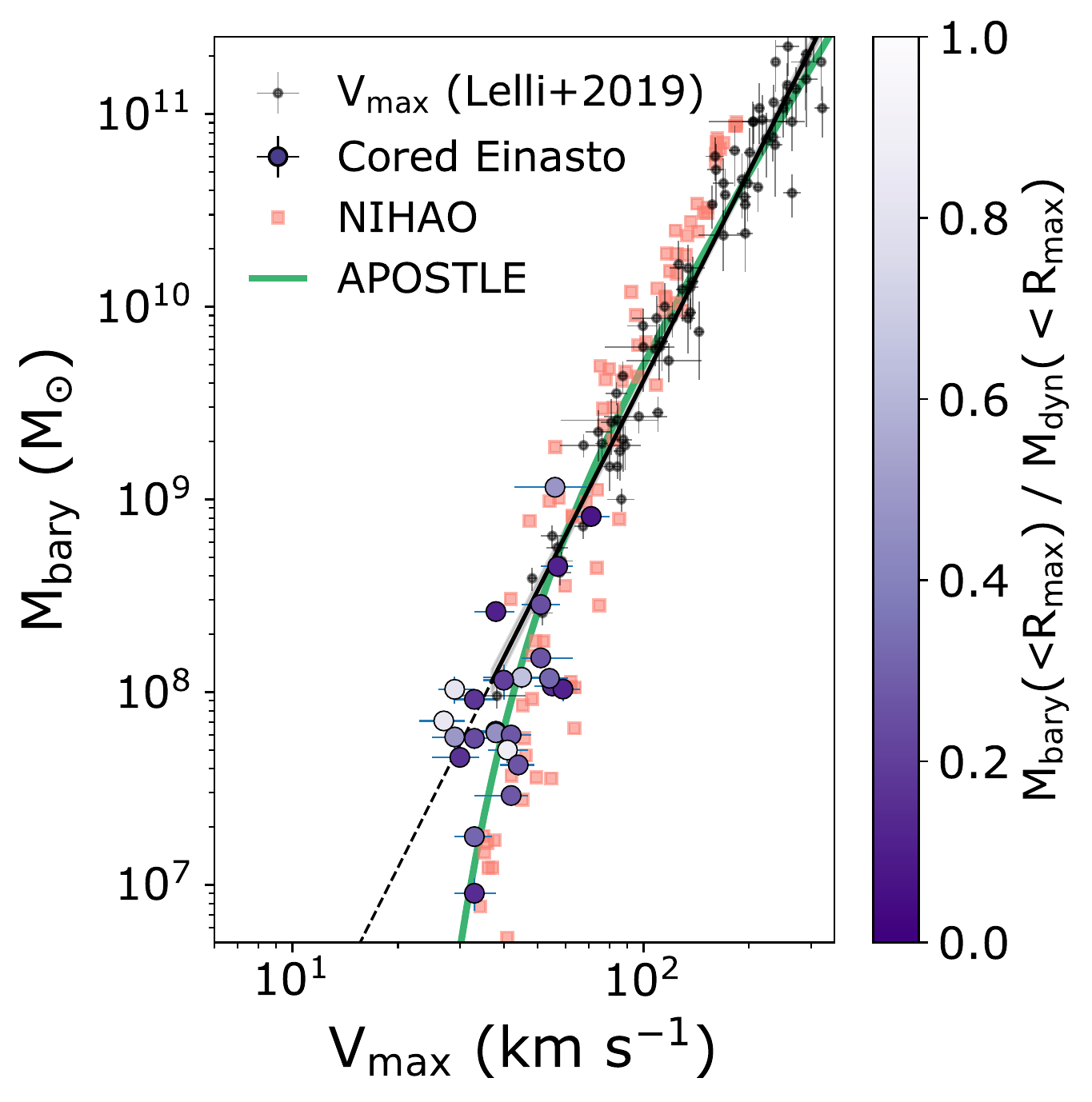}
\end{center}
\caption{The BTFR with \vmax\ estimated from fits to the Einasto rotation curves (left; purple circles) and cored Einasto rotation curves (right; purple circles) for our sample of galaxies. Similar to Figure~\ref{fig:btfr_empirical}, measurements for more massive galaxies from \citet[][black circles]{Lelli2019} are included, as well as the best-fitting line to these galaxies (solid black), 1-$\sigma$ dispersion to the fit (shaded grey), and an extrapolation of the fit to lower masses (dashed black). Predictions from the {\sc apostle} \citep[green line;][]{Sales2017} and NIHAO \citep[salmon points;][]{Dutton2017}  hydrodynamical simulations are also shown. \vmax\ estimates from the Einasto rotation curves yield unphysical values of \vmax\ for many galaxies in our sample. \vmax\ estimates from the cored Einasto rotation curves result in a turn-down in the BTFR that is consistent with predictions from the simulations shown. Points are color-coded by the ratio \mbary($<$\rmax)/\mdyn($<$\rmax), an approximate upper limit for the baryon-to-dark matter ratio based on integrating \mbary\ from the center of galaxies out to \rmax; darker shades indicate higher confidence that the galaxies are dark-matter dominated at \rmax\ (see Section~\ref{sec:Mabary_Mdyn} for details). }
\label{fig:btfr_fits}
\end{figure*}

\section{Results from Mapping the Galaxies to Dark Matter Halos}\label{sec:results}
\subsection{BTFR with Estimated \vmax\ Values}
Similar to Figure~\ref{fig:btfr_empirical}, Figure~\ref{fig:btfr_fits} shows our sample on the BTFR but now uses the \vmax\ values estimated by fitting our \vcor\ and \rmax\ values to rotation curves generated from Einasto density profiles without cores (left panel; purple circles) and with cores (right panel; purple circles). For comparison, we now also add theoretical expectations of the BTFR from two suites of high-resolution hydrodynamical simulations that extend to low-masses, namely the {\sc apostle} simulations \citep[green line;][]{Sales2017} and the NIHAO simulations \citep[salmon squares;][]{Dutton2017}. Note that \mbary\ for the {\sc apostle} simulations is based on \mbary\ within 0.15$\times R_{200}$ and assumes a slightly higher scaling for helium (i.e., \mstar\ $+$ 1.4$\times$\mhi) while \mbary\ for the NIHAO simulations is based on their reported \mbary\ within a slightly larger radius of 0.2$\times R_{200}$ with the same scaling for helium used in our calculations (i.e., \mstar\ $+$ 1.33$\times$\mhi).
 
\begin{figure*}
\begin{center}
\includegraphics[width=0.48\textwidth]{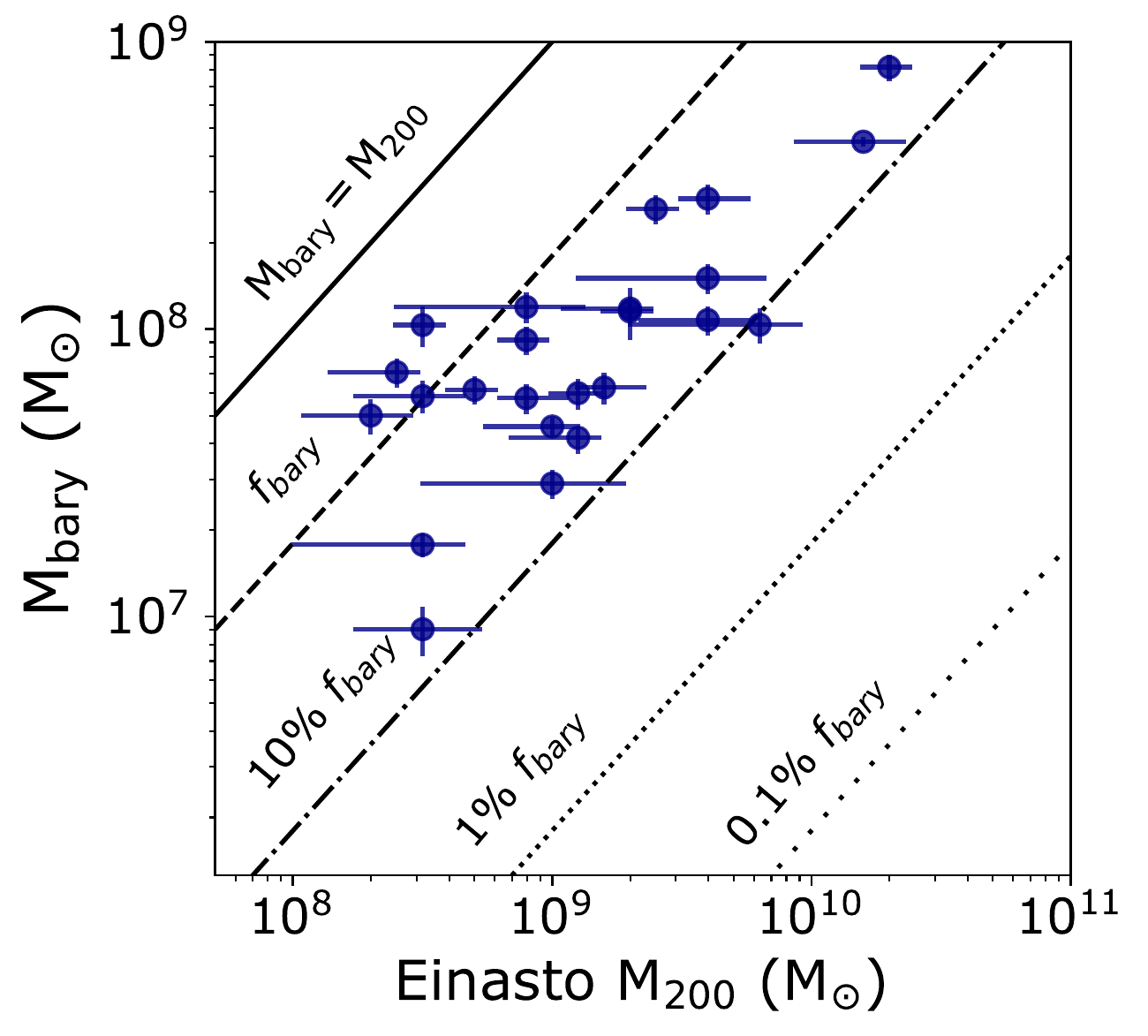}
\includegraphics[width=0.48\textwidth]{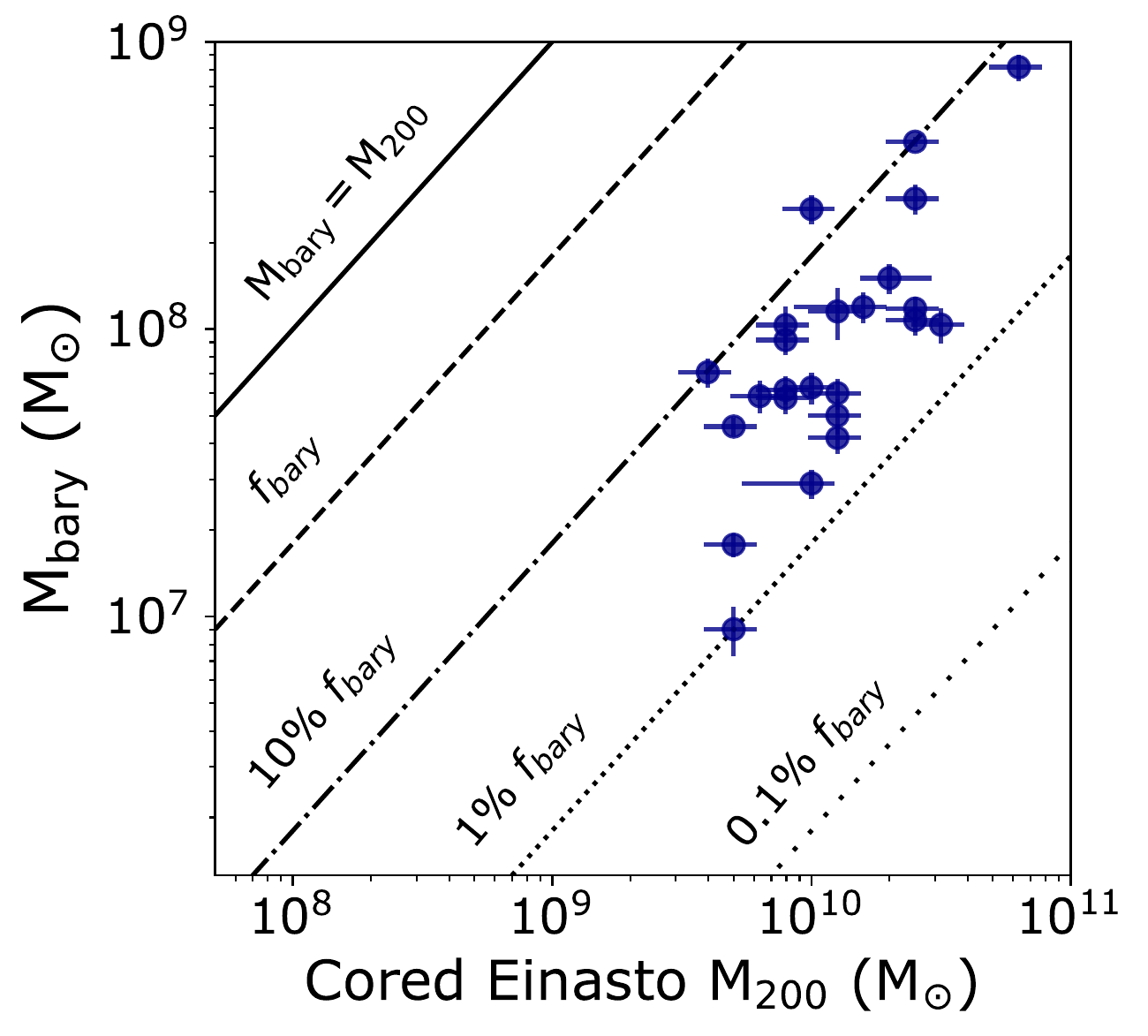}
\end{center}
\caption{\mbary\ versus \mhalo\ determined from fitting \vcor\ and \rmax\ to rotation curves generated from the Einasto profile (left) and the cored Einasto profiles (right). Diagonal lines drawn from the lower left to the upper right mark \mbary\ $=$ \mhalo\ (solid), the cosmic baryon fraction \fb\ where \mbary/\mhalo $=0.18$ (dashed), 10\% \fb\ (dashed-dotted), 1\% \fb\ (dotted), 0.1\% \fb\ (wide-dotted). }
\label{fig:baryon_fractions}
\end{figure*}

The left panel of Figure~\ref{fig:btfr_fits} shows that estimating \vmax\ based on Einasto rotation curves is a poor match to expectations. The galaxies move to only slightly higher velocities on the BTFR and still generally lie to the left of the BTFR defined based on the linear extension of the more massive galaxies. This is as expected given that \vmax\ estimated for the majority of galaxies is $\sim$\vcor\ (see Section~\ref{sec:compare_einasto} and figure set in Appendix~\ref{app:rotcurves}), which conflicts with the observation that the velocities in a majority of the galaxies are still rising.

The right panel of Figure~\ref{fig:btfr_fits} shows that estimating \vmax\ based on the cored Einasto rotation curves places galaxies with \mbary\ $>10^8$ \msun\ on the BTFR in an area that is consistent with expectations from more massive systems. In contrast, a number of galaxies with \mbary\ $<10^8$ \msun\ now lie to the right of the extrapolated best-fitting line from more massive galaxies. These points are significantly shifted to higher velocities because \vmax\ is extrapolated to be larger than \vcor, as discussed in Section~\ref{sec:compare_einasto},  which is consistent with the rising rotation curves seen in the PV slices.

Interestingly, \vmax\ estimated from the cored Einasto profiles results in a turn-down in the BTFR that is similar to the predictions from two different hydrodynamical simulations, although the turn-down appears to begin at a slightly lower value of \mbary. The rough agreement is somewhat surprising given the number of assumptions in the analysis and that the predictions are based on different star formation and feedback prescriptions used in the FIRE simulations (upon which the shape of our cored Einasto profiles are based), in the NIHAO simulations, and in the {\sc apostle} simulations. Note that the {\sc apostle} simulations do not produce cores in galaxy density profiles \citep{Sawala2016}, but, as these simulations show a sharp drop in galaxy formation efficiencies in dwarf galaxies and broadly reproduce the observed galaxy mass function and the sizes of galaxies as a function of mass, they recover a steep BTFR at low-masses \citep{Sales2017, Oman2016}. However, there is a notable difference in the gas fractions in the {\sc apostle} galaxies in the sense that they have lower $M_{gas}$/\mstar\ by a factor of 3 to 4 on average compared with observations of galaxies over the range \mstar $\sim10^6-10^{10}$ \msun\ \citep{Sales2017}. The low gas fractions may impact the radius at which the velocity of the simulated galaxies is measured, but it is unclear if this is the case, or how this impacts our comparison.

\subsection{Baryon Fractions}\label{sec:baryon_fractions}
The relative amount of baryonic to dark matter mass in the universe, or the cosmic baryon fraction (\fb), is a fundamental constraint on the distribution of matter. Detailed measurements of the comic baryon budget yield estimates of \fb\ to be $\sim18$\% \citep{Planck2018}. Yet, only a fraction of the available baryons end up in galaxies. Here, we use the estimates of \mhalo\ for our sample to constrain the baryonic fraction in low-mass galaxies. 

Figure~\ref{fig:baryon_fractions} presents the measurements of \mbary\ as a function of \mhalo\ based on the fits to the Einasto (left) and cored Einasto (right) rotation curves. Baryon fractions based on the Einasto profiles fall predominantly between 10-100\% \fb. These values are similar to the baryon fraction of  $\sim$20\% in spiral galaxies at the low-end and reach \fb\ at the high-end. Such high baryon fractions are unrealistic as there is no sensible mechanism to explain why such low mass halos would be able to accrete and retain such a high fraction of baryons, reinforcing how poor the fits are to a cuspy Einasto profile. In contrast, because the cored Einasto profile mapped the galaxies to more massive halos, the baryon fractions based on the cored Einasto profiles are lower, falling mainly between 1-10\% \fb, with a steep decline at \mhalo\ $\sim10^{10}$ \msun. 

\subsection{Stellar Mass - Halo Mass and Baryon Mass - Halo Mass Comparisons}\label{sec:smhm}
Figure~\ref{fig:smhm_bmhm} compares \mstar\ with \mhalo\ for our galaxies (i.e., the SMHM relation) based on fits to the cored Einasto rotation curves. The ratio of \mstar/\mhalo\ falls below $10^{-2}$ for the sample and lies closer to $10^{-3}$ for the majority of galaxies. The distribution of points is in agreement with previous work on the SMHM relation for gas-rich dwarfs \citep[e.g.,][]{Read2017}. 

In Figure~\ref{fig:smhm_bmhm} we also add a comparison of \mbary\ to \mhalo, which can be thought of as the baryonic mass - halo mass (BMHM) relation. Points are color-coded by the log of the gas fraction; see color bar for range. The BMHM relation sits mainly above the SMHM, which is simply a result of adding in the gas mass. Note, also, that $M_{gas}$/\mstar\ ranges from 0.8 to 21 for our sample, resulting in a slight increase in the distribution of galaxy mass at a given \mhalo.

\section{Discussion}\label{sec:discuss}
We have combined robust measurements of \mbary\ with constraints on the gas kinematics in the largest sample of galaxies below \mbary\ $\sim10^8$ \msun\ studied to date to explore where galaxies lie at the low-mass end of the BTFR. Our empirically determined values of the \hi\ rotational velocities, \vrot, were measured at relatively small radii and are thus lower limits on the maximum velocity, \vmax, for many of the galaxies. When placed on the BTFR after correcting for asymmetric drift (\vcor; Figure~\ref{fig:btfr_empirical}), these points largely lie at higher \mbary\ and/or lower \vmax\ than expectations based on both extrapolation from higher mass galaxies and from galaxy formation models. This mis-match between the measured and expected values is as expected given that the velocities trace only the inner gravitational potentials. 

To estimate \vmax\ for the galaxies, we have matched the measured velocities and spatial extents of each system to families of theoretical rotation curves based on an Einasto and a cored Einasto profile. The best-fitting rotation curve yields estimates of \mhalo, \vmax, and, when combined with measurements of \mbary, constraints on the baryon fraction. 

The results based on fitting to the Einasto rotation curves that are not significantly modified by baryon physics and remain cuspy are clearly a poor match to the data: \vmax\ is estimated to be $\sim$ \vcor\ for a number galaxies, which we know is not correct based on the galaxies' PV slices. When placed on the BTFR (Figure~\ref{fig:btfr_fits}, left panel), the galaxies lie to the left of expectations without a viable physical explanation for why lower mass halos would be better at accreting and retaining baryons. The resulting baryon fractions are consistent with or higher than values determined for $L_*$ galaxies (i.e., $\sim$20\% \fb); in a few cases, the estimated baryon fractions approach the cosmic baryon fraction of $\sim18$\%. As previously mentioned, the Einasto profile was chosen as a reference point for fitting to cuspy dark matter halo profiles. Given the strong evidence that stellar feedback can dramatically impact the properties of low-mass galaxies, it is expected that fitting the galaxies to rotation curves that remain cuspy produces unphysical results.

\begin{figure}
\begin{center}
\includegraphics[width=0.5\textwidth]{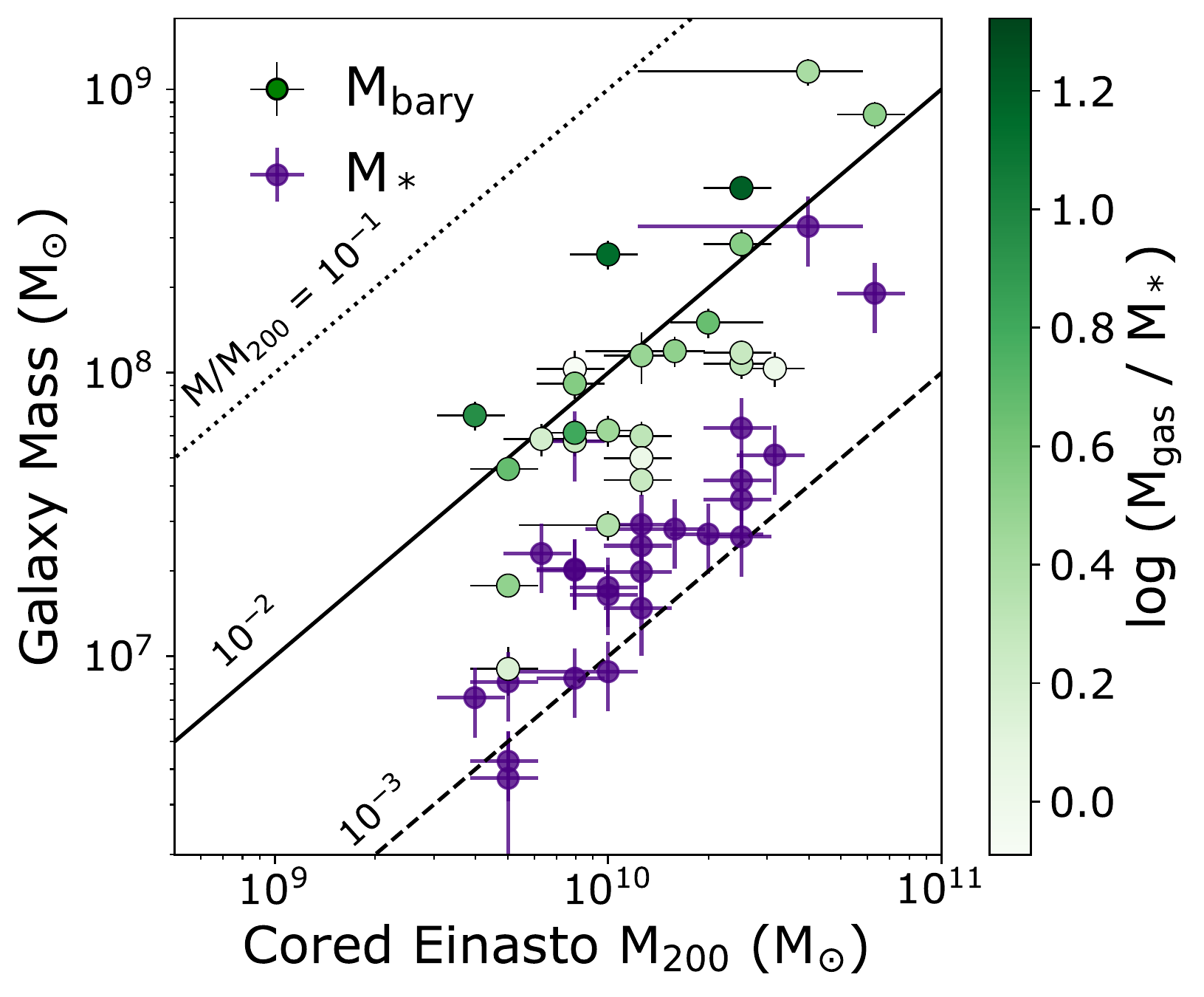}
\end{center}
\caption{The SMHM and BMHM relations showing \mstar\ (purple) and \mbary\ (green) versus \mhalo\ determined from fitting \vcor\ and \rmax\ to rotation curves generated from the cored Einasto profiles. The BMHM points are color-coded by log ($M_{gas}$/\mstar).  Diagonal lines drawn from the lower left to the upper right mark Mass/\mhalo\ ratios of $10^{-1}$ (dotted), $10^{-2}$ (solid), and $10^{-3}$ (dashed). The BMHM shows a larger vertical spread than the SMHM, driven by the different gas fractions in the galaxies. The larger range of \mbary\  at \mhalo$\sim10^{10}$ \msun\ is just the turndown in the BTFR seen now in \mhalo\ instead of \vmax. In addition, galaxies with higher \mbary/\mhalo\ ratios tend to have ``under-sized'' stellar disks relative to their gas content.}
\label{fig:smhm_bmhm}
\end{figure}

Using cored Einasto profiles, we find a turn-down in the BTFR that is consistent with theoretical expectations based on simulations that include baryon physics and CDM. The main difference is in the mass at which the turn-down begins. The predictions from the high-resolution NIHAO simulations \citep{Dutton2017} and {\sc apostle} simulations \citep{Sales2017} show a turn-down in the BTFR at $\sim45$ \kms\ at \mbary\ $\sim10^8$ \msun, with \vrot\ declining steeply to $\sim30-35$ km s$^{-1}$ for \mbary\ $\sim10^7$ \msun. The turn-down suggested by our analysis begins at a slightly lower \mbary, although the small number of points with \mbary\ $<10^{7.5}$ \msun\ limits the current comparison. 

Such a turn-down in the BTFR translates to a decreasing baryon fraction at low masses. Few robust constraints on \fb\ exist below \mbary\ $\sim10^8$ \msun\ due to the challenges in mapping low-mass galaxies to their halo masses, although values as low as 2\% have been suggested for slightly more massive galaxies from the ALFALFA survey \citep{Papastergis2012}. Our results, shown in Figure~\ref{fig:baryon_fractions},  suggest that galaxies in the mass regime \mbary\ $\sim10^7-10^9$ \msun\ have a fairly narrow range in their baryon fraction from $\ltsimeq$ 1-10\% \fb, with a steep drop in the baryon-to-dark matter mass below \mbary $\sim10^8$ \msun. Qualitatively, this steep drop is expected at low galaxy masses, resulting from the combined impact of a suppressed rate of baryon accretion at early times due to reionization and a lower retention rate over cosmic timescales mainly due to baryonic physics \citep[e.g.,][]{Hoeft2006, Guo2010, Munshi2013, Trujillo-Gomez2018, Benitez-Llambay2020, Katz2020, Munshi2021}. In this scenario, below some halo mass scale, gas accretion and star formation is suppressed and/or gas mass is lost as a result of stellar feedback and heating from reionization, which brings about a decline in the baryon fraction as a function of halo mass and a turn-down in the BTFR, analogous to a change in slope, or `knee', in the SMHM relation. Our analysis of the BTFR in Figure~\ref{fig:btfr_fits} suggests a lower limit to \vmax\ of $\sim30$ \kms, corresponding to \mhalo$\sim10^{10}$ \msun. 

We also investigated the SMHM and BMHM relations for the sample in Figure~\ref{fig:smhm_bmhm}, where the BMHM is similar to the BTFR but plotted with \mhalo\ instead of \vmax. While the SMHM is more readily measured from optical surveys and used in abundance matching of galaxies, the BMHM provides a more fundamental approach to connecting baryons in gas-rich galaxies to their halos. The BMHM relation shows a slight increase in the range in galaxy masses that occupy a given halo mass over the SMHM in Figure~\ref{fig:smhm_bmhm}, driven by the different gas fractions in the galaxies. This vertical spread is another representation of the turndown in the BTFR. Interestingly, the galaxies with the highest gas fraction also tend to have the highest ratio of \mbary/\mhalo. In other words, galaxies with the largest reservoirs of \hi\ have less massive stellar components and reside in halos with less dark matter mass per unit baryon mass, compared to systems with lower gas fractions. It is an open question as to why some low-mass galaxies are less efficient in forming stars even though they are well-fueled with gas. There is some observational evidence that lower efficiency may occur in galaxies with a high angular momentum distribution \citep[and a subsequently less concentrated halo;][]{ManceraPina2021}. The lower mass stellar disks would also suggest less overall feedback-driven baryon loss, which may help explain the higher baryon fractions in these systems, although less feedback would also mean there is less energy injected into the ISM to prevent stars from readily forming. 

Note that our results are based on just one realization of fitting to DM halos. Given the diversity in rotation curves and the limited extent of the detected \hi\ from which we are measuring the velocities (i.e., at small radii where there is less diagnostic power from the rotation curves in matching to a halo), the interpretation is not straightforward. Small changes in the shape of the real rotation curve can result in systematic uncertainties in \mhalo\ and \vmax\ when mapping to inherently uniform families of theoretical rotation curves. Thus, it is unclear whether \mhalo\ and the corresponding values of \vmax\ are the true values for individual galaxies. Furthermore, detailed kinematic analysis of simulated dwarfs that are slightly more massive than our sample connect the diversity in rotation curves to the effects of random, non-circular motions in the central regions of galaxies \citep{Oman2019, SantosSantos2020, Roper2022}. These motions introduce errors that impact the interpretation of the observed velocities, and allow cuspy density profiles to appear cored, which suggests an alternative interpretation of our results.

Our current sample of 25 galaxies begins to constrain the \mbary\ $-$ \mhalo\ relation, but our conclusions are limited by the caveats listed above. Future work will include additional galaxies from the SHIELD program with new high-resolution \hi\ synthesis imaging from the VLA with resolved rotation curves and mass decomposition analysis that can more directly measure \mhalo\ and \vmax. With a larger sample of galaxies with \mbary\ $<10^8$ \msun\ and resolved rotation curves, the mean relation between \mbary\ and \mhalo\ should be more definitive. It would be very useful to increase the sample with low \hi\ masses and extended \hi\ distributions.

\begin{acknowledgments}
KBWM thanks Yao-Yuan Mao for helpful discussions on dark matter density profiles. EAKA is supported by the WISE research programme, which is financed by the Dutch Research Council (NWO). JMC, JF, and JLI are supported by NSF/AST-2009894. MPH acknowledges support from NSF/AST-1714828 and grants from the Brinson Foundation. This research has made use of NASA Astrophysics Data System Bibliographic Services and the NASA/IPAC Extragalactic Database (NED), which is operated by the Jet Propulsion Laboratory, California Institute of Technology, under contract with the National Aeronautics and Space Administration. 
\end{acknowledgments}

\software{This research made use of Astropy,\footnote{http://www.astropy.org} a community-developed core Python package for Astronomy \citep{astropy:2013, astropy:2018}, and Colossus \citep{Diemer2018}.}

\renewcommand\bibname{{References}}
\bibliographystyle{apj}
\bibliography{ms.bbl}

\appendix
\vspace{-0.25in}
\section{Atlas of PV Diagrams and \vpv\ and \dpv\ Measurements }\label{app:pv}
\begin{figure*}[h!]
\begin{center}
\includegraphics[width=0.48\textwidth]{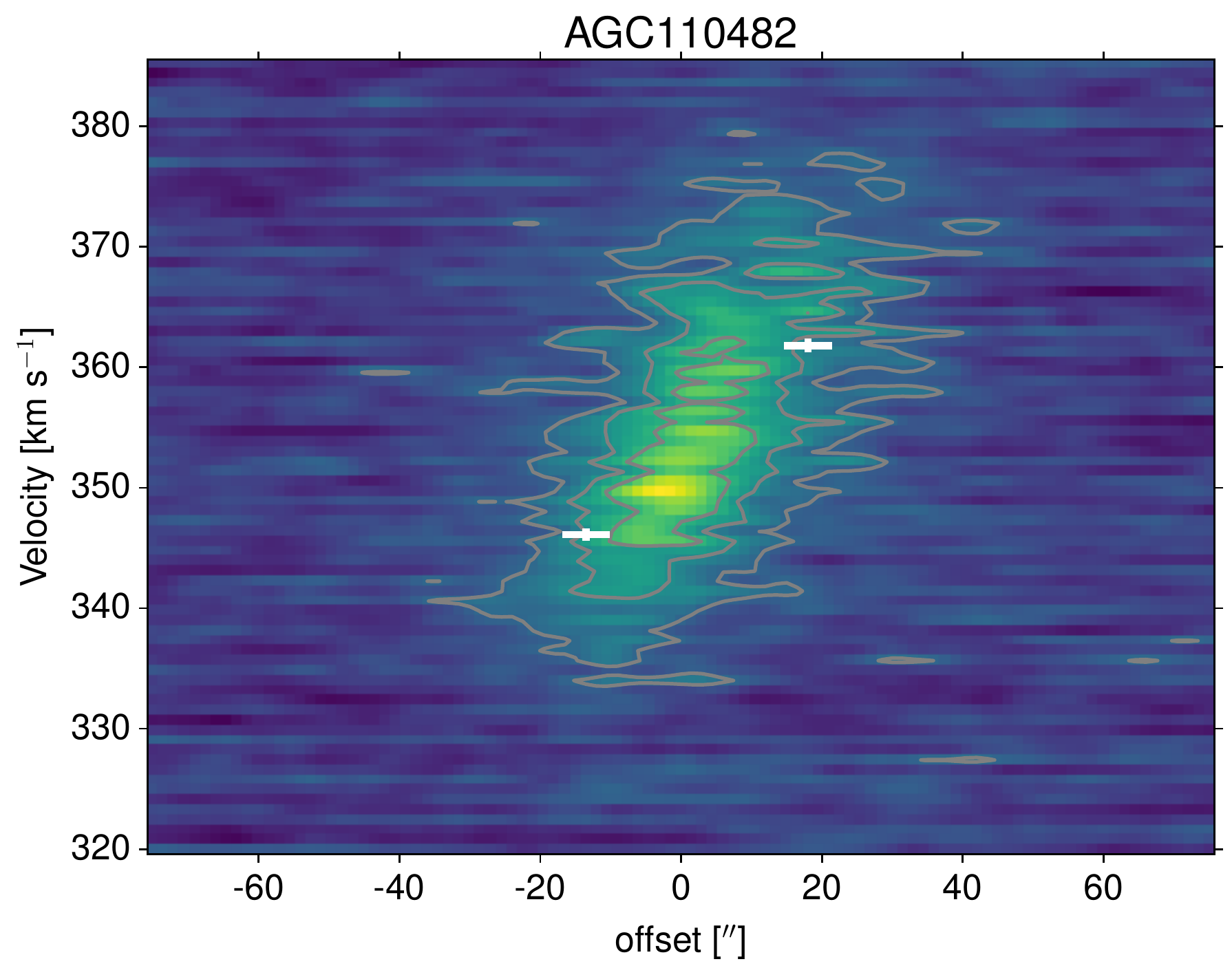}
\includegraphics[width=0.48\textwidth]{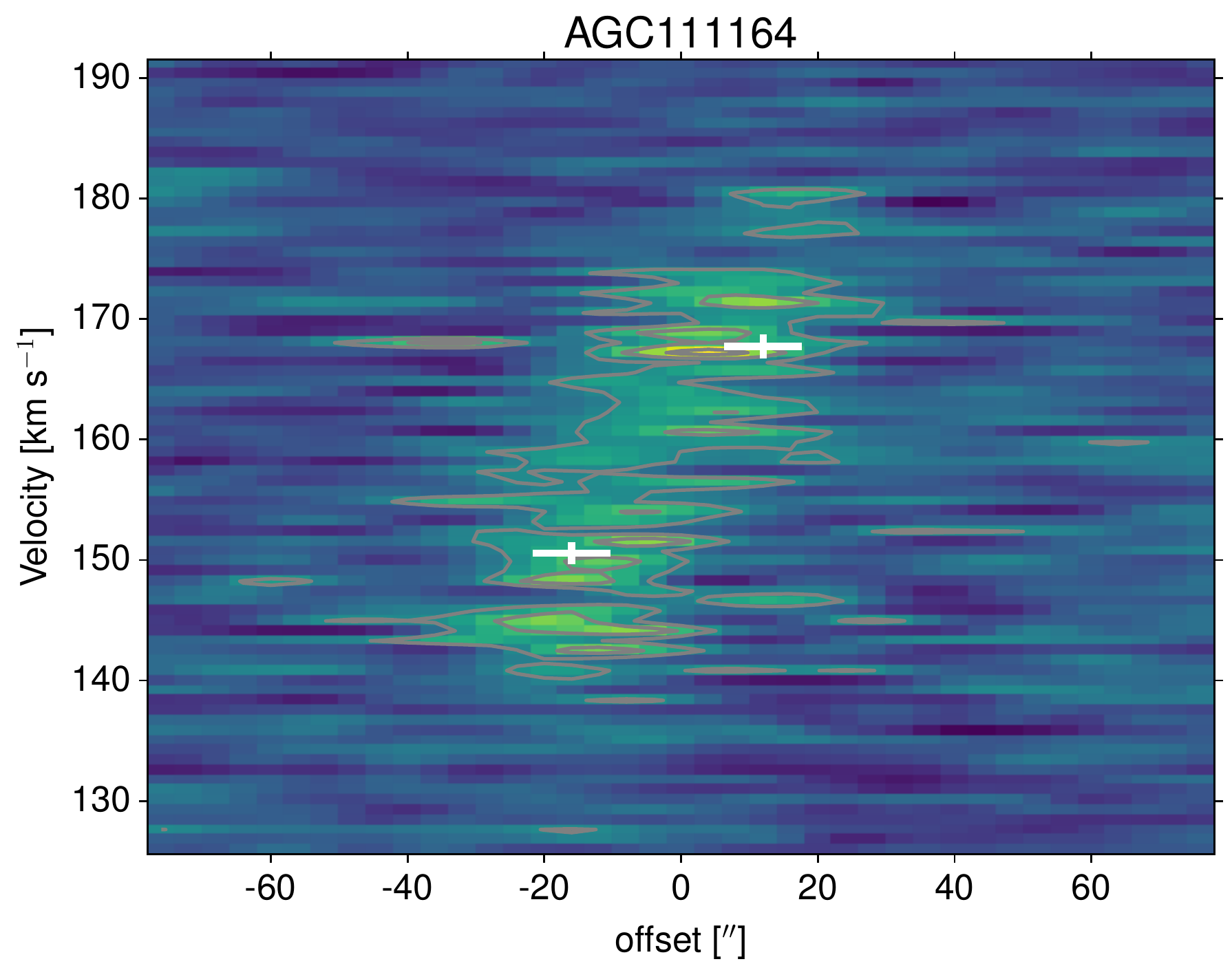}
\includegraphics[width=0.48\textwidth]{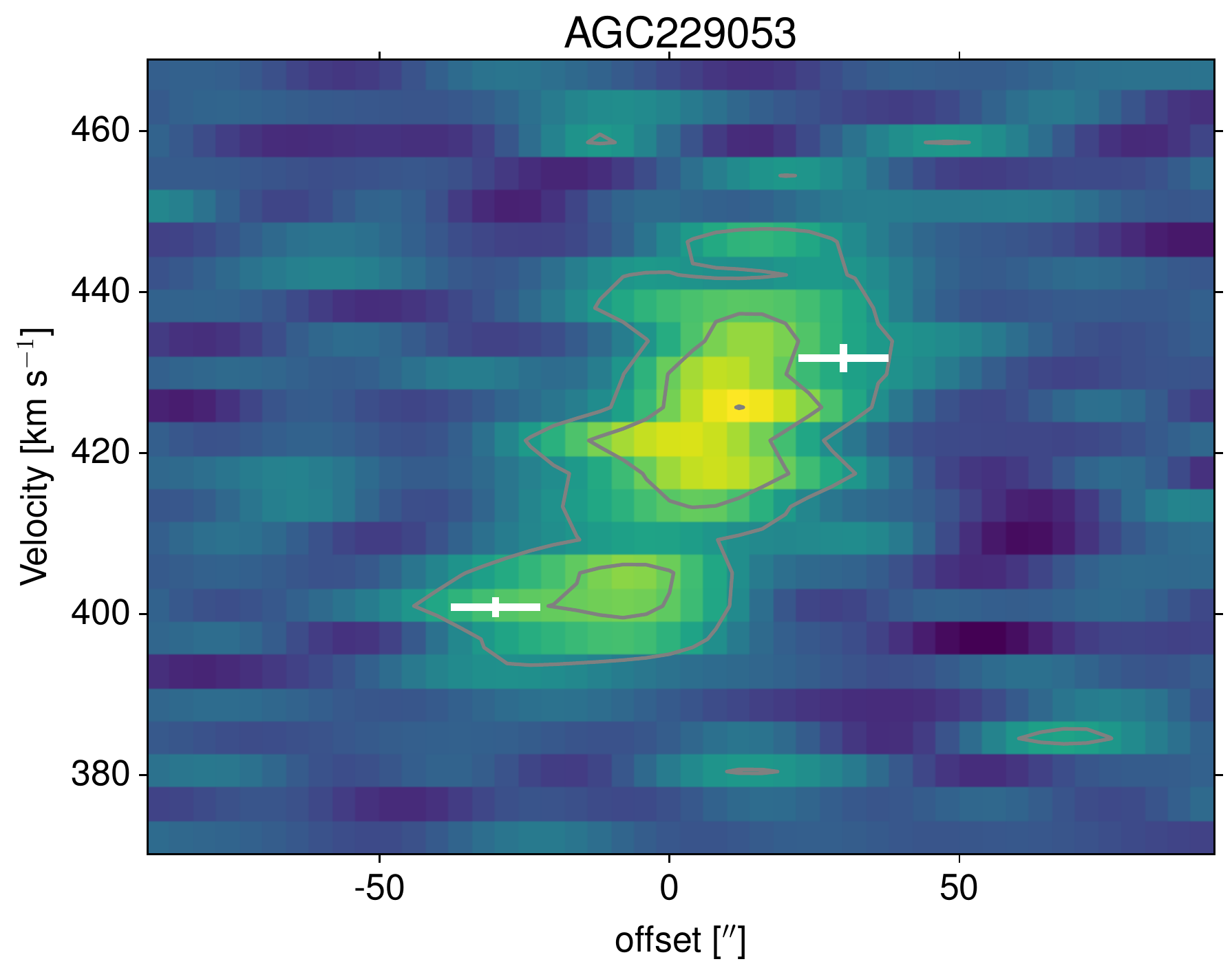}
\includegraphics[width=0.48\textwidth]{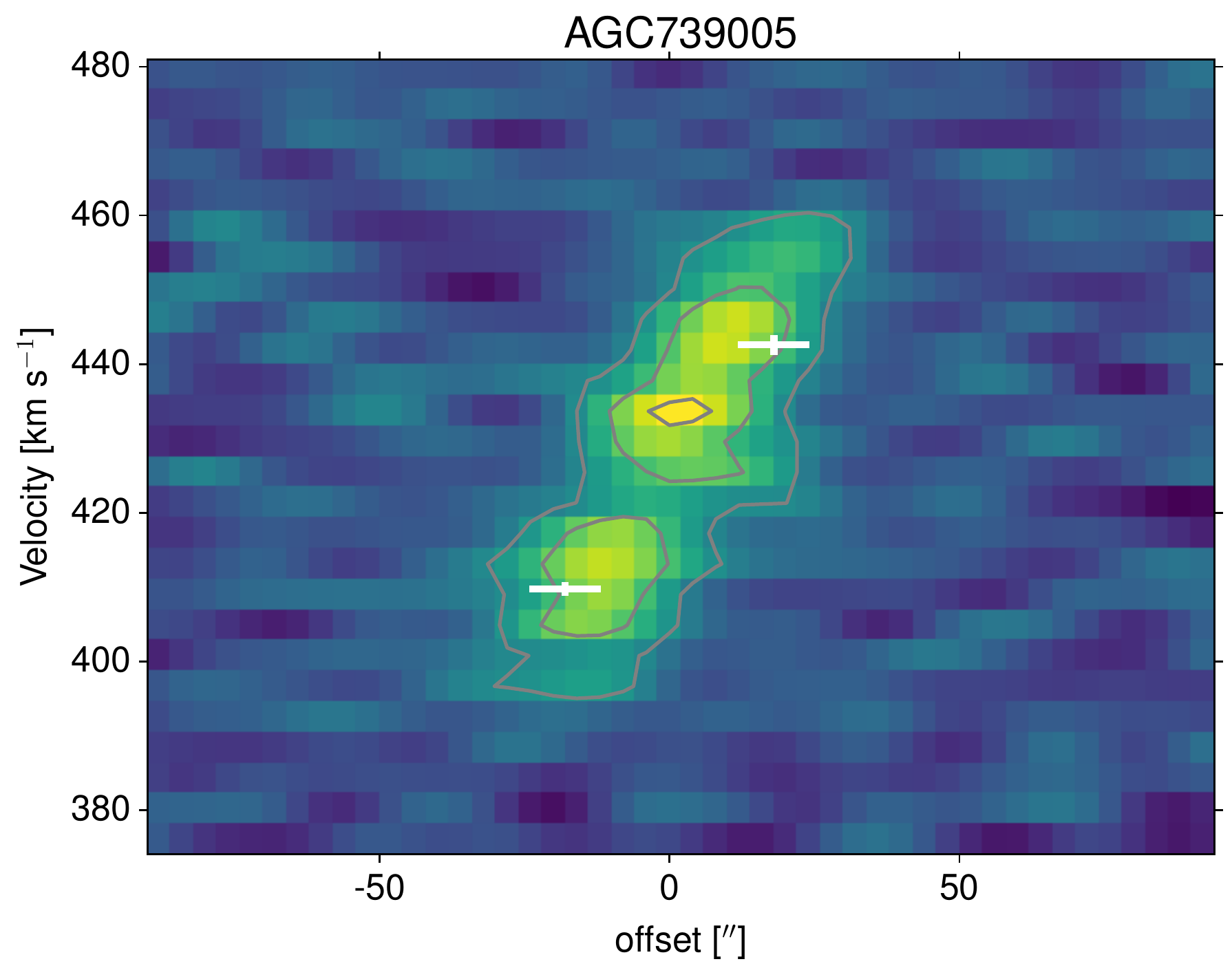}
\includegraphics[width=0.48\textwidth]{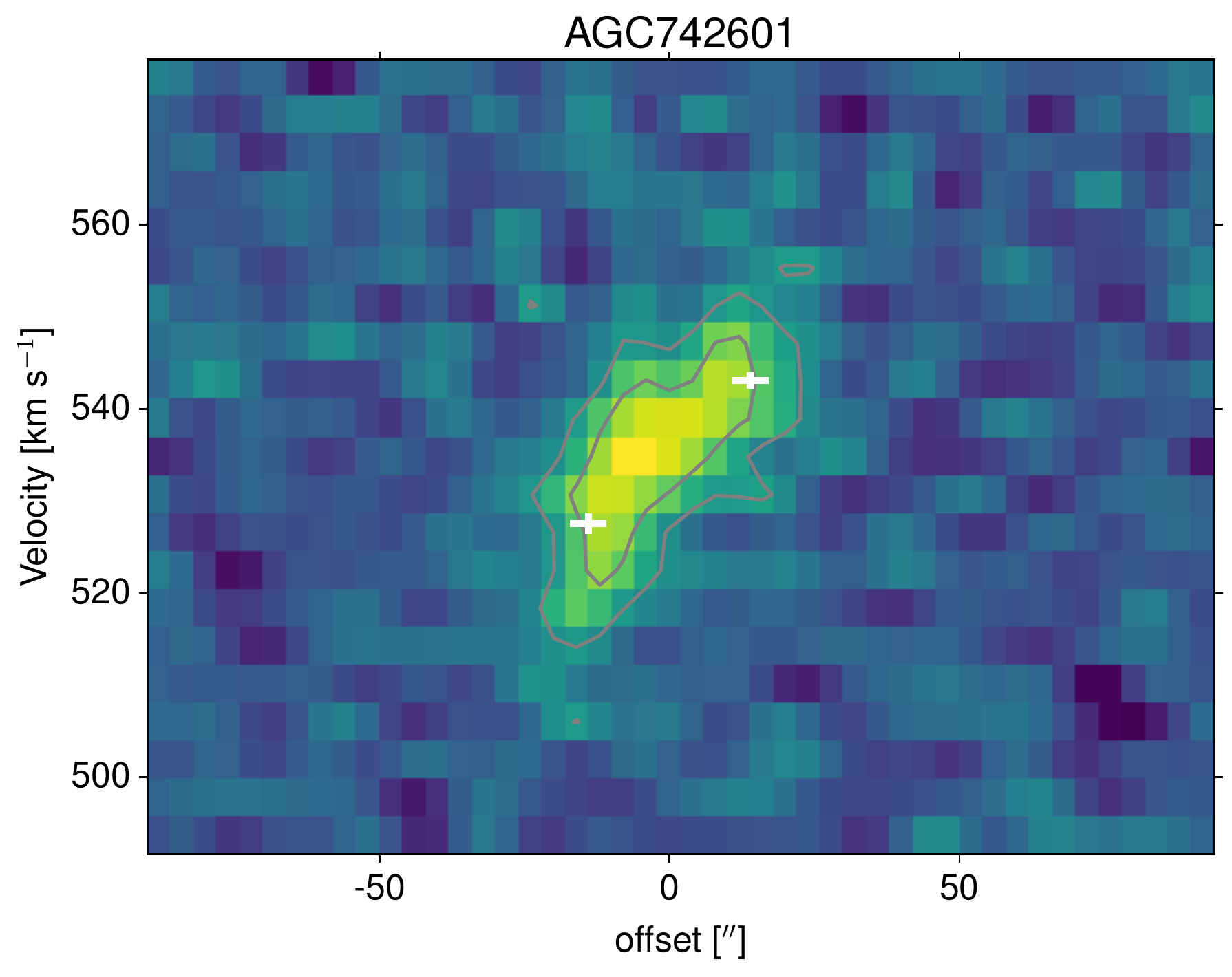}
\includegraphics[width=0.48\textwidth]{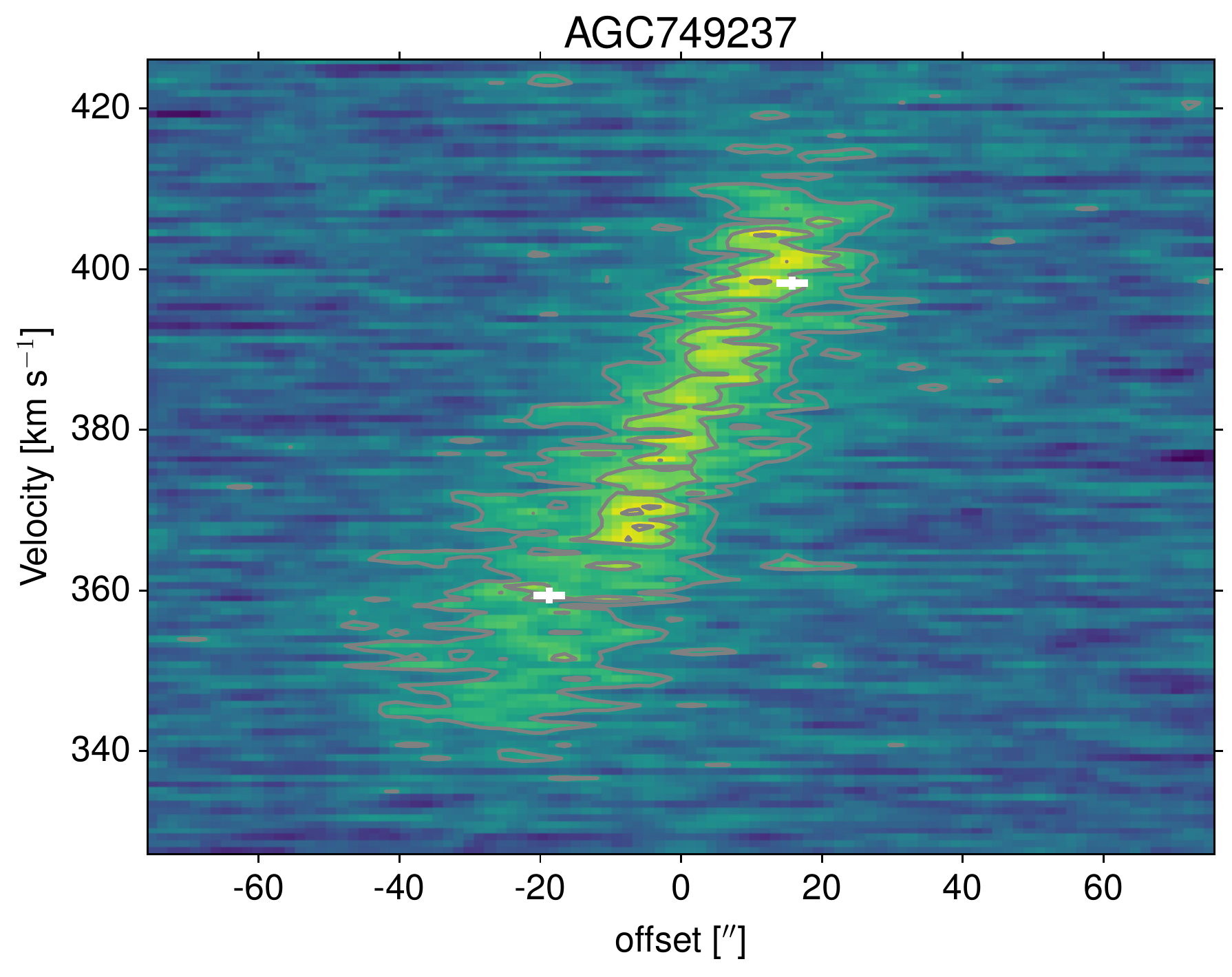}
\end{center}
\caption{PV diagrams for SHIELD galaxies AGC110482, AGC111164, AGC229053, AGC7393005, AGC742601, and AGC749237 with the location of the best-fitting \vpv\ and \dpv\ marked in white.}
\label{fig:pv_2}
\end{figure*}

\begin{figure*}
\begin{center}
\includegraphics[width=0.48\textwidth]{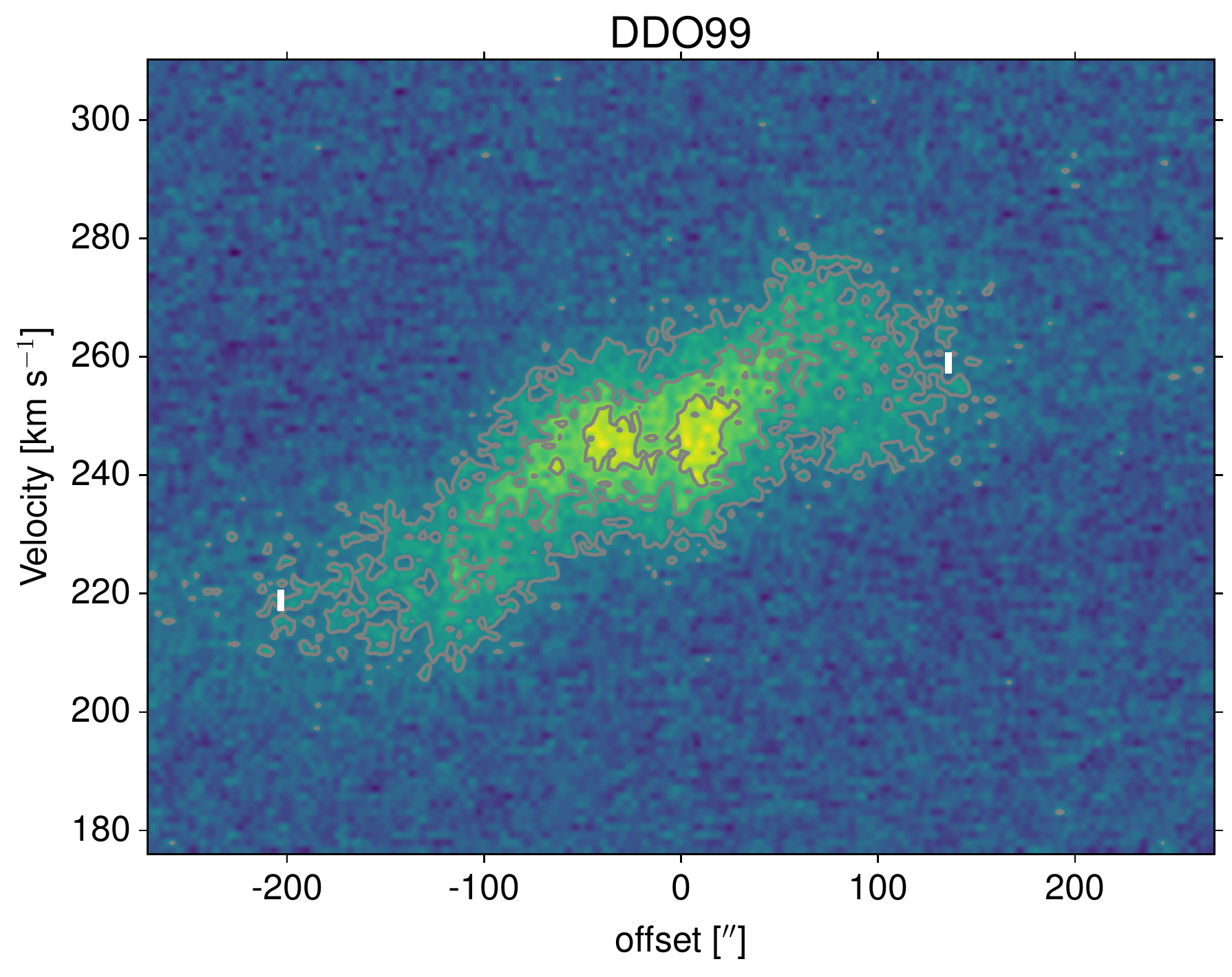}
\includegraphics[width=0.48\textwidth]{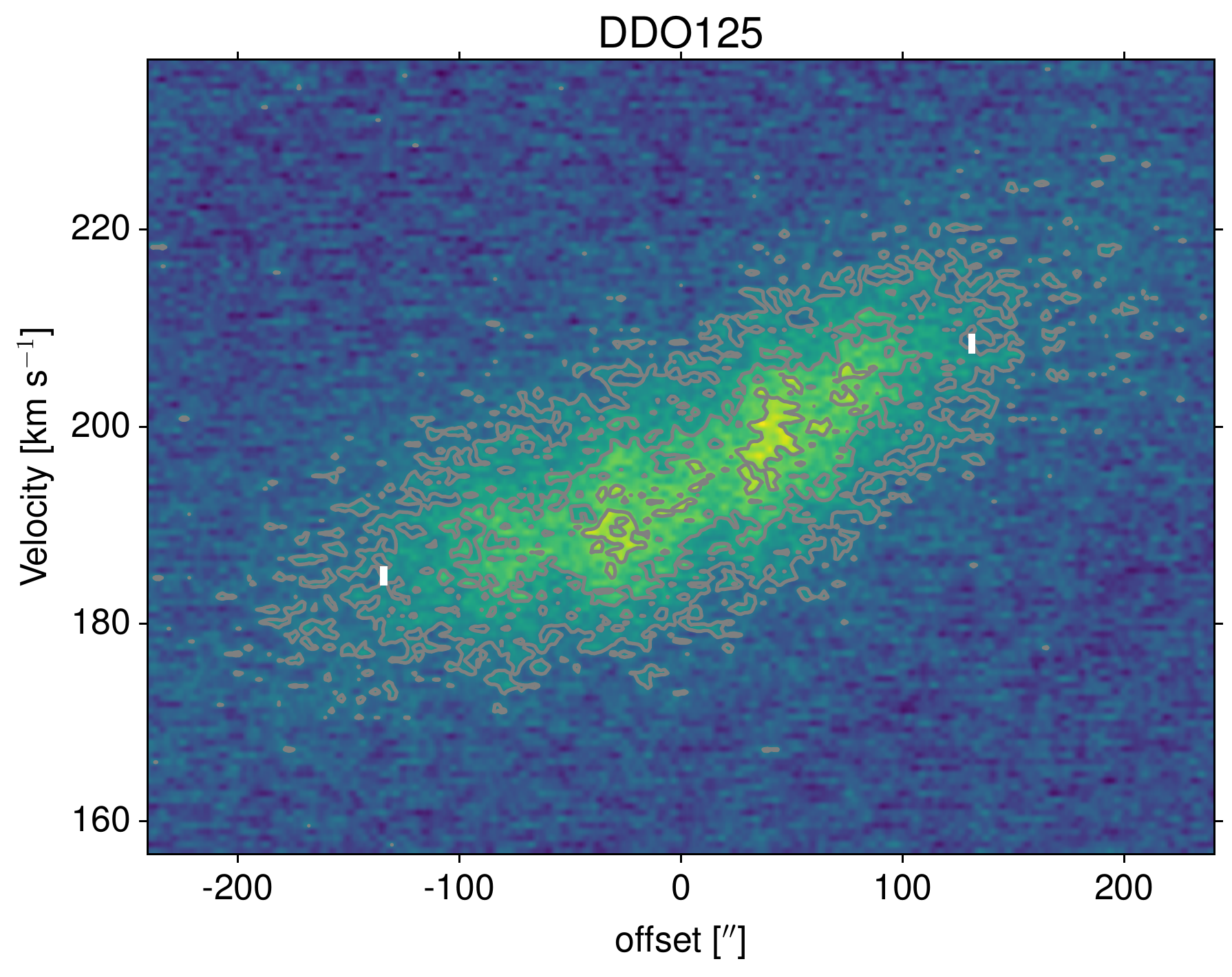}
\includegraphics[width=0.48\textwidth]{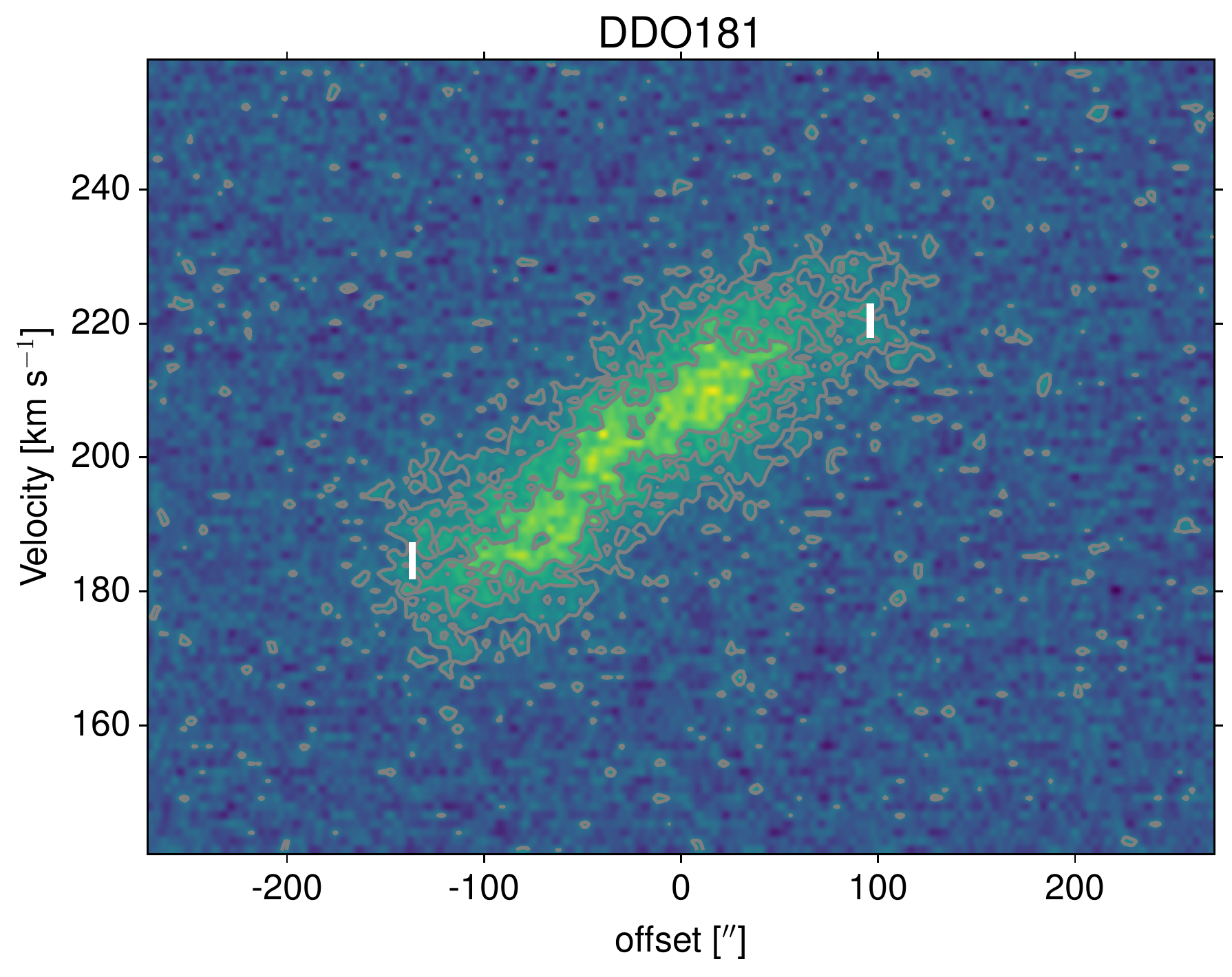}
\includegraphics[width=0.48\textwidth]{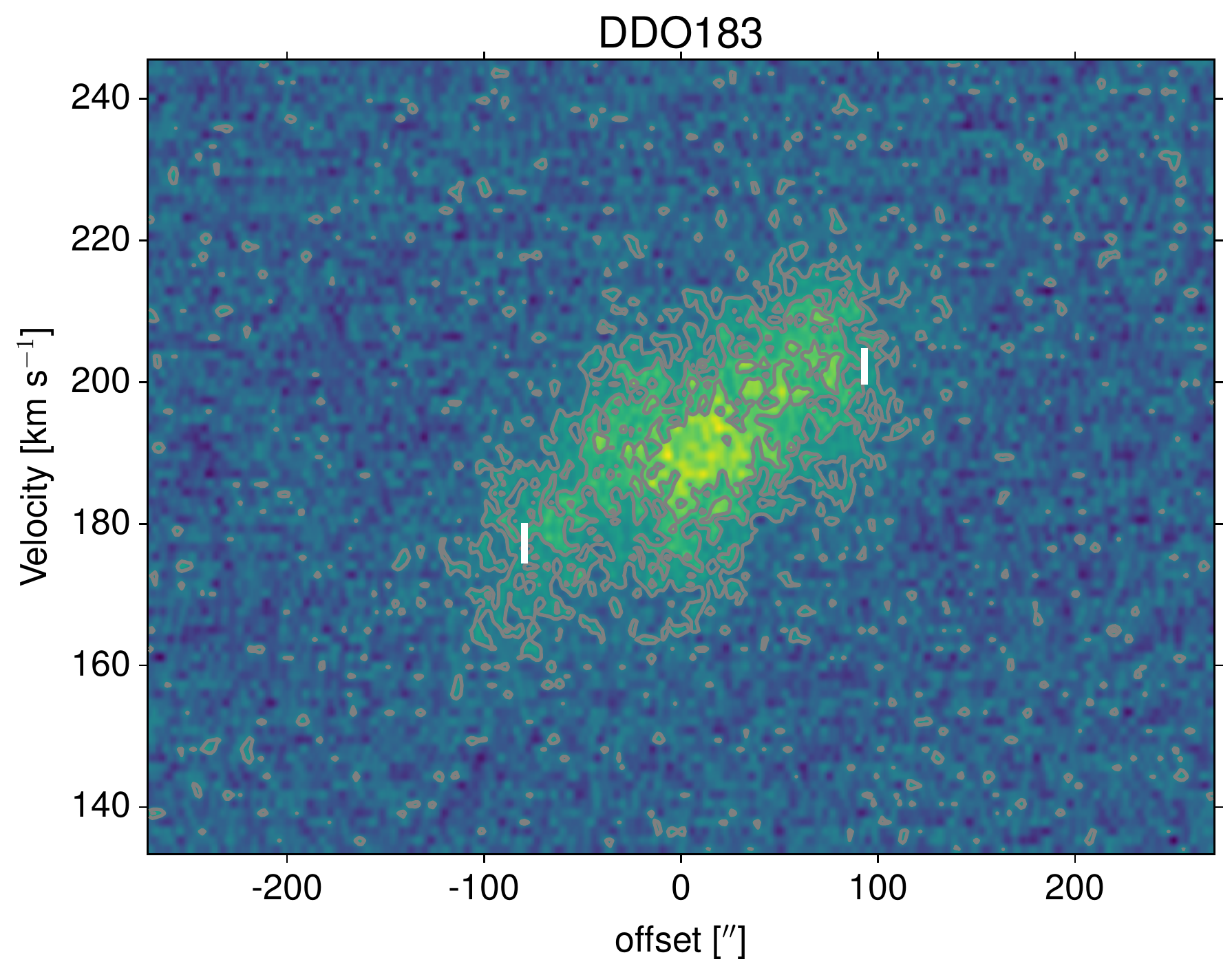}
\includegraphics[width=0.48\textwidth]{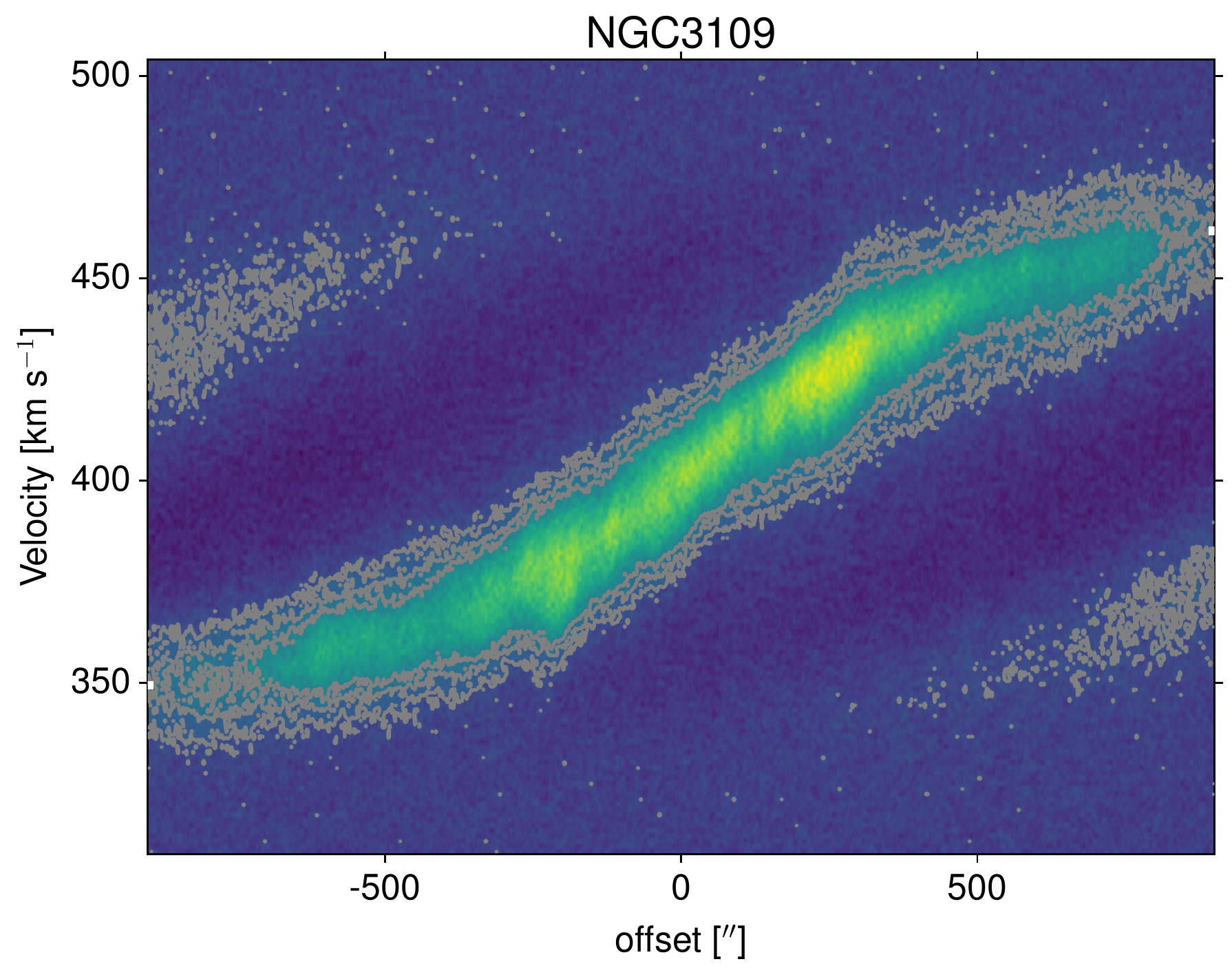}
\includegraphics[width=0.48\textwidth]{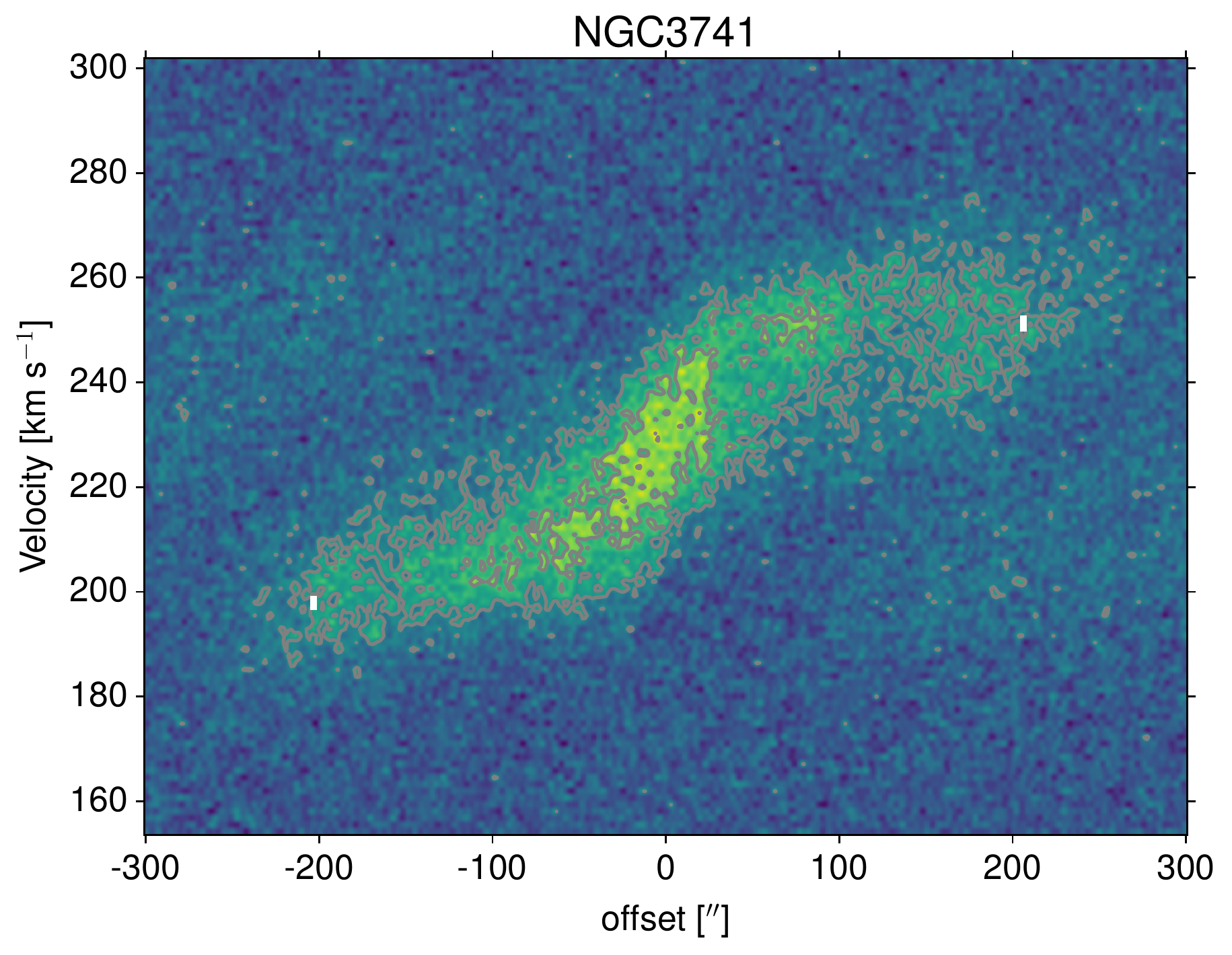}
\end{center}
\caption{PV diagrams for VLA-ANGST galaxies DDO99, DDO125, DDO181, DDO183, NGC3109, and NGC3741 with the location of the best-fitting \vpv\ and \dpv\ marked in white. Both NGC3109 and NGC3741 have cleaning ``bowl'' effects visible, but these do not impact our measurements of \vpv\ and \dpv.}
\label{fig:pv_3}
\end{figure*}

\begin{figure*}
\begin{center}
\includegraphics[width=0.48\textwidth]{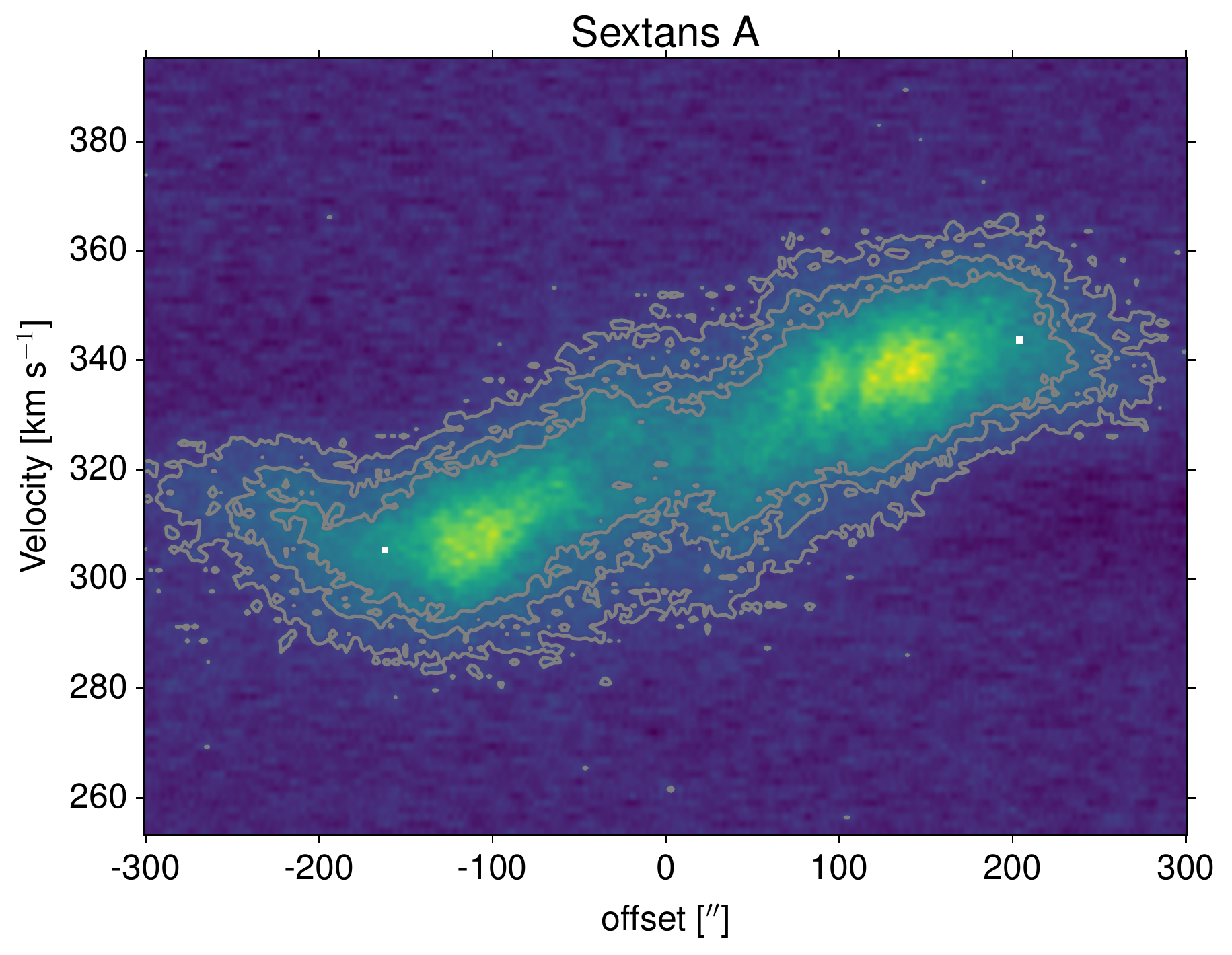}
\includegraphics[width=0.48\textwidth]{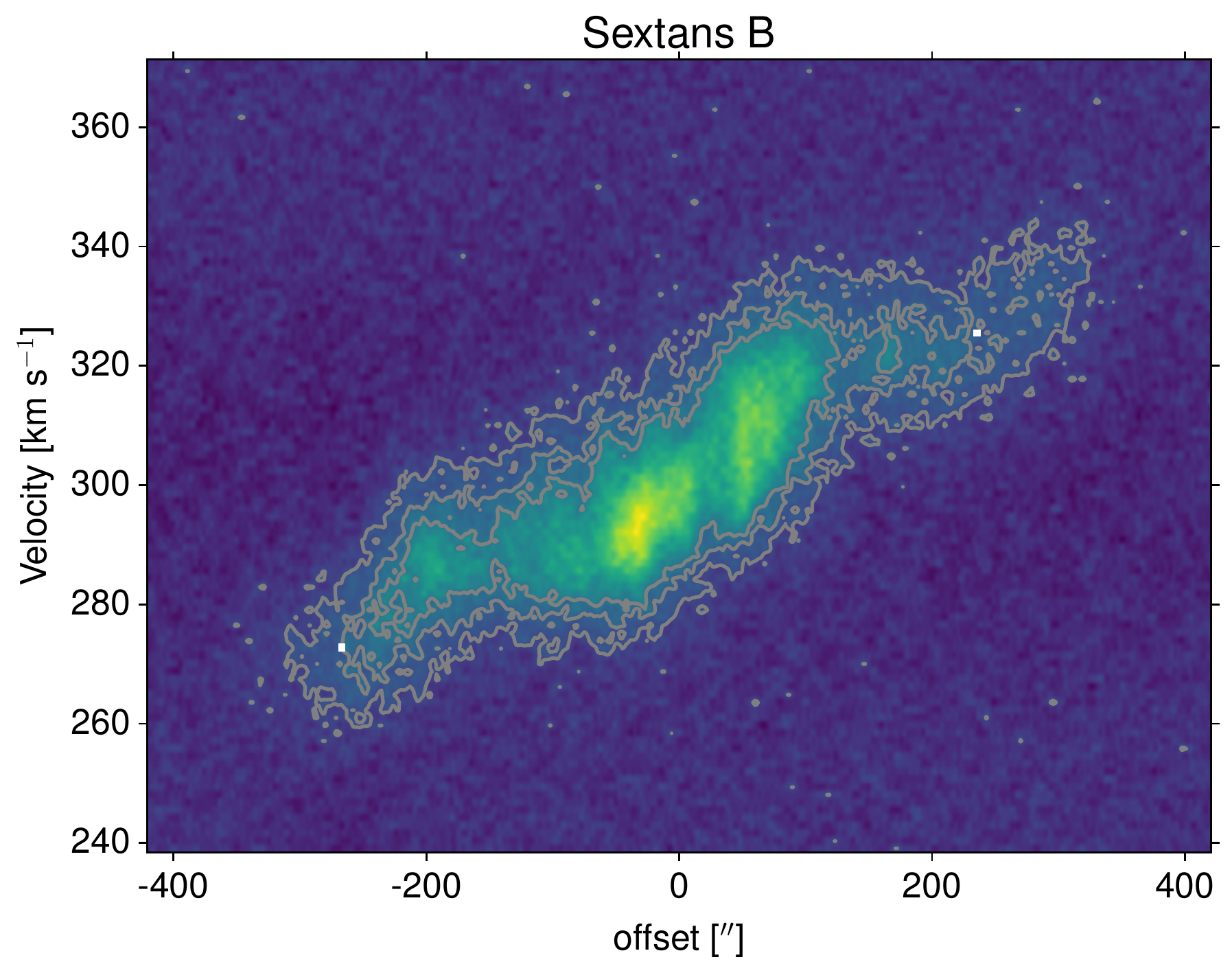}
\includegraphics[width=0.48\textwidth]{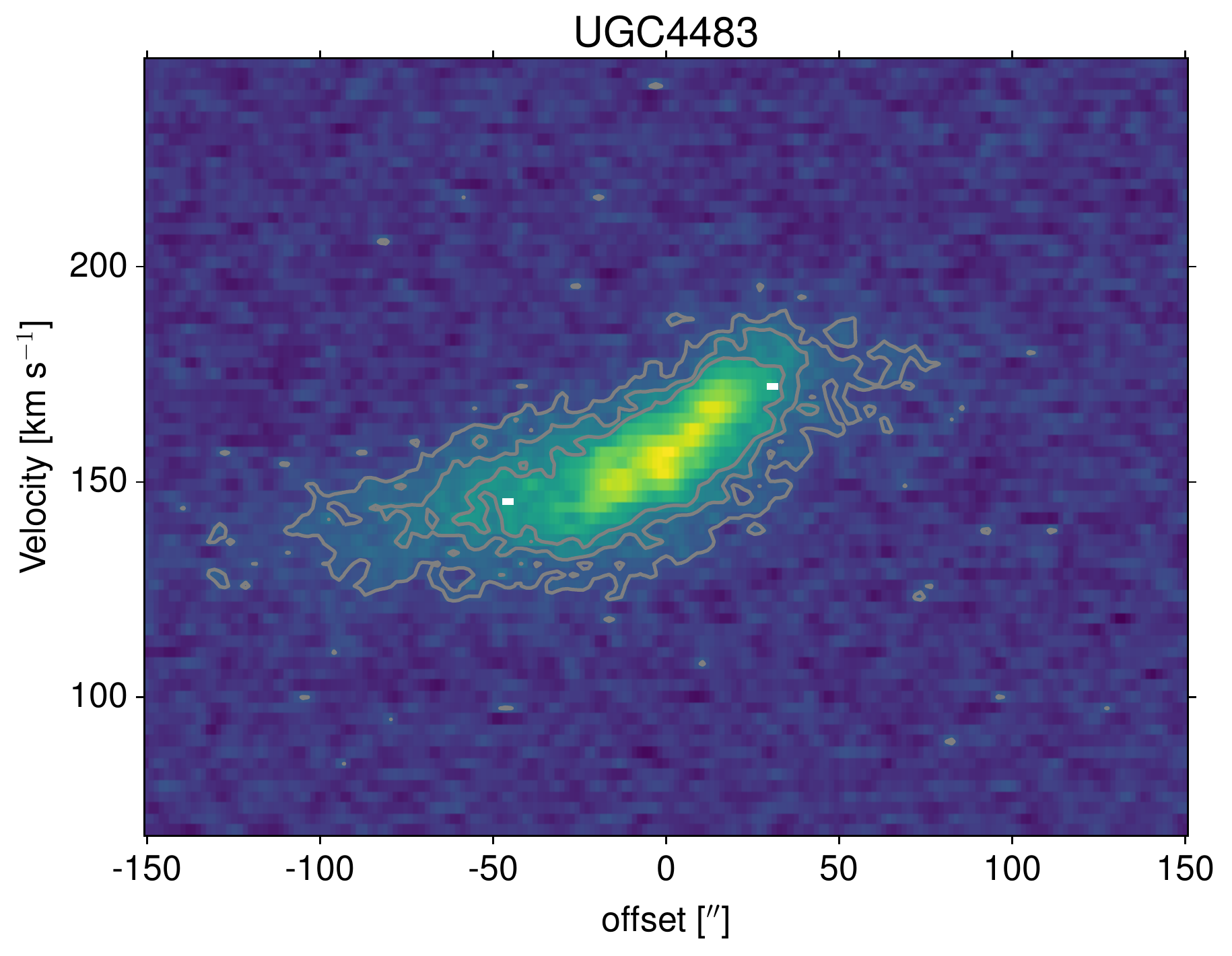}
\includegraphics[width=0.48\textwidth]{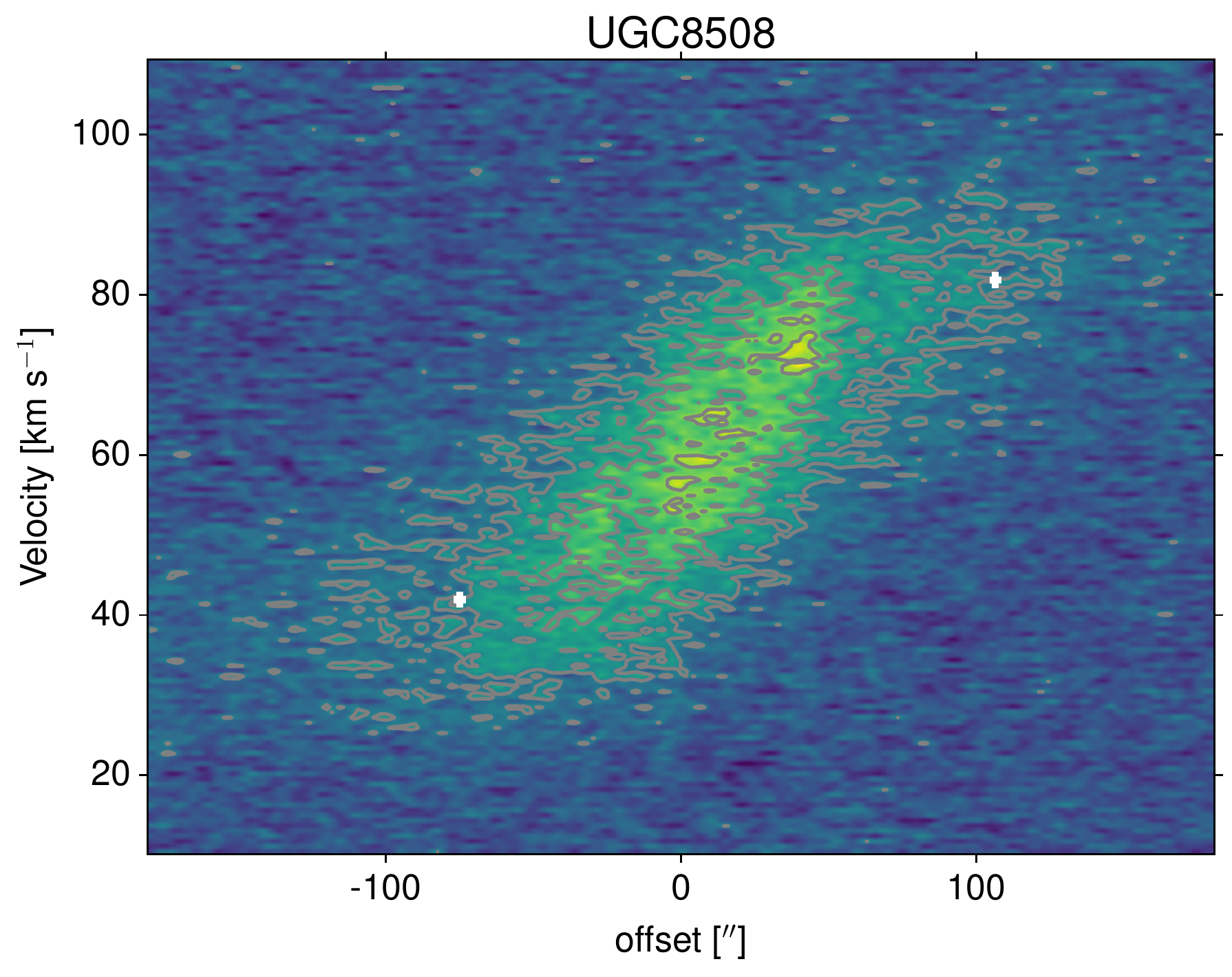}
\includegraphics[width=0.48\textwidth]{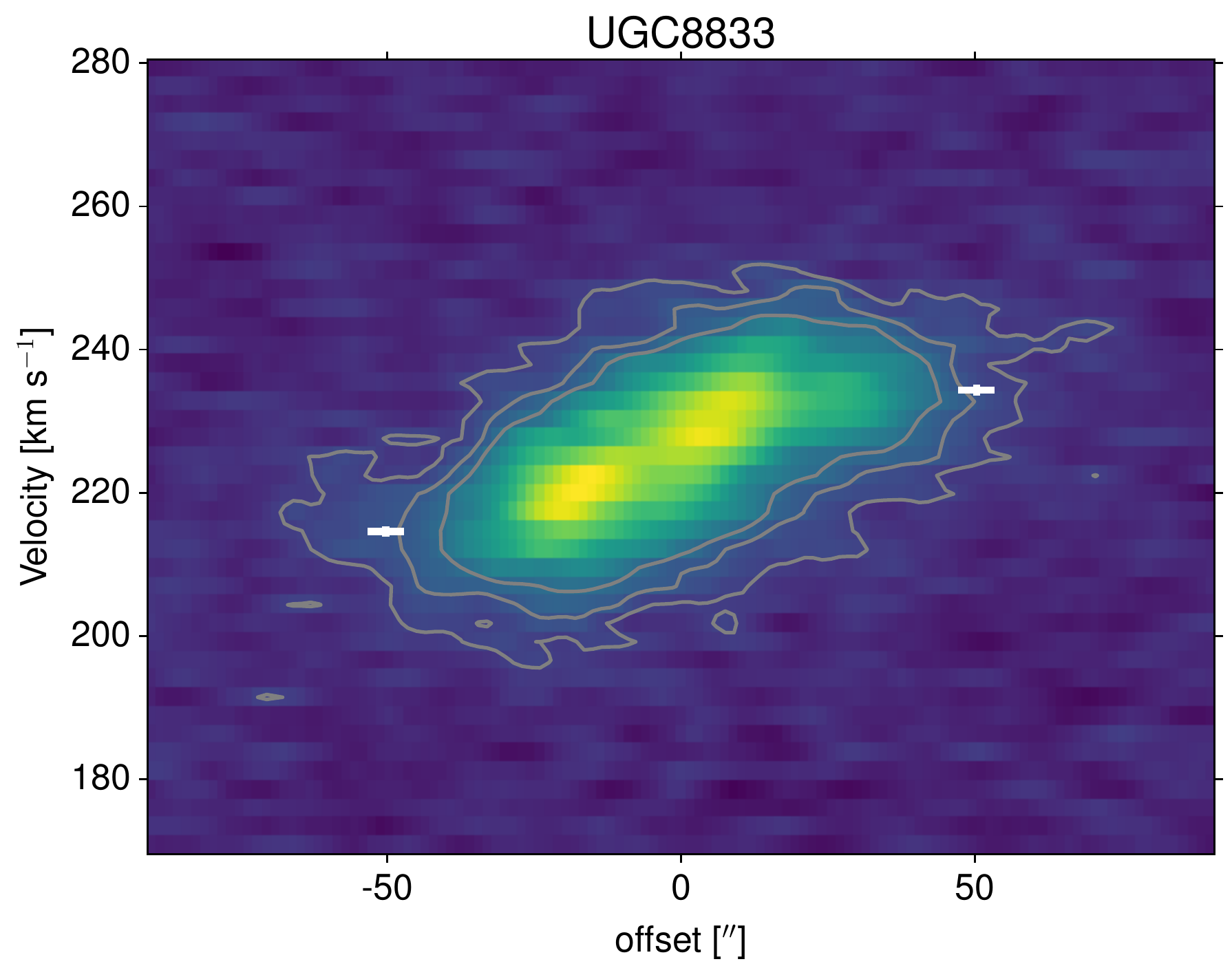}
\includegraphics[width=0.48\textwidth]{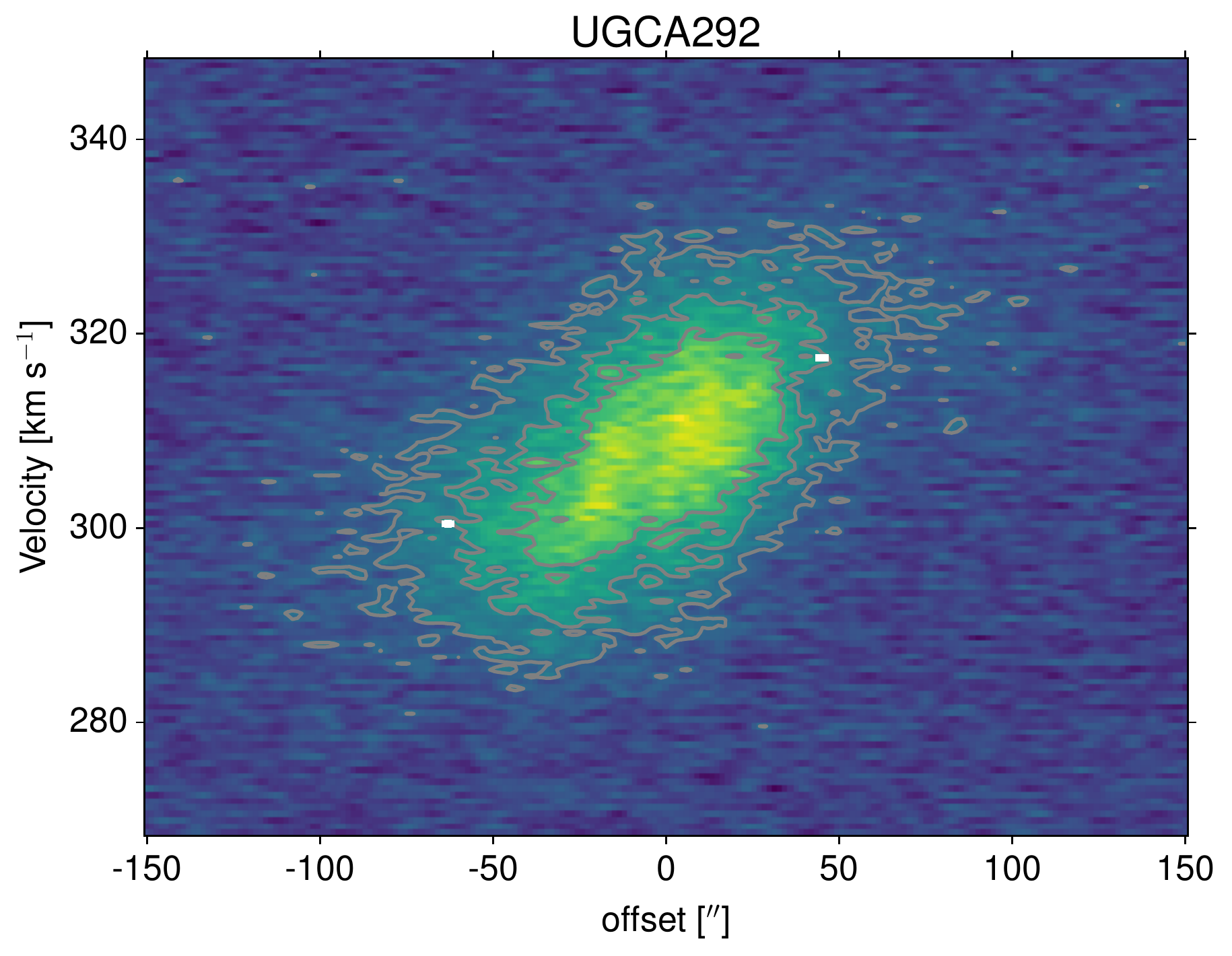}
\end{center}
\caption{PV diagrams for VLA-ANGST galaxies Sextans~A, Sextans~B, UGC04483, UGC08508, UGC08833, and UGCA292 with the location of the best-fitting \vpv\ and \dpv\ marked in white.}
\label{fig:pv_4}
\end{figure*}

\begin{figure*}
\begin{center}
\includegraphics[width=0.48\textwidth]{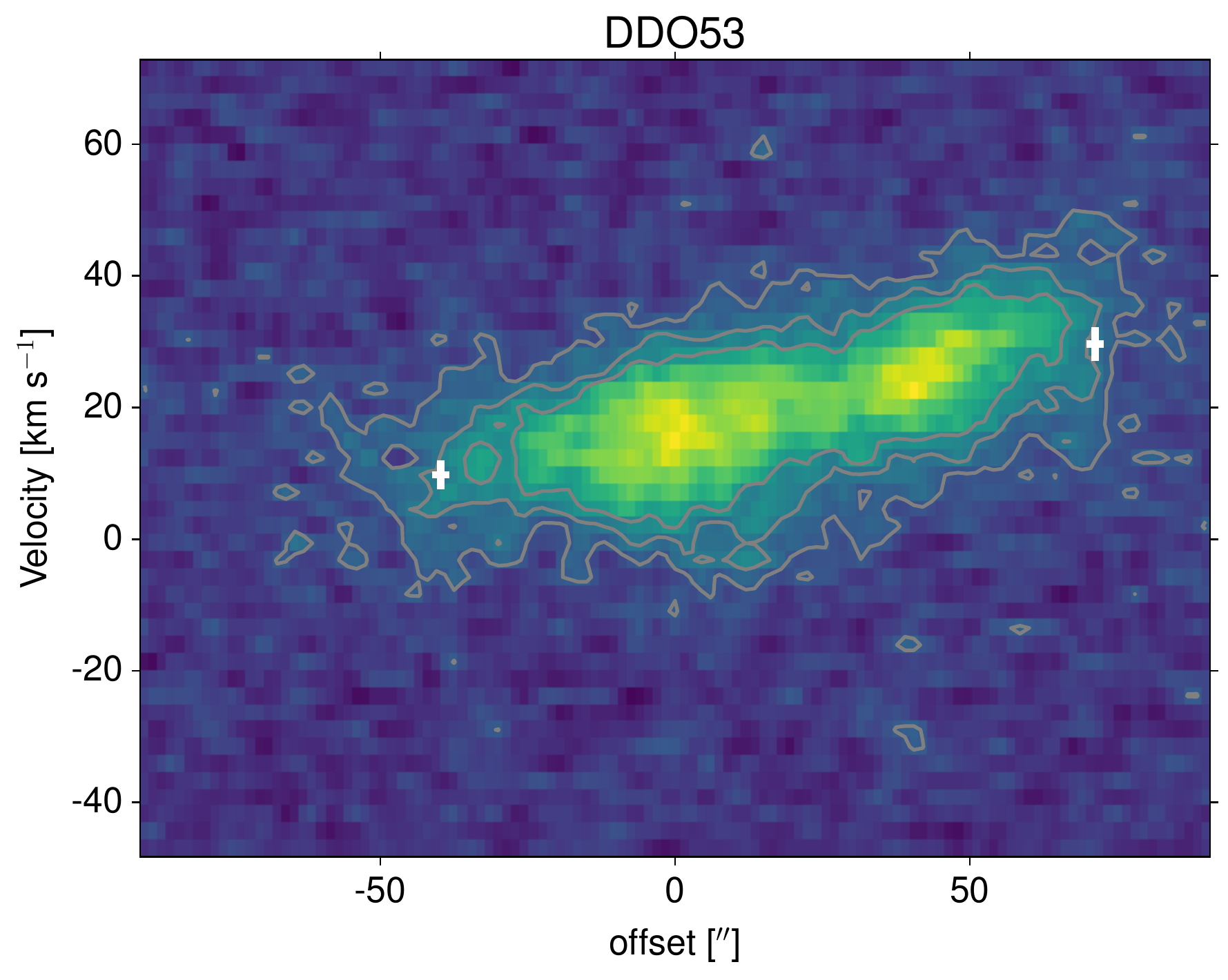}
\includegraphics[width=0.48\textwidth]{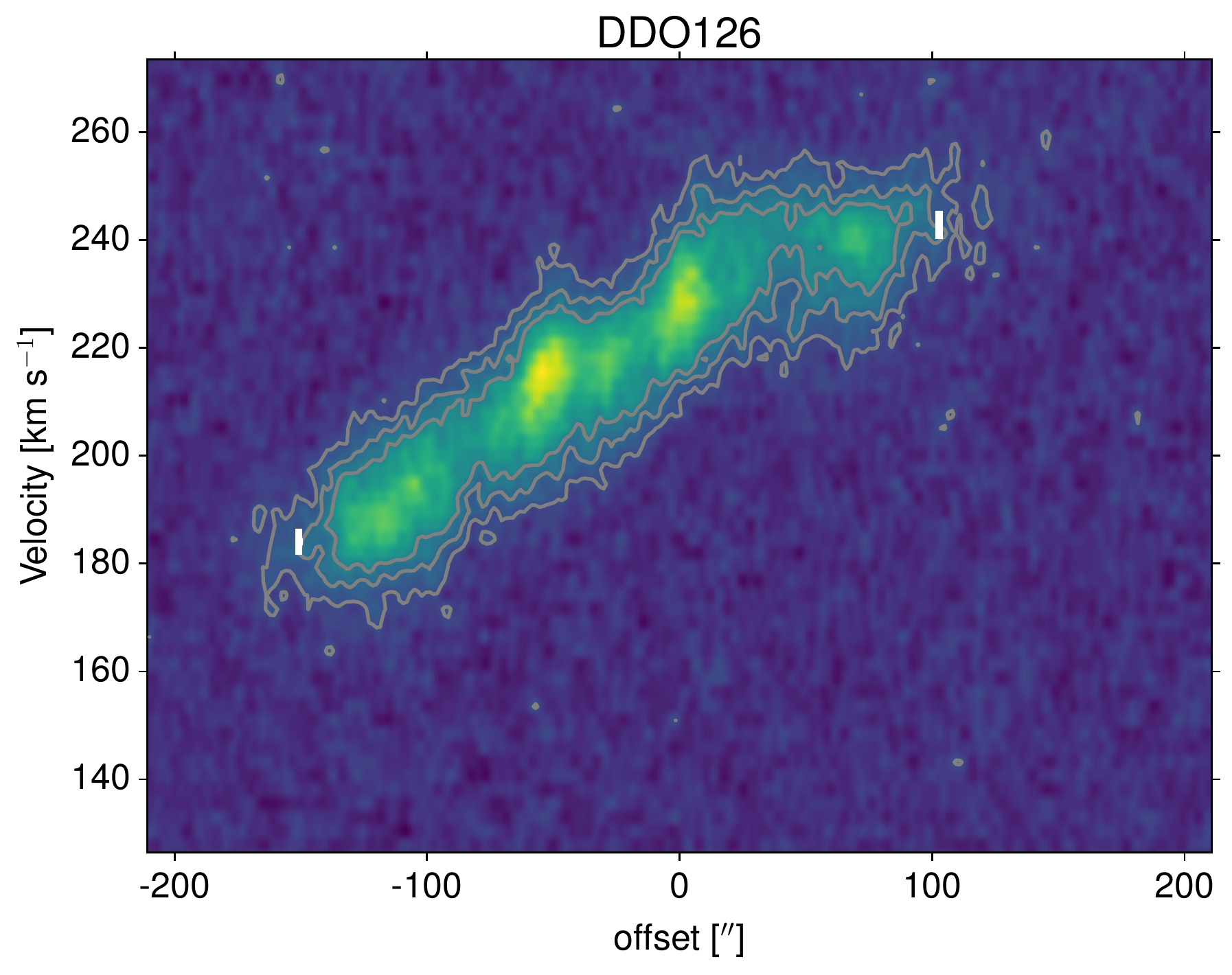}
\includegraphics[width=0.48\textwidth]{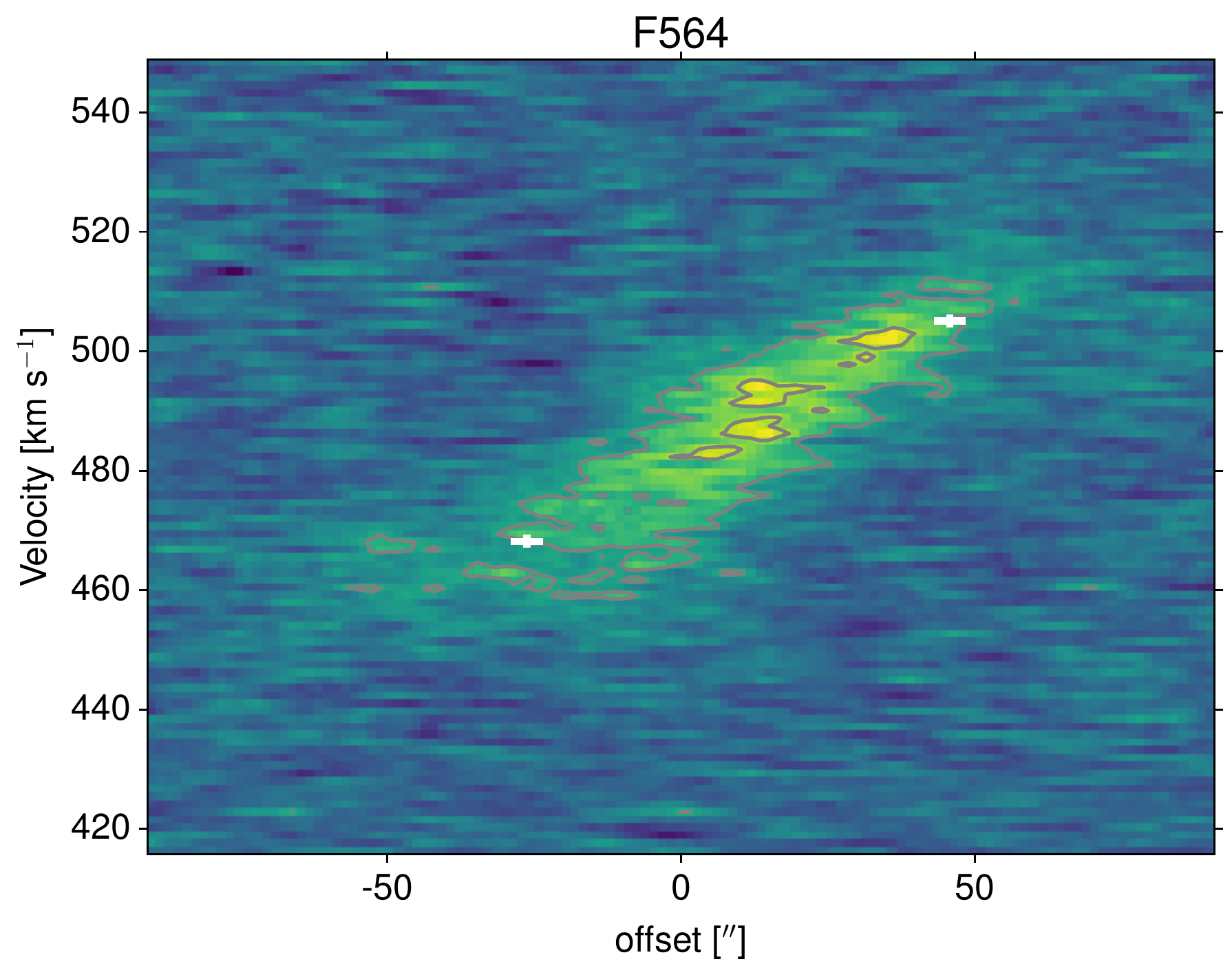}
\includegraphics[width=0.48\textwidth]{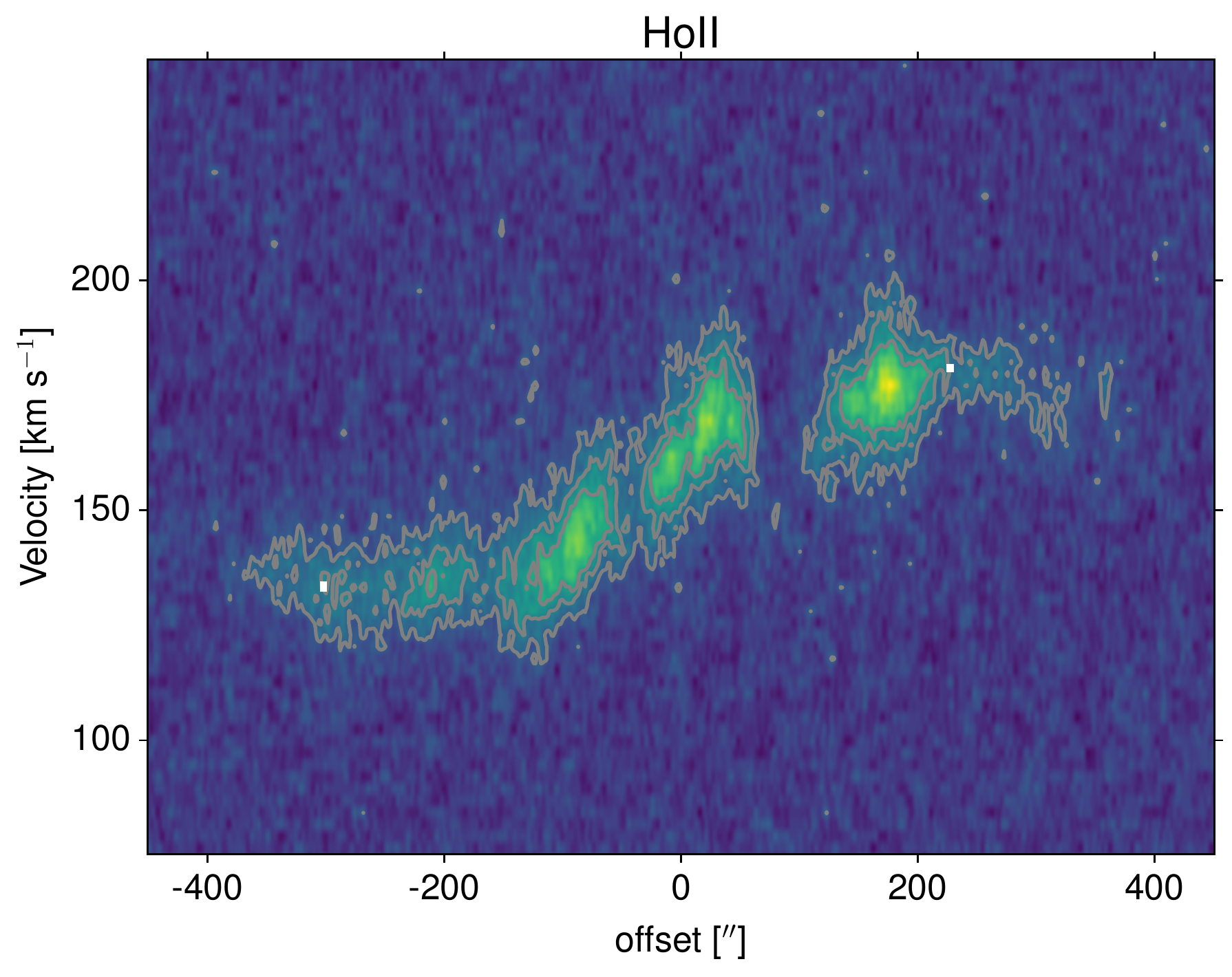}
\includegraphics[width=0.48\textwidth]{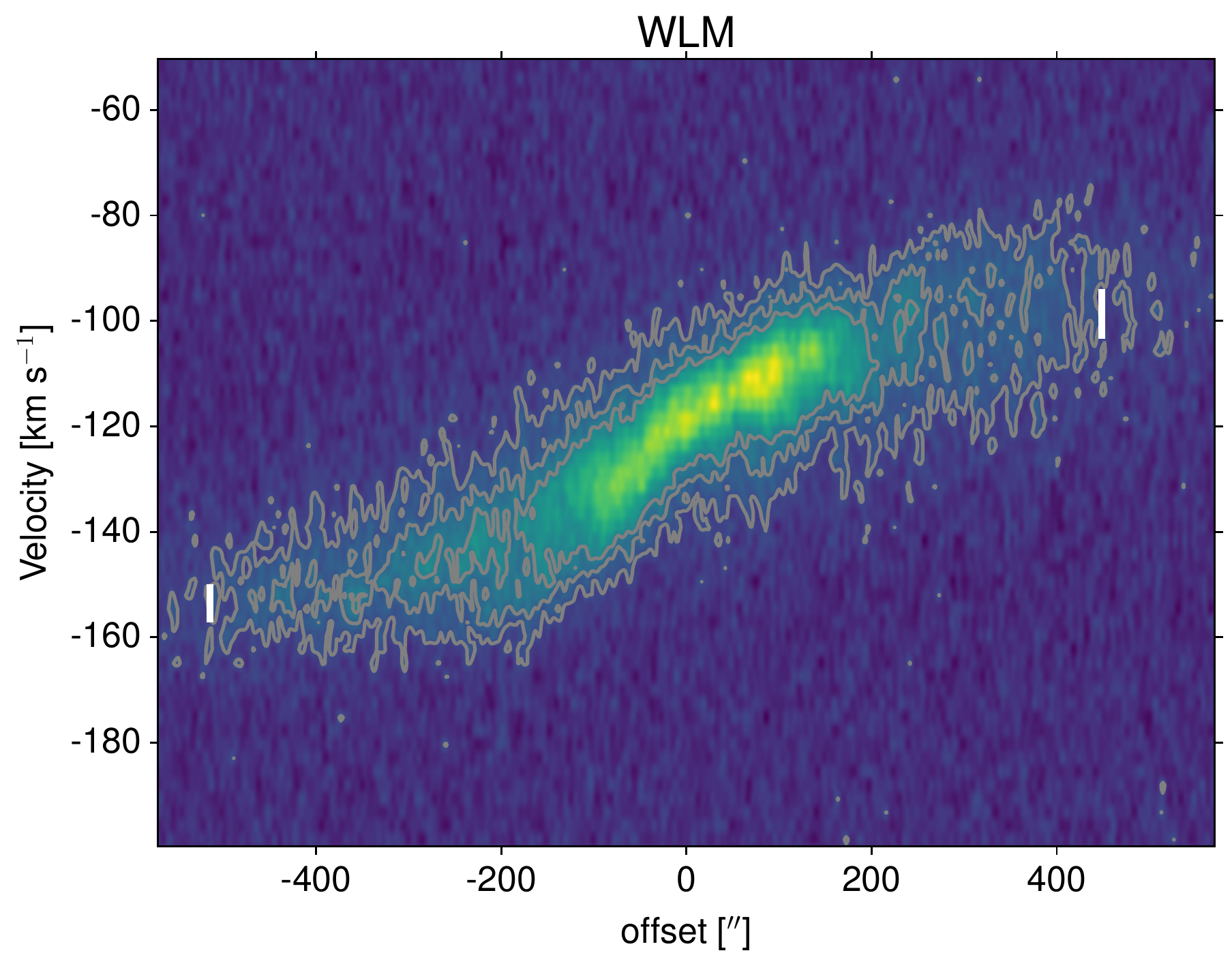}
\end{center}
\caption{PV diagrams for LITTLE THINGS galaxies DDO53, DDO126  F564v3, HoII, and WLM with the location of the best-fitting \vpv\ and \dpv\ marked in white.}
\label{fig:pv_5}
\end{figure*}

Here, we provide the details on the \hi\ data cubes in Table~\ref{tab:hi_data} and show the PV diagrams with the location of the best-fitting \vpv\ and \dpv\ values for the remainder of the sample. Figures~\ref{fig:pv_2}-\ref{fig:pv_5} follow the format shown in Figure~\ref{fig:pv_examples}; see Section~\ref{sec:pv} for details. At the end, we also present a comparison of our measured velocities after correcting for asymmetric drift (but not inclination angle) and \rmax\ values for galaxies in LITTLE THINGS with the velocities  (also without inclination angle corrections) and radial extents determined from rotation curve fitting in \citet{Iorio2017}.

\begin{table*}[h!]
\begin{center}
\caption{Properties of the \hi\ Data Cubes and PV Slices}
\label{tab:hi_data}
\end{center}
\vspace{-15pt}
\begin{center}
\begin{tabular}{lccrr}
\hline 
\hline 
Galaxy & Velocity Res.	& Beam Size 			& rms of Cube& rms of PV slice \\
	& (\kms)		& (\arcsec $\times$ \arcsec)	& ($10^{-3}$)	&	\\
\hline 
\multicolumn{5}{c}{SHIELD Galaxies}  \\ 
\hline
AGC110482 & 0.82 & 14.1$\times$12.0 & 1.00 & 0.5 \\
AGC111164 & 0.82 & 28.5$\times$22.5 & 1.40 & 0.8 \\
AGC229053 & 4.10 & 31.4$\times$13.6 & 1.20 & 0.5 \\
AGC731921 & 4.10 & 26.6$\times$12.7 & 1.34 & 0.7 \\
AGC739005 & 4.10 & 39.3$\times$12.2 & 1.32 & 0.7 \\
AGC742601 & 4.10 & 34.0$\times$12.4 & 1.40 & 0.8 \\
AGC749237 & 0.82 & 9.8$\times$8.9 & 0.79 & 0.3 \\
\hline
\multicolumn{5}{c}{VLA-ANGST Galaxies}  \\    
\hline
DDO99 & 1.3 & 7.7$\times$5.2 & 1.06 & 0.18 \\
DDO125 & 0.6 & 6.2$\times$5.4 & 1.44 & 0.2 \\
DDO181 & 1.3 & 7.5$\times$5.5 & 1.01 & 0.1 \\
DDO183 & 1.3 & 7.6$\times$6.1 & 1.10 & 0.1 \\
NGC3109 & 1.3 & 7.6$\times$5.0 & 1.62 & 0.18 \\
NGC3741 & 1.3 & 5.5$\times$4.7 & 1.07 & 0.2 \\
SextansA & 1.3 & 7.3$\times$6.0 & 1.24 & 0.2 \\
SextansB & 1.3 & 9.4$\times$7.5 & 0.95 & 0.17 \\
UGC04483 & 2.6 & 7.6$\times$5.7 & 0.63 & 0.18 \\
UGC08508 & 0.6 & 8.1$\times$6.3 & 1.36 & 0.2 \\
UGC08833 & 2.6 & 12.3$\times$11.1 & 0.64 & 0.25 \\
UGCA292 & 0.6 & 6.9$\times$4.9 & 1.51 & 0.3 \\
\hline
\multicolumn{5}{c}{LITTLE THINGS Galaxies}  \\ 
\hline
DDO53 & 2.6 & 6.3$\times$5.6 & 0.59 & 0.1 \\
DDO126 & 2.6 & 6.9$\times$5.6 & 0.46 & 0.1 \\
DDO154 & 2.6 & 7.9$\times$6.2 & 0.54 & 0.1 \\
F564v3 & 2.6 & 12.5$\times$8.1 & 0.68 & 0.26 \\
HoII & 2.6 & 7.0$\times$6.1 & 1.02 & 0.19 \\
WLM & 2.6 & 7.5$\times$5.07 & 0.76 & 0.06 \\
\hline   
\hline           
\end{tabular}
\end{center}
\tablecomments{Properties of the \hi\ data including velocity resolution, beam size, and the rms of the data cubes and PV slices.}
\end{table*}   

\begin{figure}
\begin{center}
\includegraphics[width=0.48\textwidth]{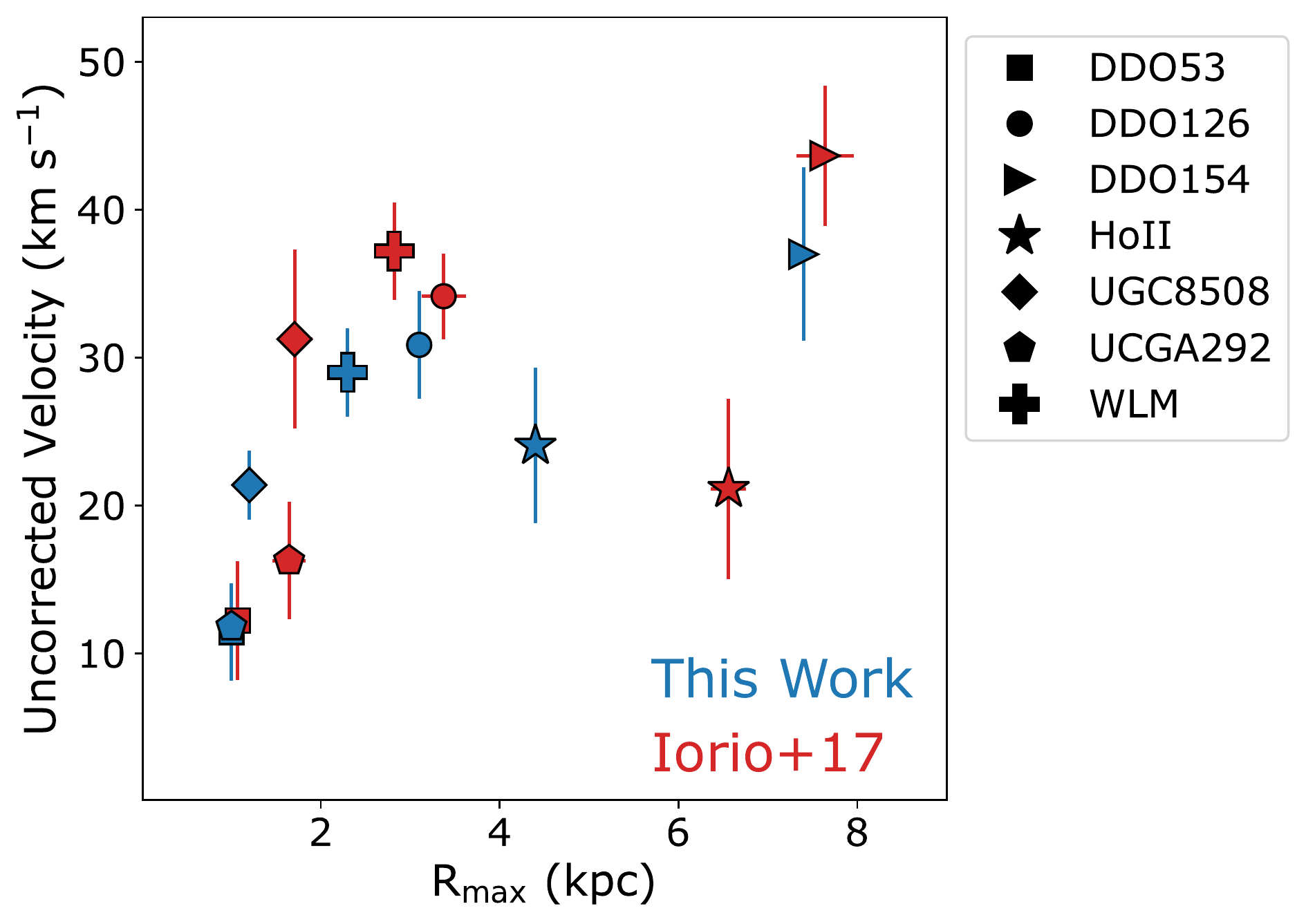}
\end{center}
\caption{Comparison of the measured velocities (before correcting for inclination angles) and the radius at which they were measured from the PV diagrams using our new technique (blue points) and from rotation curve fitting from \citet[][red points]{Iorio2017}. Each galaxy has a unique plot symbol, as defined in the legend. The values are in overall good agreement.}
\label{fig:compare_velocities}
\end{figure}

\clearpage
In Figure~\ref{fig:compare_velocities}, we provide a consistency check on our velocities measured at \rmax\ by comparing them with values determined from fitting \hi\ rotation curves for galaxies that overlap with the LITTLE THINGS sample analyzed in \citet{Iorio2017}. To make a direct comparison, we use the asymmetric drift corrected velocities \citep[\vcor\ in our work; $V_o$ from Table~2 in][]{Iorio2017} but remove the correction for inclination angles, and re-calculate \rmax\ in kpc from the angular extents reported in \citet{Iorio2017} by adopting the distances listed in our Table~\ref{tab:properties}. Each galaxy is shown with a different plot symbol, as noted in the legend. Our measurements are determined at slighter smaller radii and have slightly smaller velocities, but are overall consistent with the values from the rotation curve analysis. For the WLM galaxy, our velocity was measured at a notably smaller radius than in \citet{Iorio2017}, but the velocity agrees within the uncertainties; this is expected as the \hi\ data reaches the flat part of the rotation curve.

\clearpage
\section{Atlas of the Fits to Rotation Curves}\label{app:rotcurves}
\begin{figure*}[h!]
\begin{center}
\includegraphics[width=0.7\textwidth]{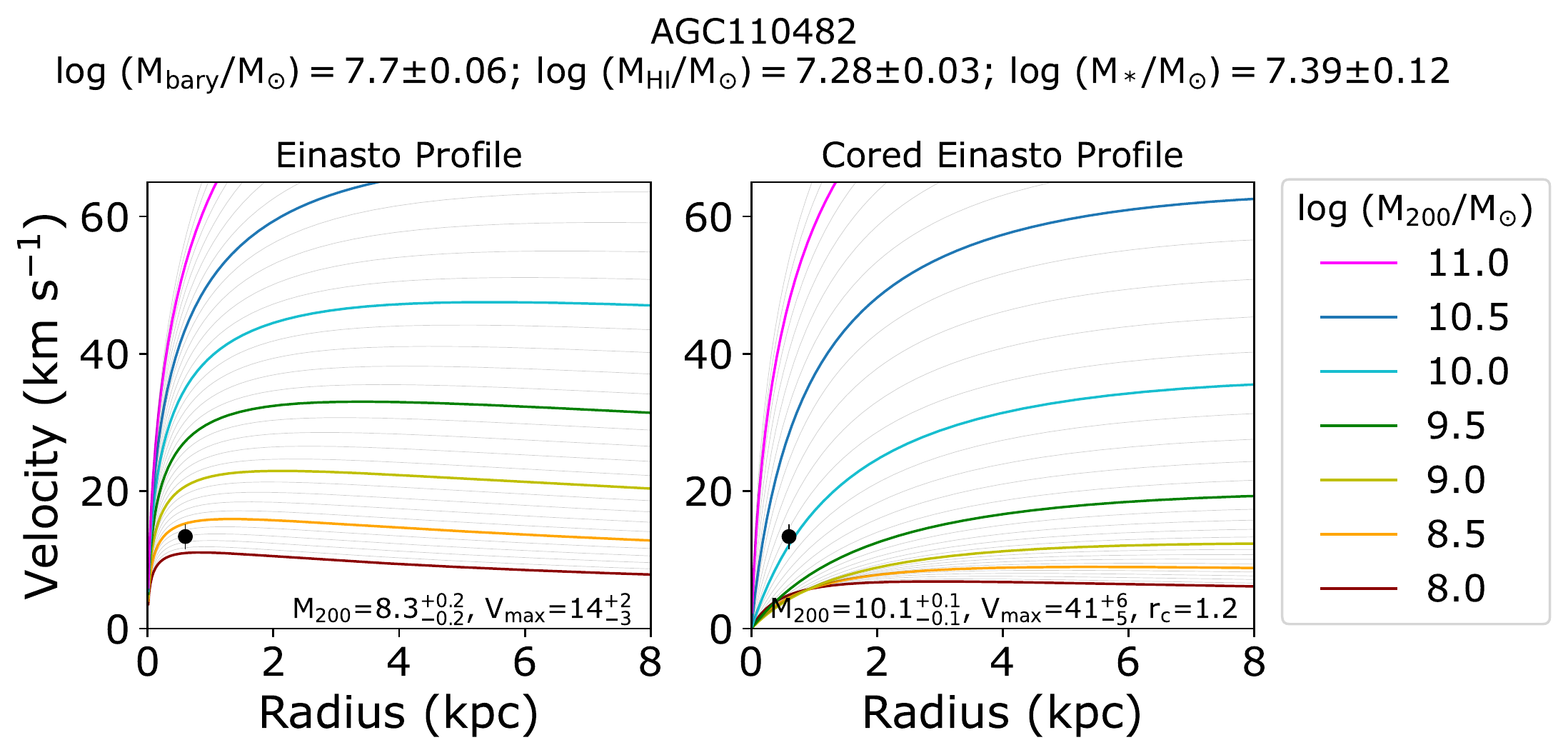}
\vspace{-0.1in}
\includegraphics[width=0.7\textwidth]{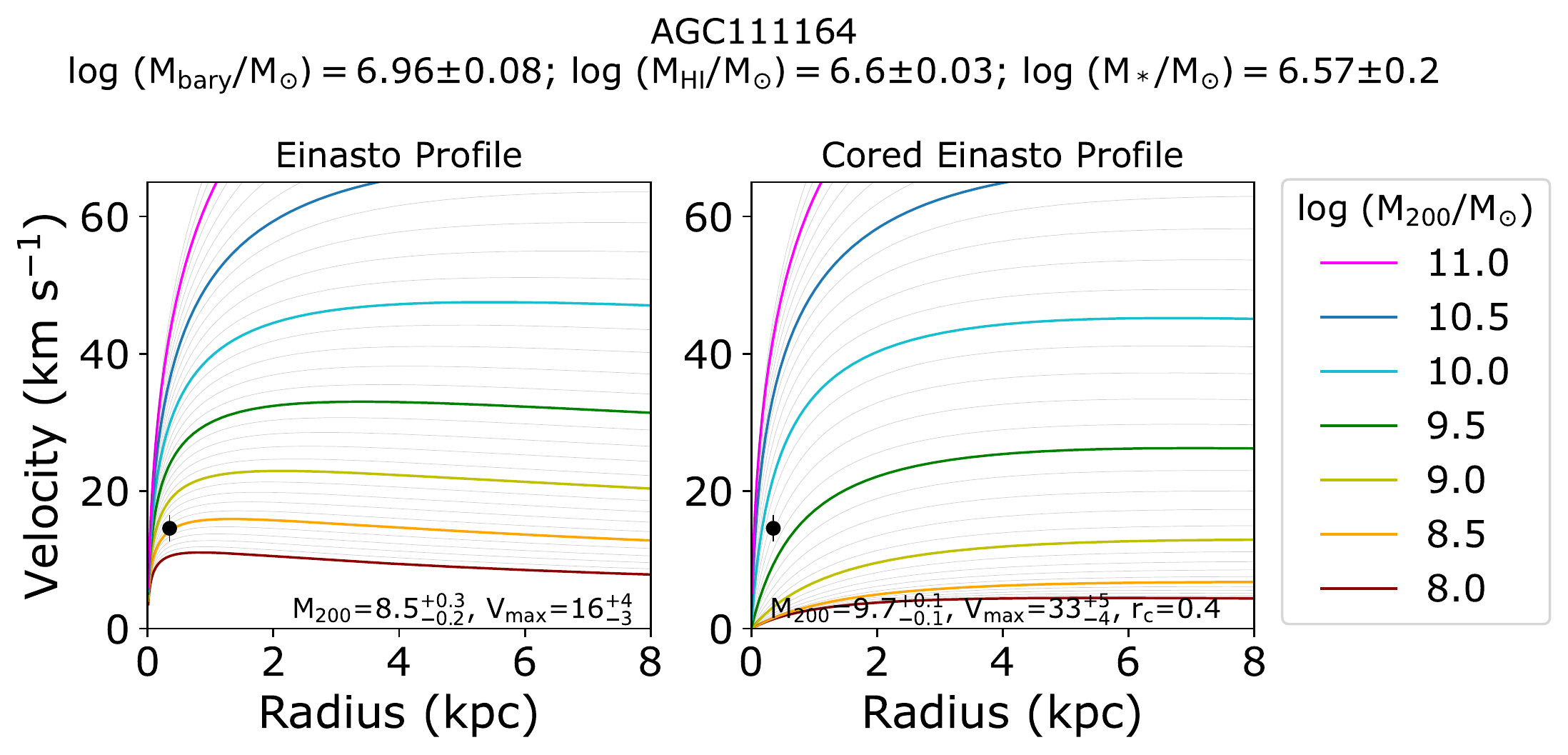}
\vspace{-0.1in}
\includegraphics[width=0.7\textwidth]{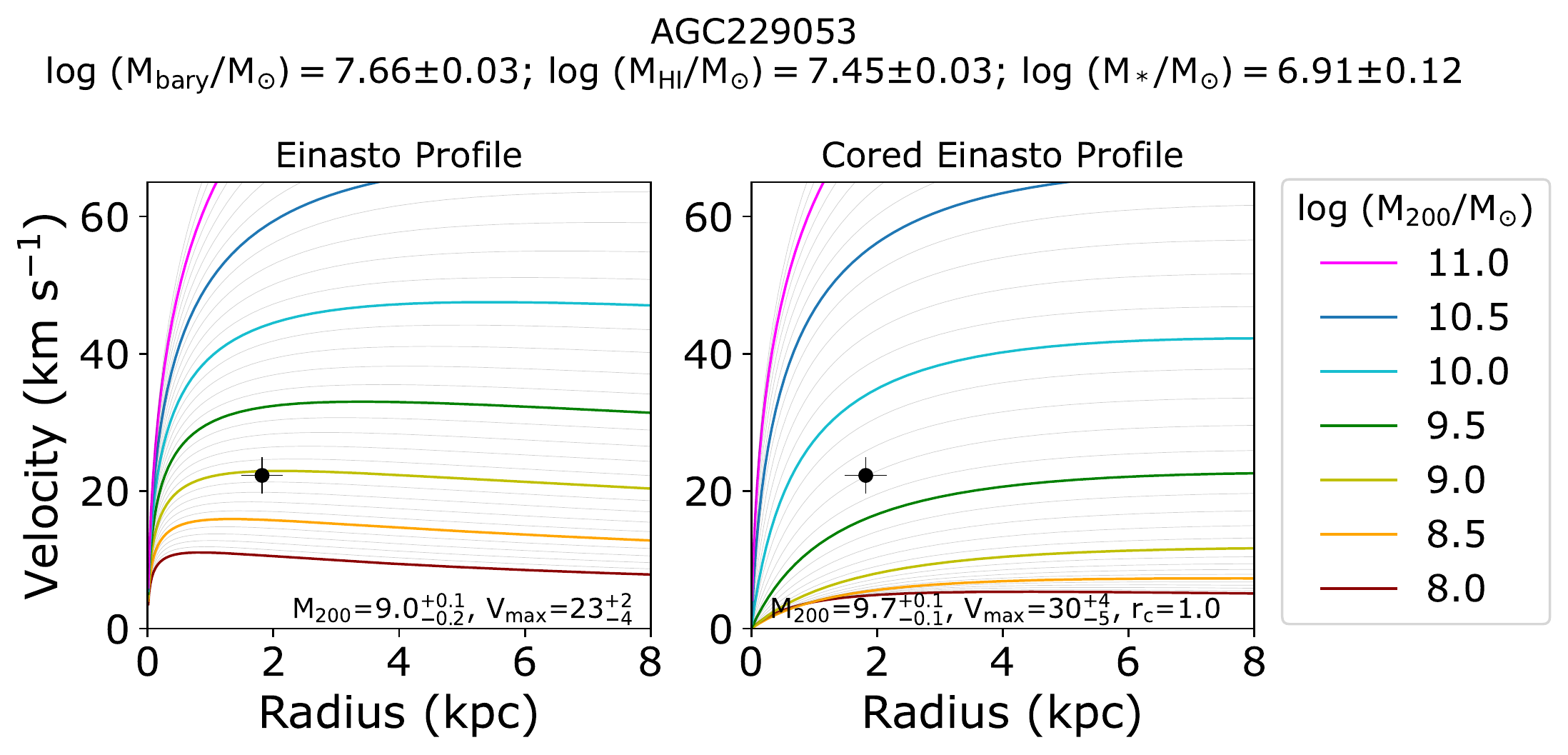}
\end{center}
\vspace{-0.1in}
\caption{Theoretical rotation curves for the SHIELD galaxies AGC110482, AGC111164, and AGC229053 with the measured velocities corrected for asymmetric drift overplotted at the radii at which the velocity was measured. See Figure~\ref{fig:rot_curve_example} caption and Section~\ref{sec:rot_curves}.}
\label{fig:rot_curve_2}
\end{figure*}

\begin{figure*}
\begin{center}
\includegraphics[width=0.8\textwidth]{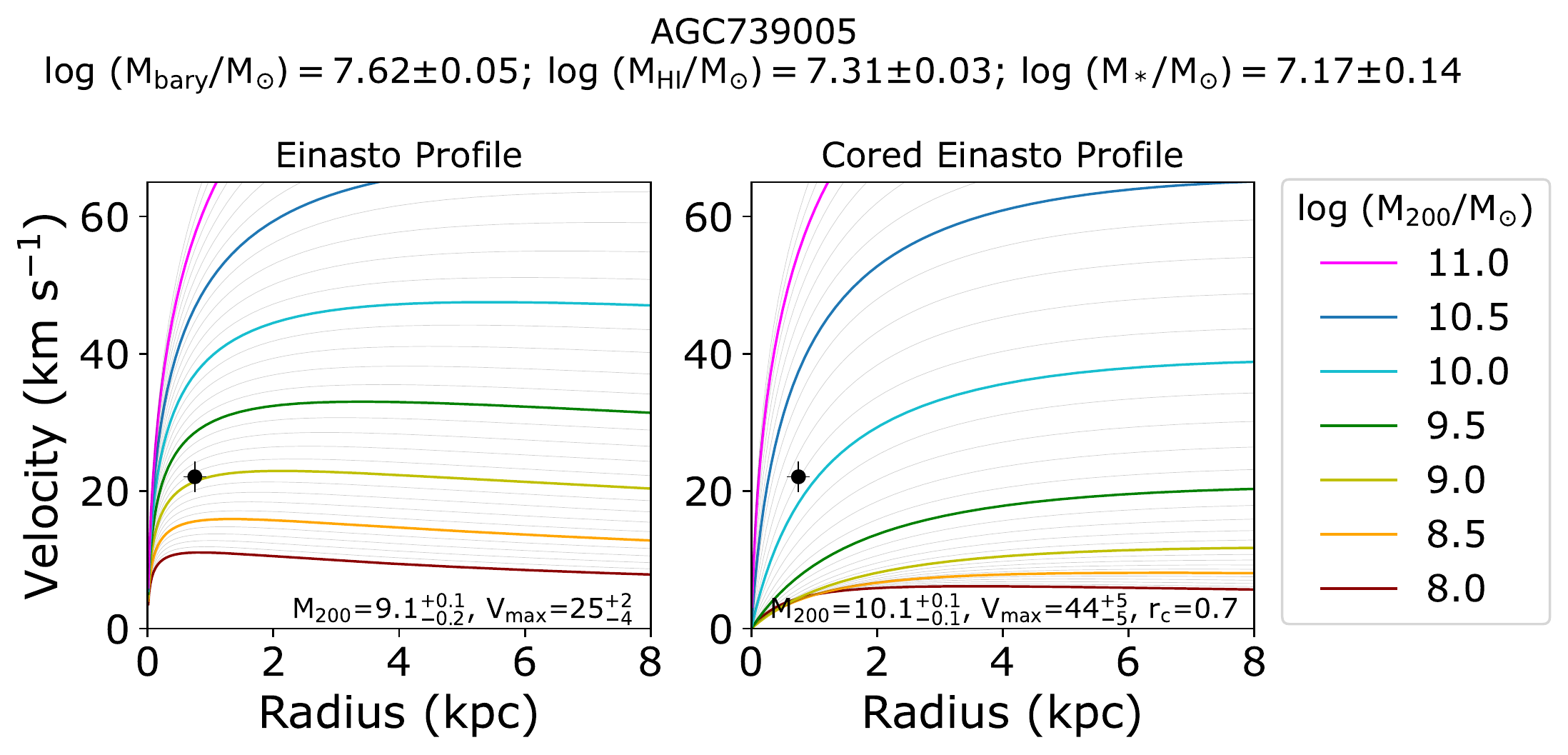}
\includegraphics[width=0.8\textwidth]{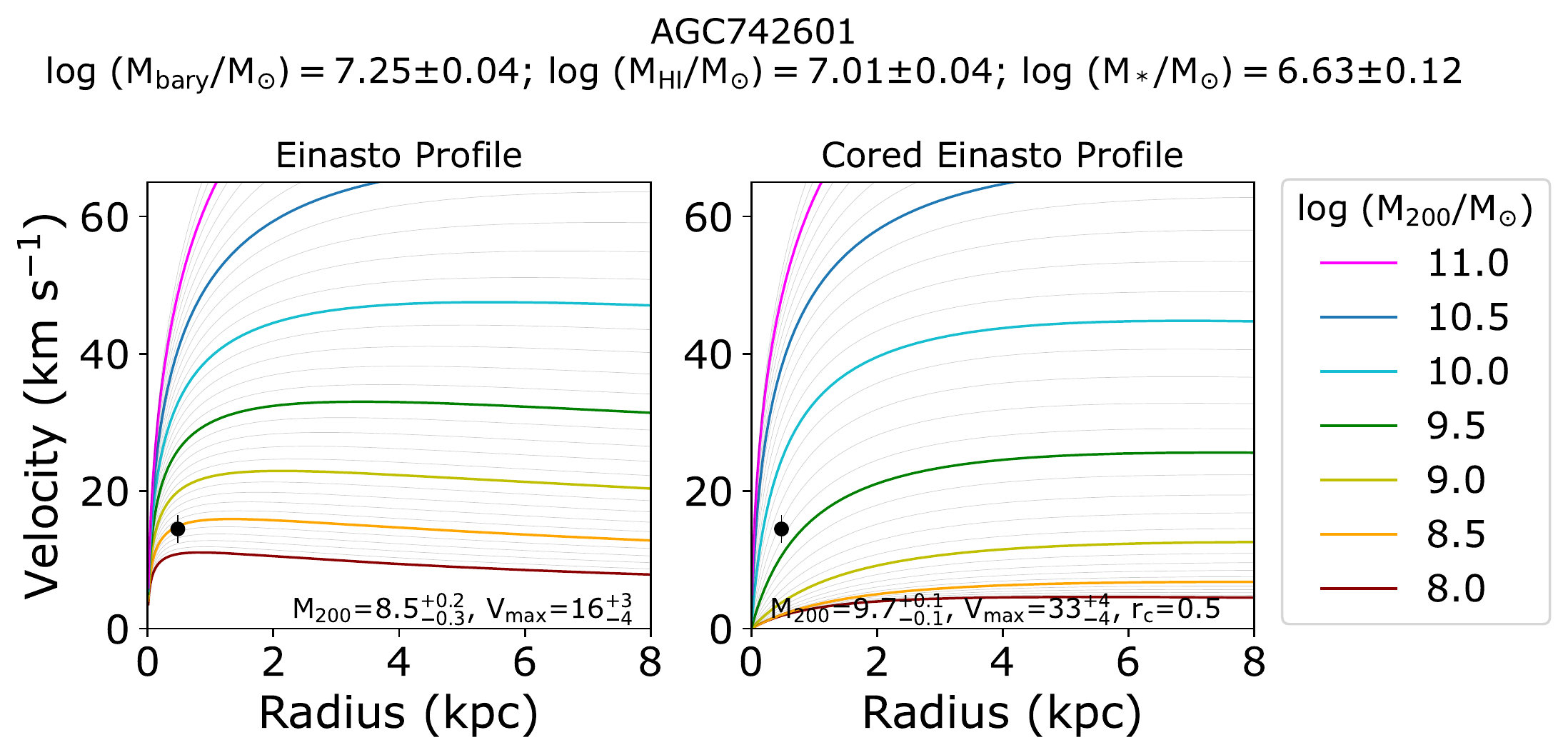}
\includegraphics[width=0.8\textwidth]{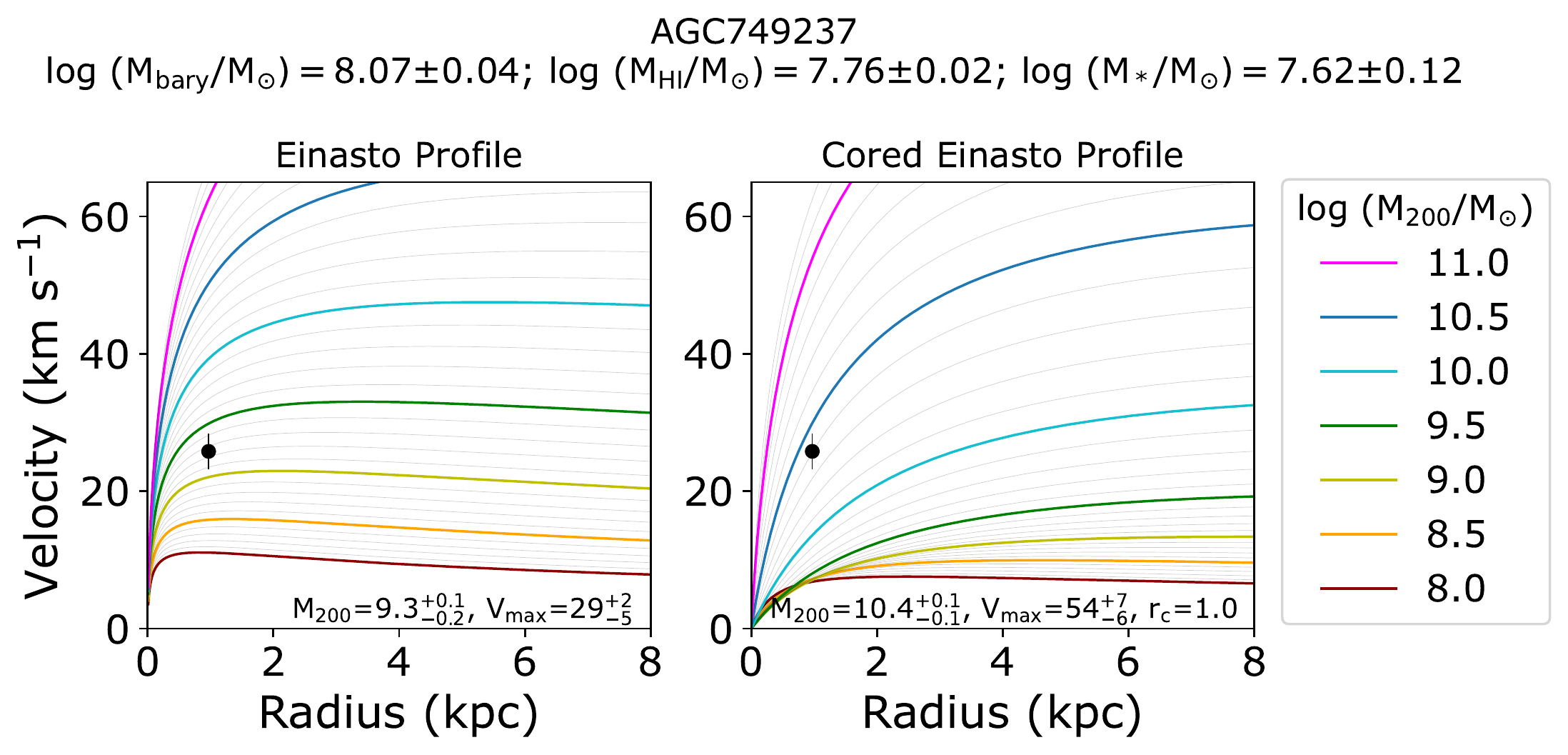}
\end{center}
\caption{Theoretical rotation curves for the SHIELD galaxies AGC7393005, AGC742601, and AGC 749237 with the measured velocities corrected for asymmetric drift overplotted at the radii at which the velocity was measured. See Figure~\ref{fig:rot_curve_example} caption and Section~\ref{sec:rot_curves}.}
\label{fig:rot_curve_3}
\end{figure*}

\begin{figure*}
\begin{center}
\includegraphics[width=0.8\textwidth]{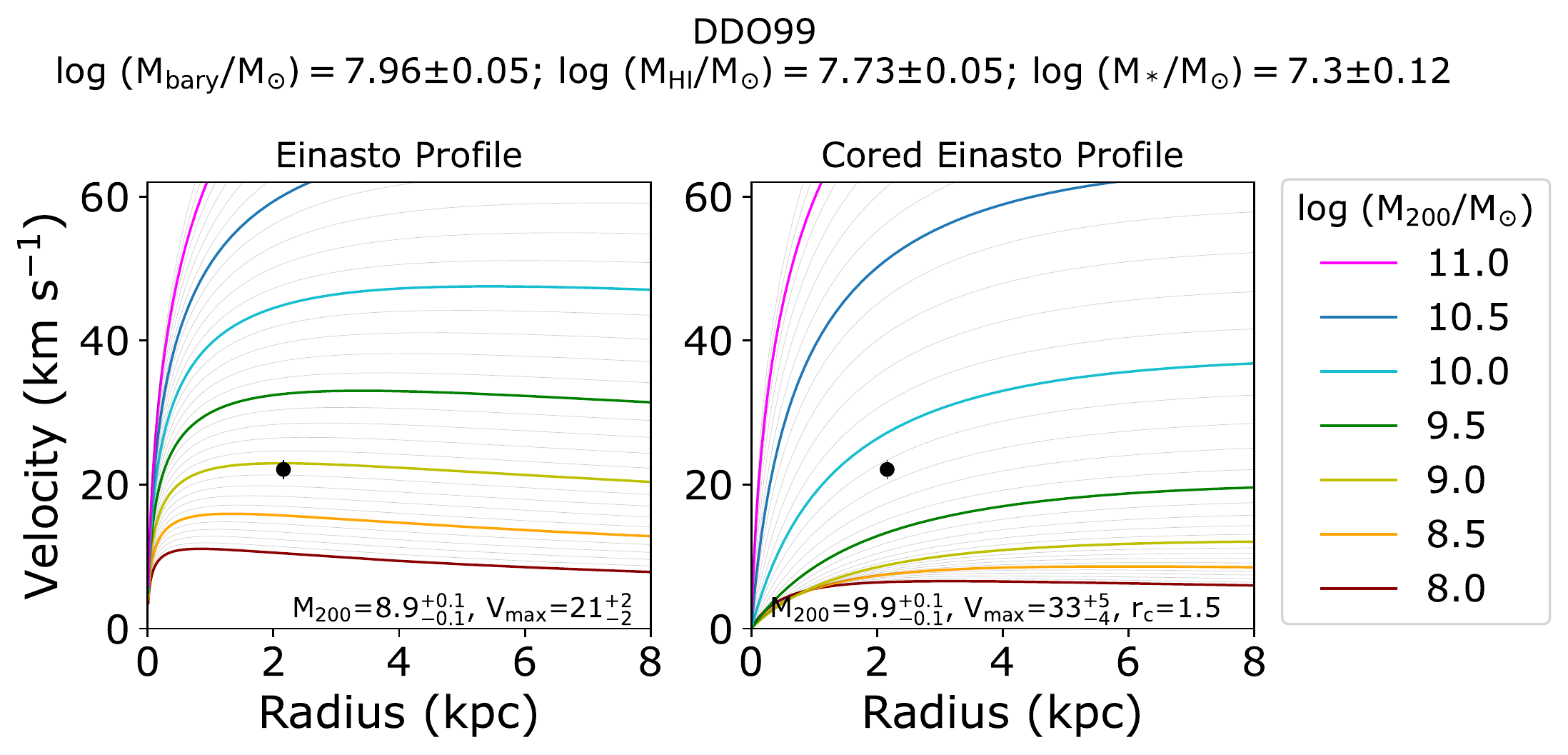}
\includegraphics[width=0.8\textwidth]{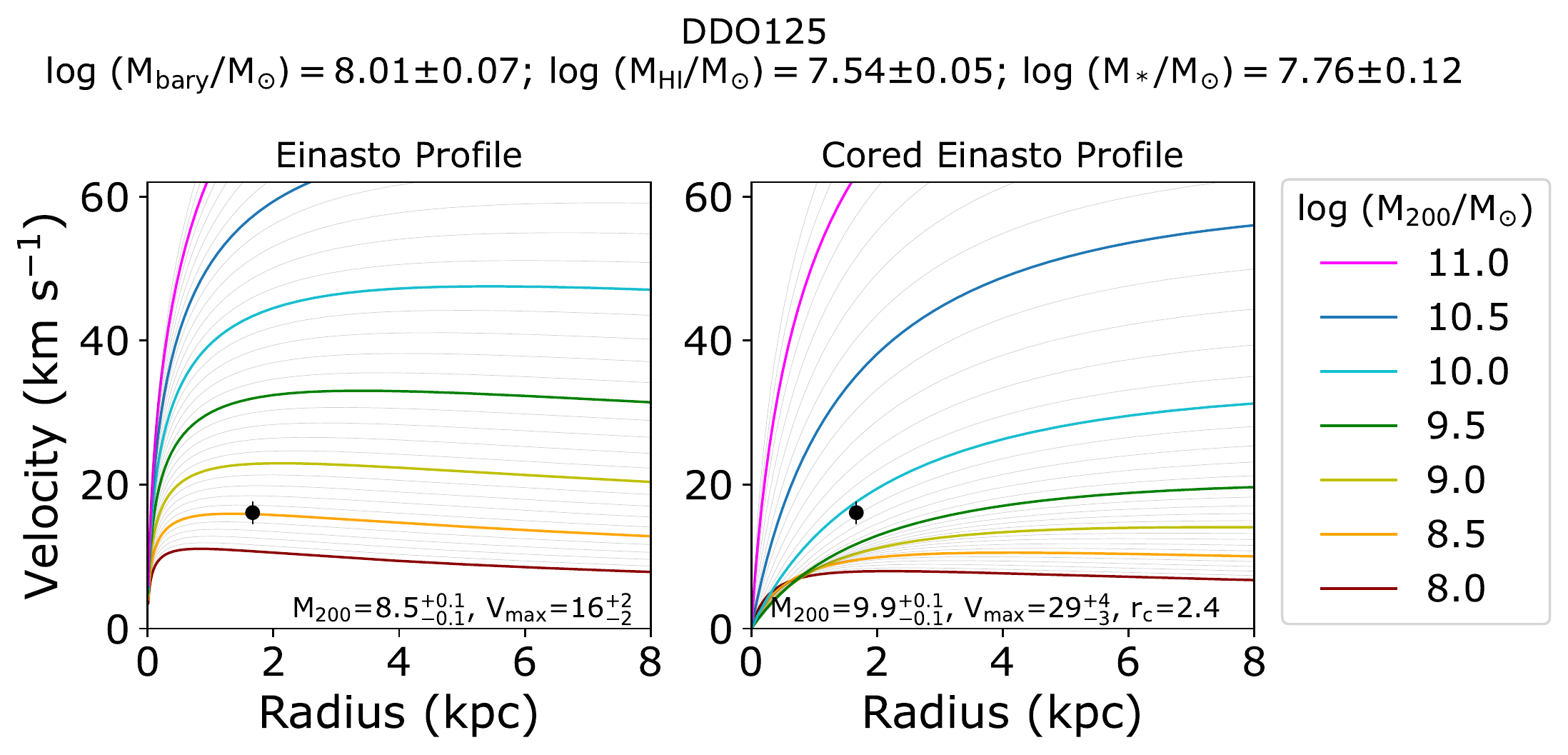}
\includegraphics[width=0.8\textwidth]{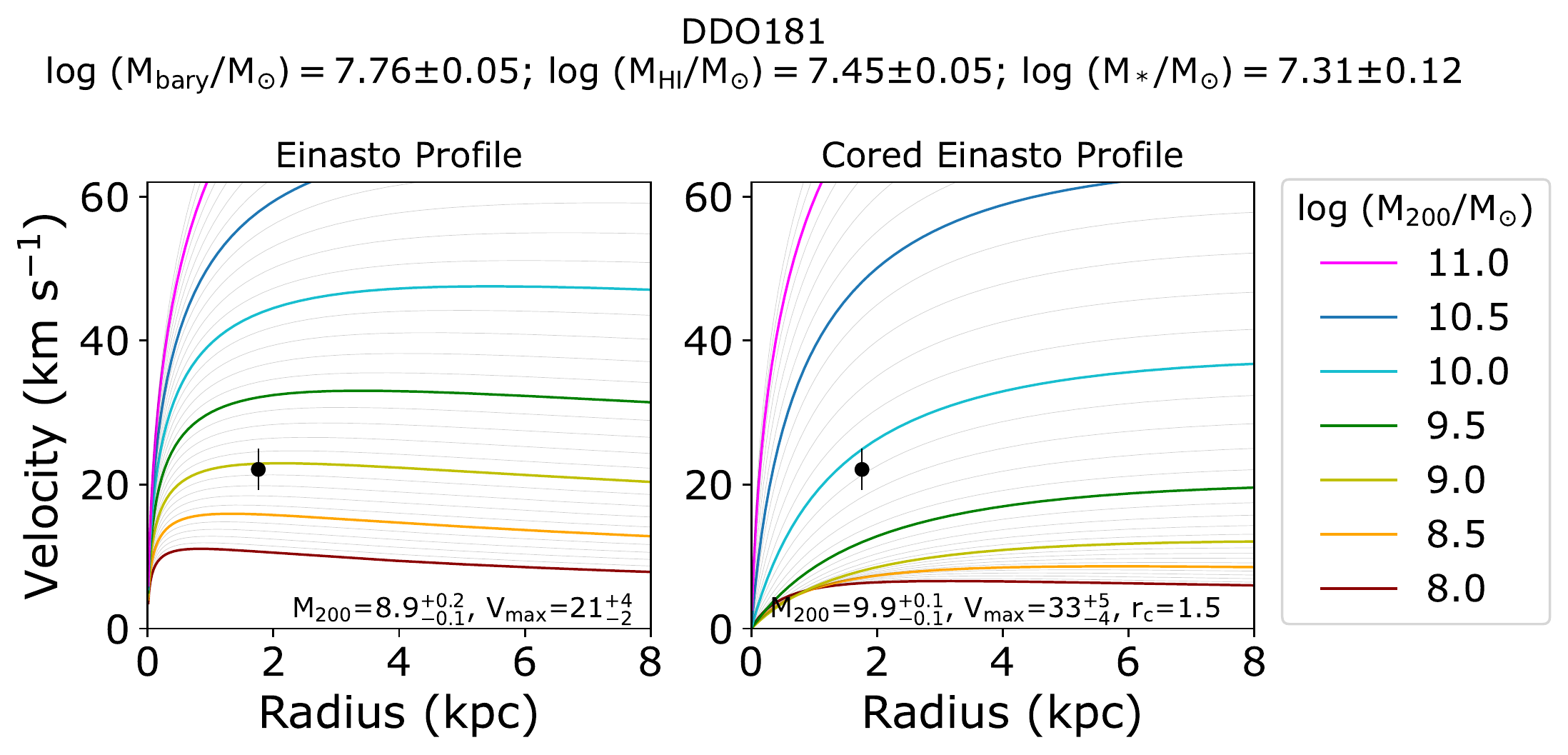}
\end{center}
\caption{Theoretical rotation curves for the VLA-ANGST galaxies DDO99, DDO125, and DDO181 with the measured velocities corrected for asymmetric drift overplotted at the radii at which the velocity was measured. See Figure~\ref{fig:rot_curve_example} caption and Section~\ref{sec:rot_curves}.}
\label{fig:rot_curve_4}
\end{figure*}

\begin{figure*}
\begin{center}
\includegraphics[width=0.8\textwidth]{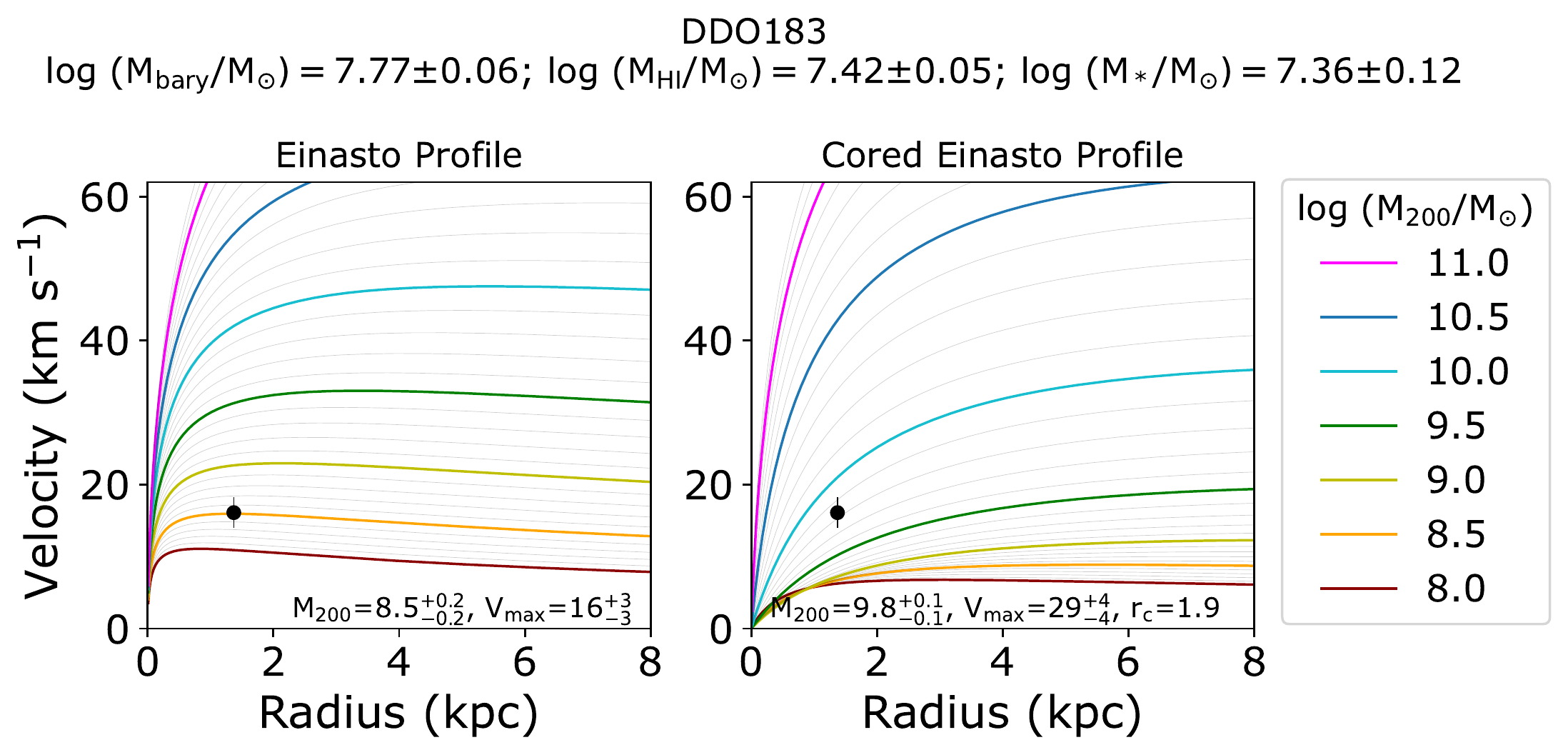}
\includegraphics[width=0.8\textwidth]{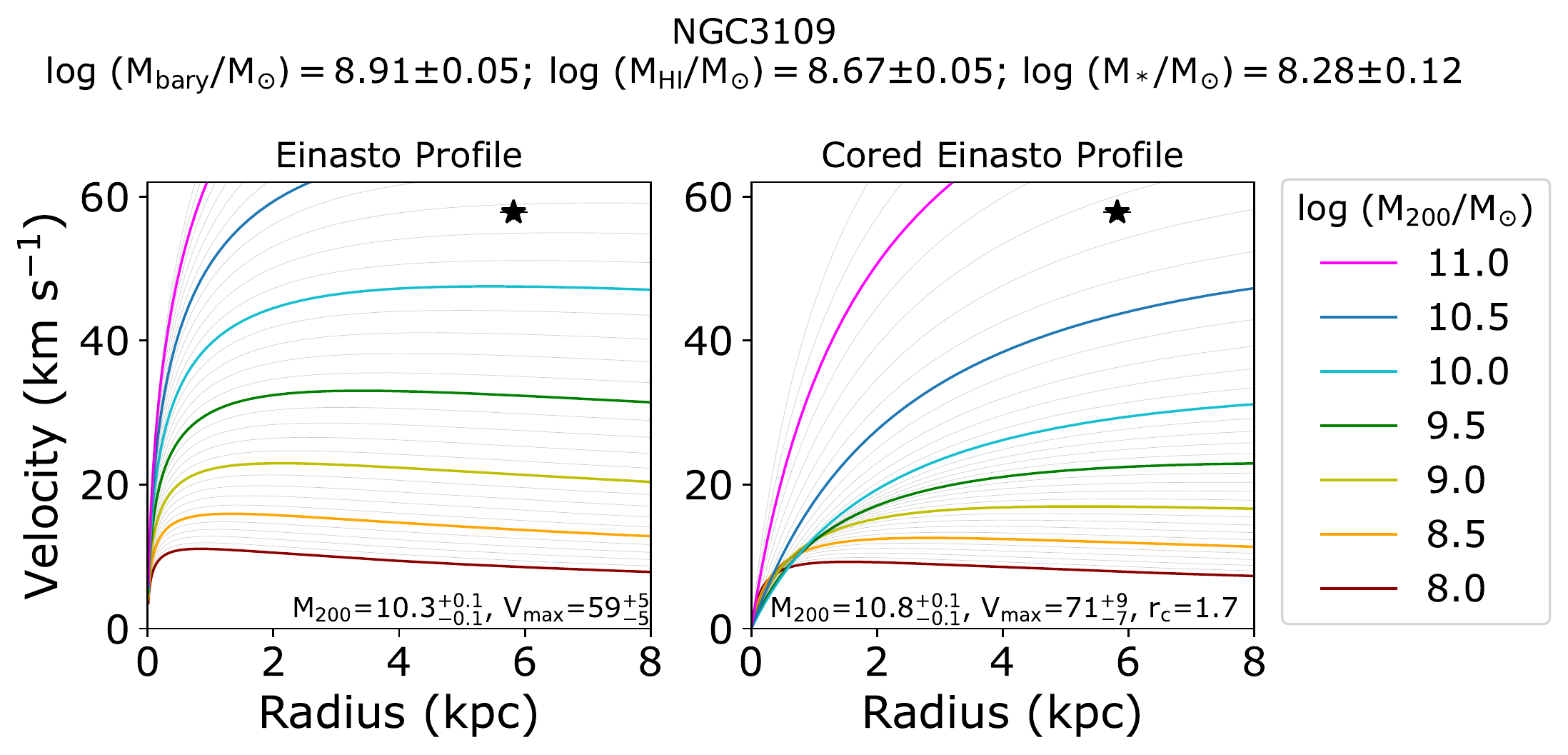}
\includegraphics[width=0.8\textwidth]{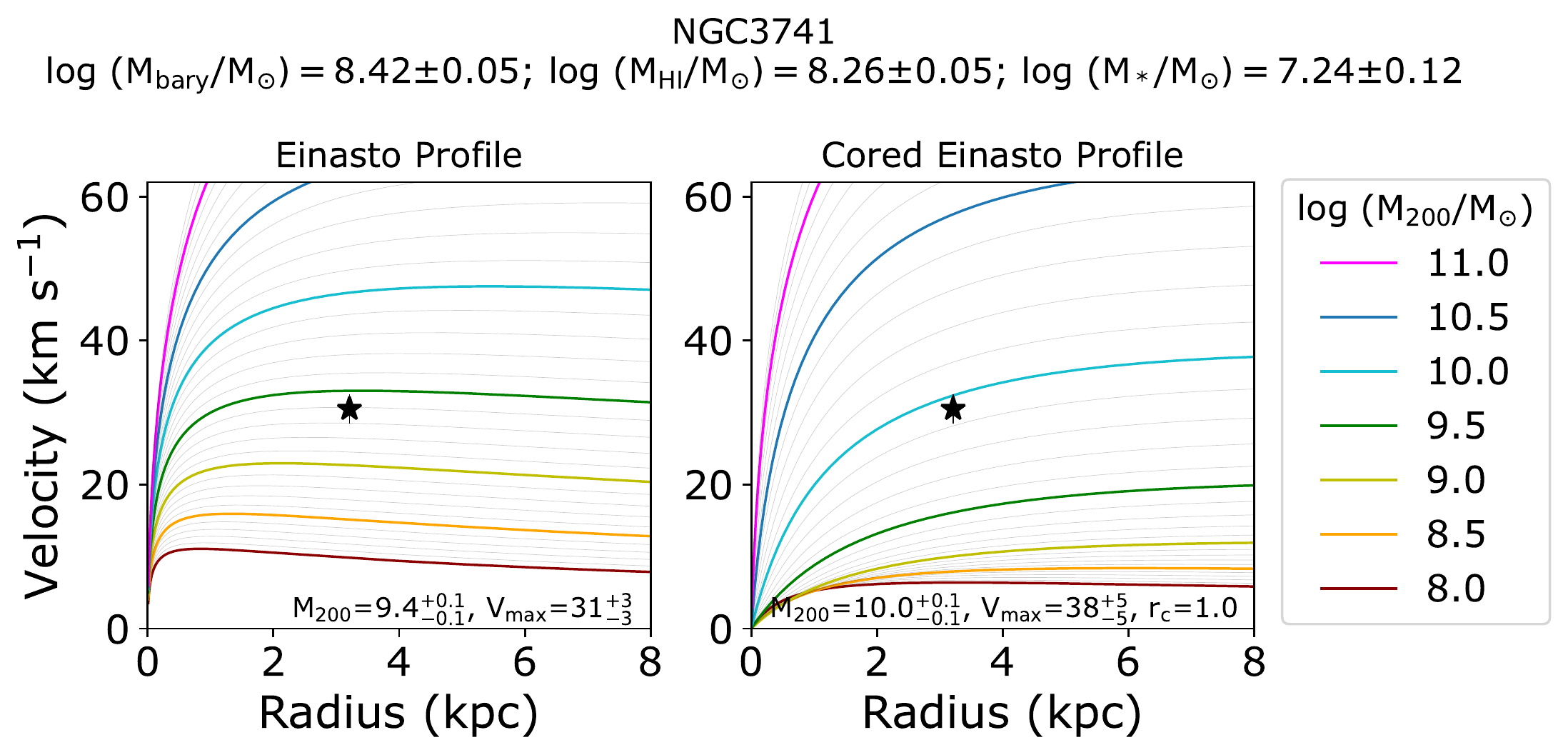}
\end{center}
\caption{Theoretical rotation curves for the VLA-ANGST galaxies DDO183, NGC3109, and NGC3741 with the measured velocities corrected for asymmetric drift overplotted at the radii at which the velocity was measured. See Figure~\ref{fig:rot_curve_example} caption and Section~\ref{sec:rot_curves}.}
\label{fig:rot_curve_5}
\end{figure*}

\begin{figure*}
\begin{center}
\includegraphics[width=0.8\textwidth]{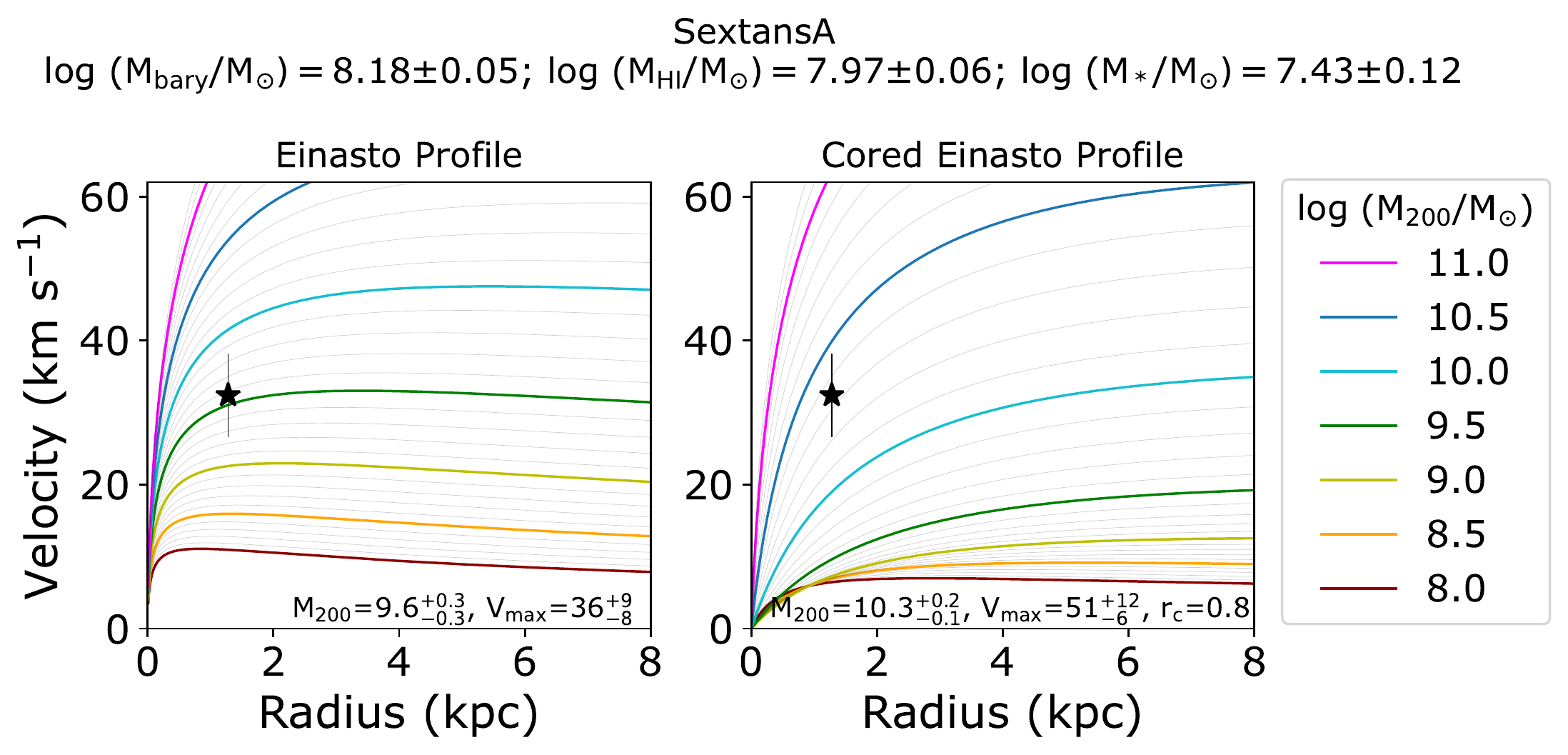}
\includegraphics[width=0.8\textwidth]{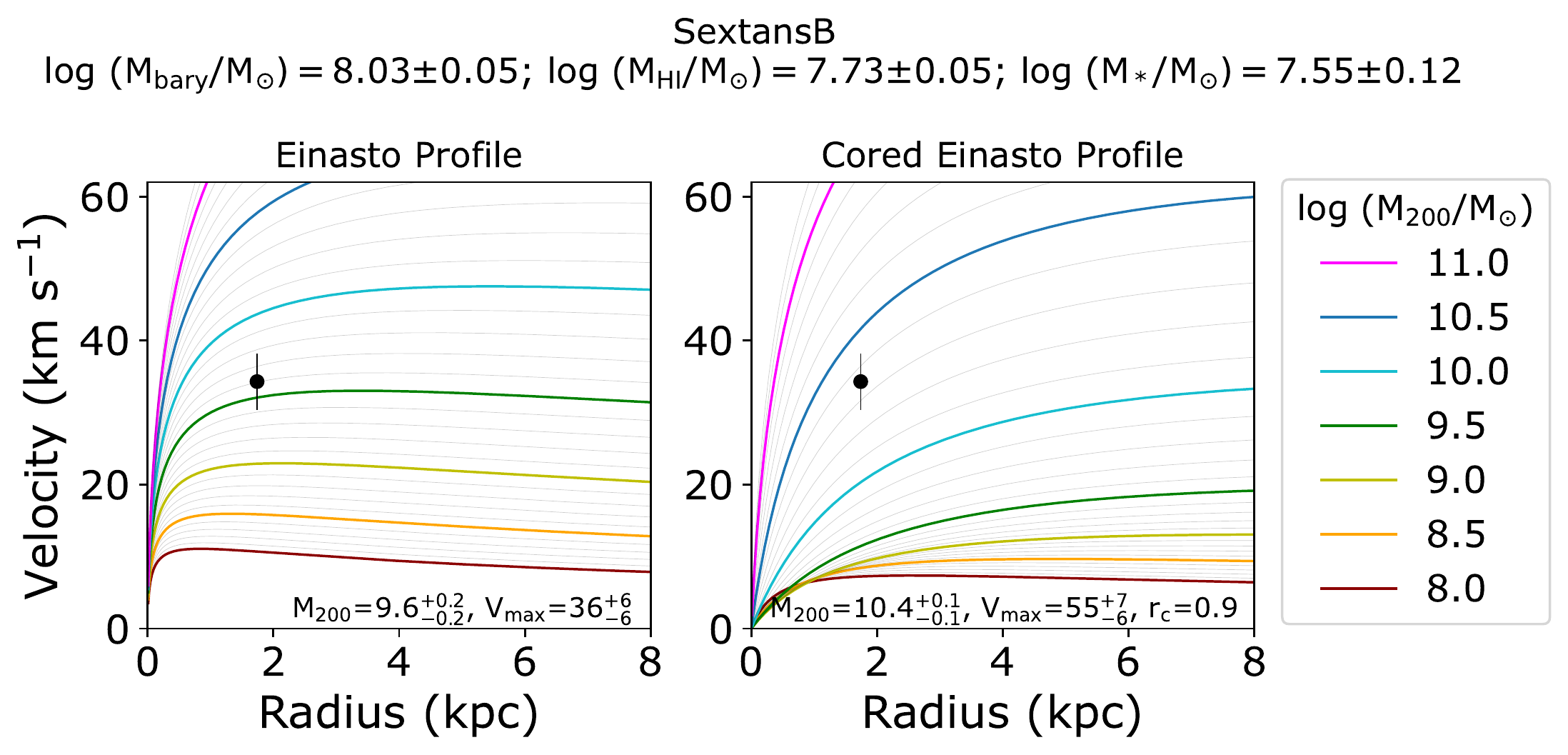}
\includegraphics[width=0.8\textwidth]{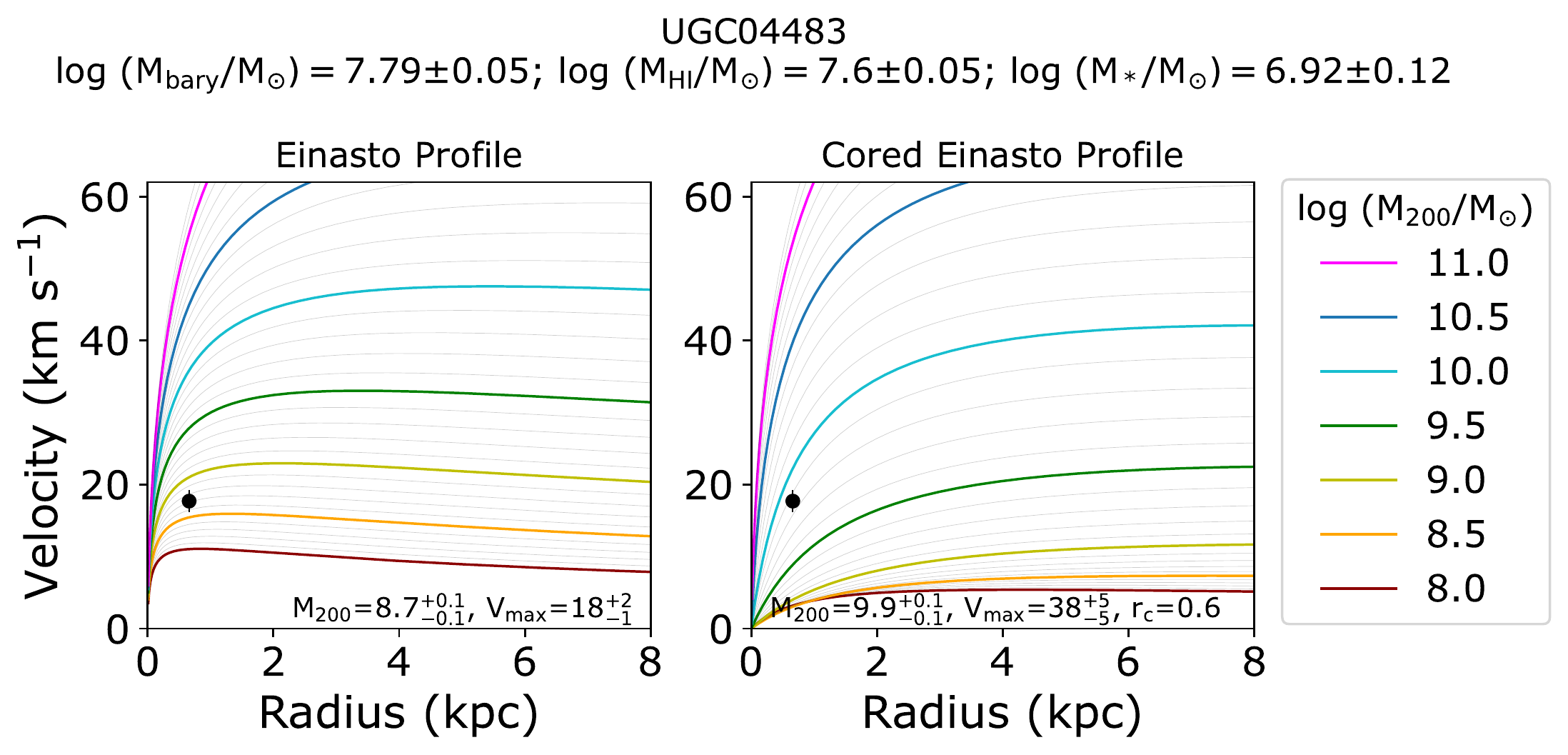}
\end{center}
\caption{Theoretical rotation curves for the VLA-ANGST galaxies Sextans~A, Sextans~B, and UGC04483 with the measured velocities corrected for asymmetric drift overplotted at the radii at which the velocity was measured. See Figure~\ref{fig:rot_curve_example} caption and Section~\ref{sec:rot_curves}.}
\label{fig:rot_curve_6}
\end{figure*}

\begin{figure*}
\begin{center}
\includegraphics[width=0.8\textwidth]{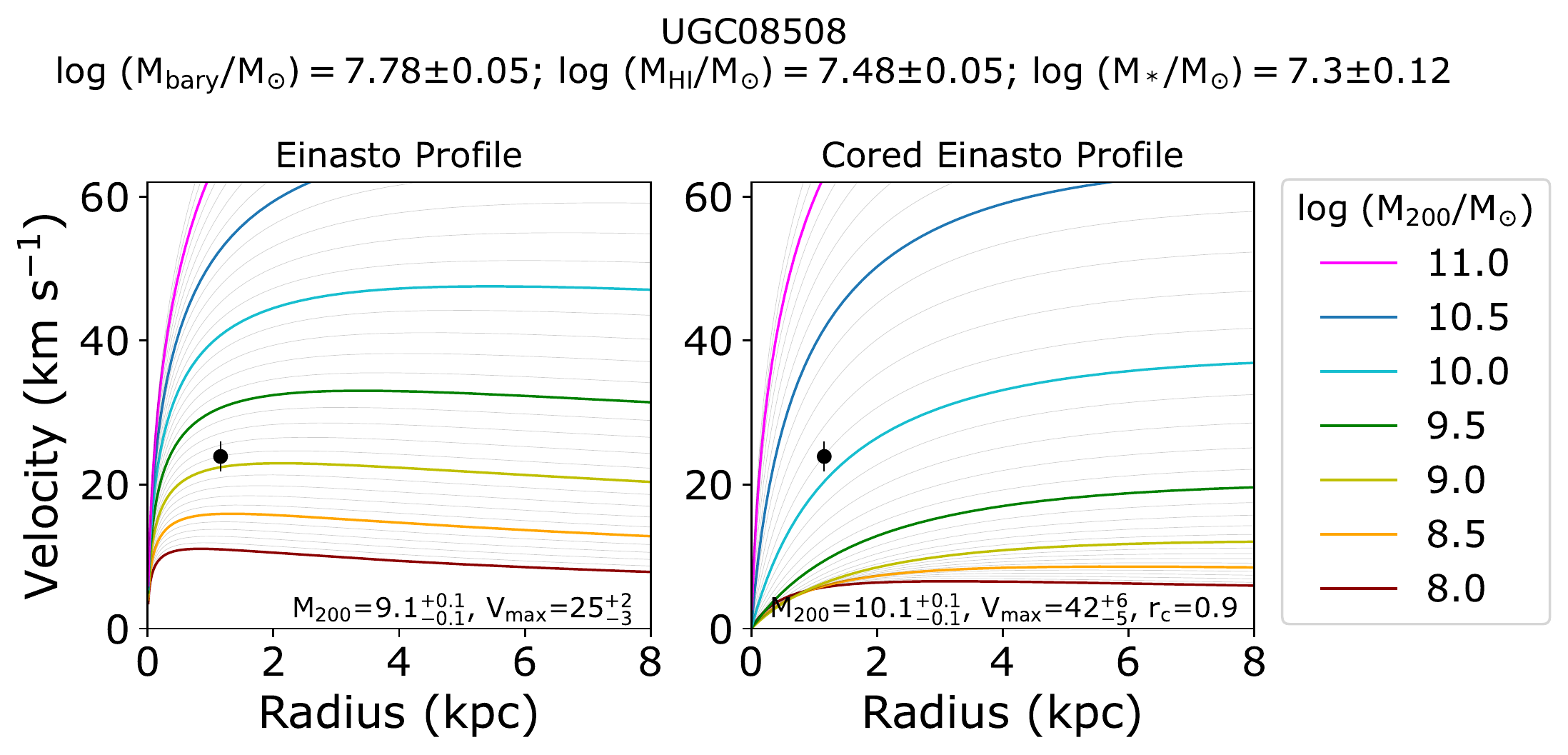}
\includegraphics[width=0.8\textwidth]{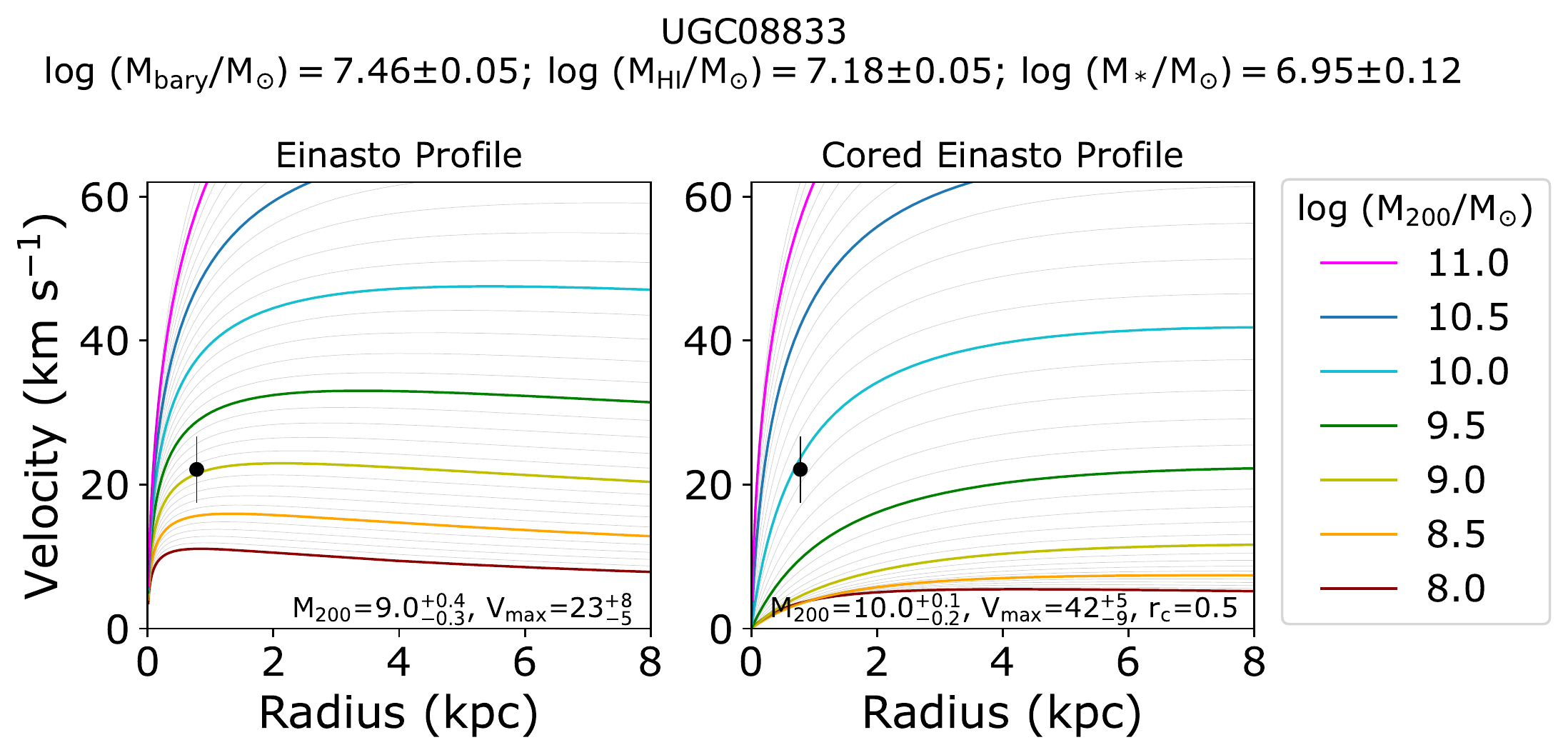}
\includegraphics[width=0.8\textwidth]{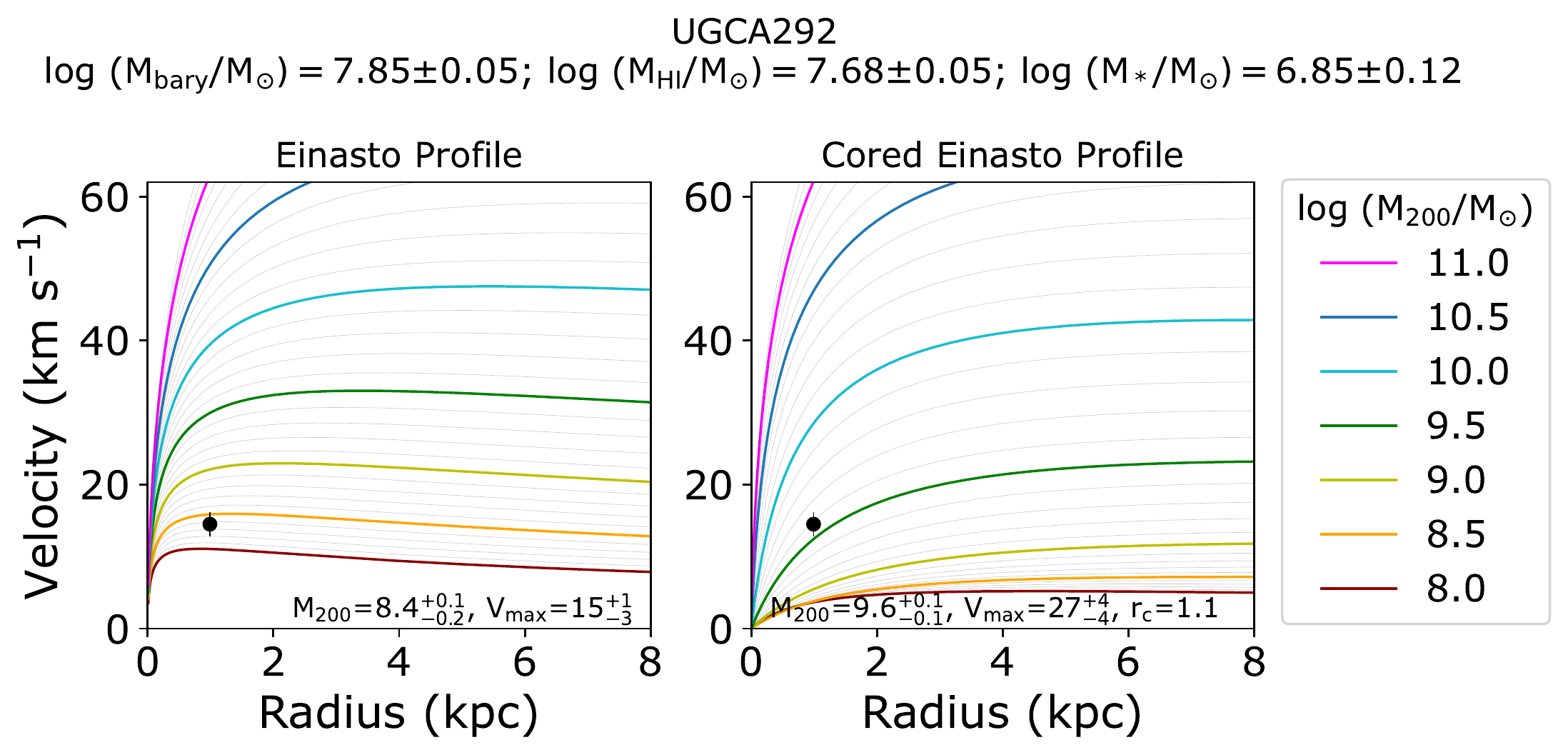}
\end{center}
\caption{Theoretical rotation curves for the VLA-ANGST galaxies UGC08508, UGC08833, and UGCA292 with the measured velocities corrected for asymmetric drift overplotted at the radii at which the velocity was measured. See Figure~\ref{fig:rot_curve_example} caption and Section~\ref{sec:rot_curves}.}
\label{fig:rot_curve_7}
\end{figure*}

\begin{figure*}
\begin{center}
\includegraphics[width=0.8\textwidth]{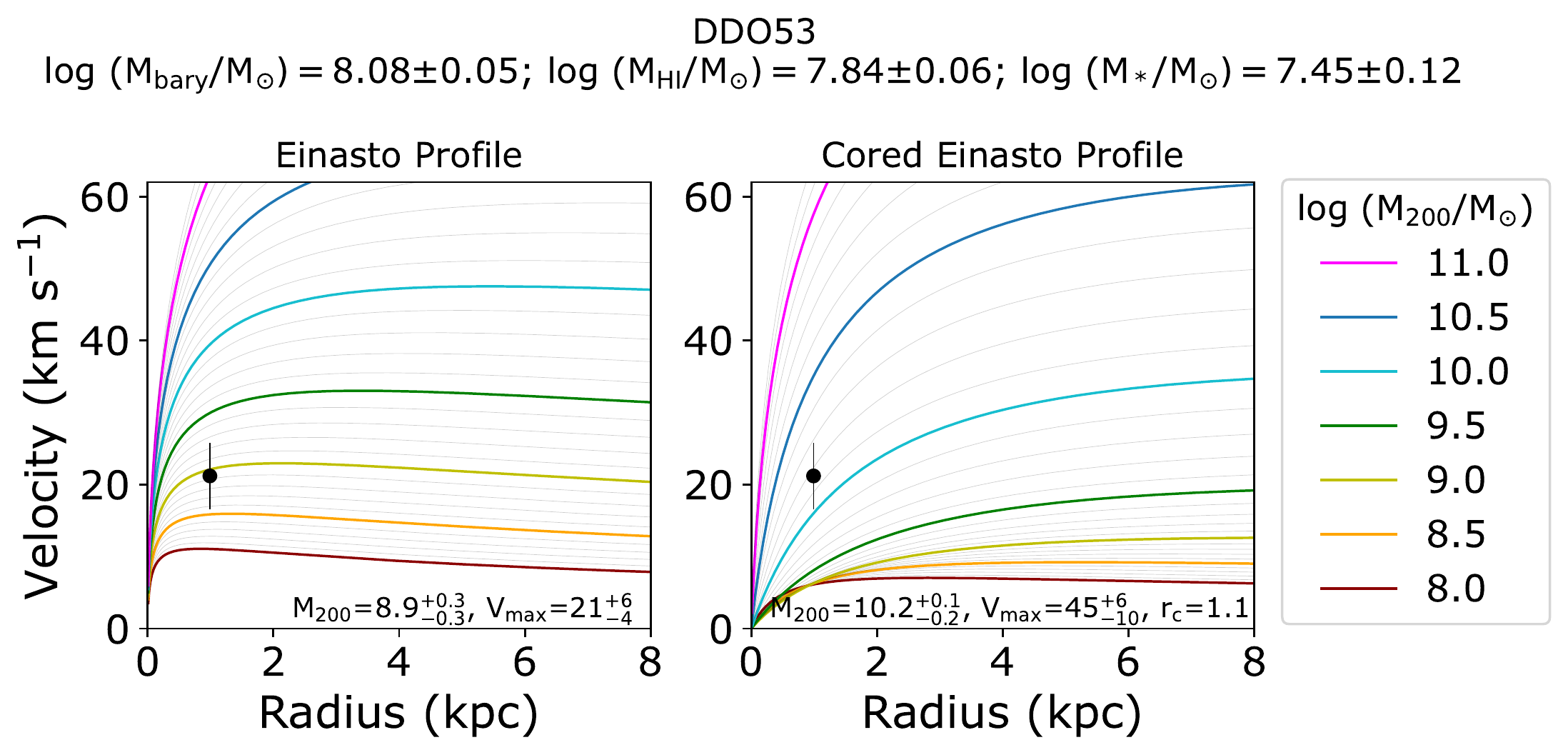}
\includegraphics[width=0.8\textwidth]{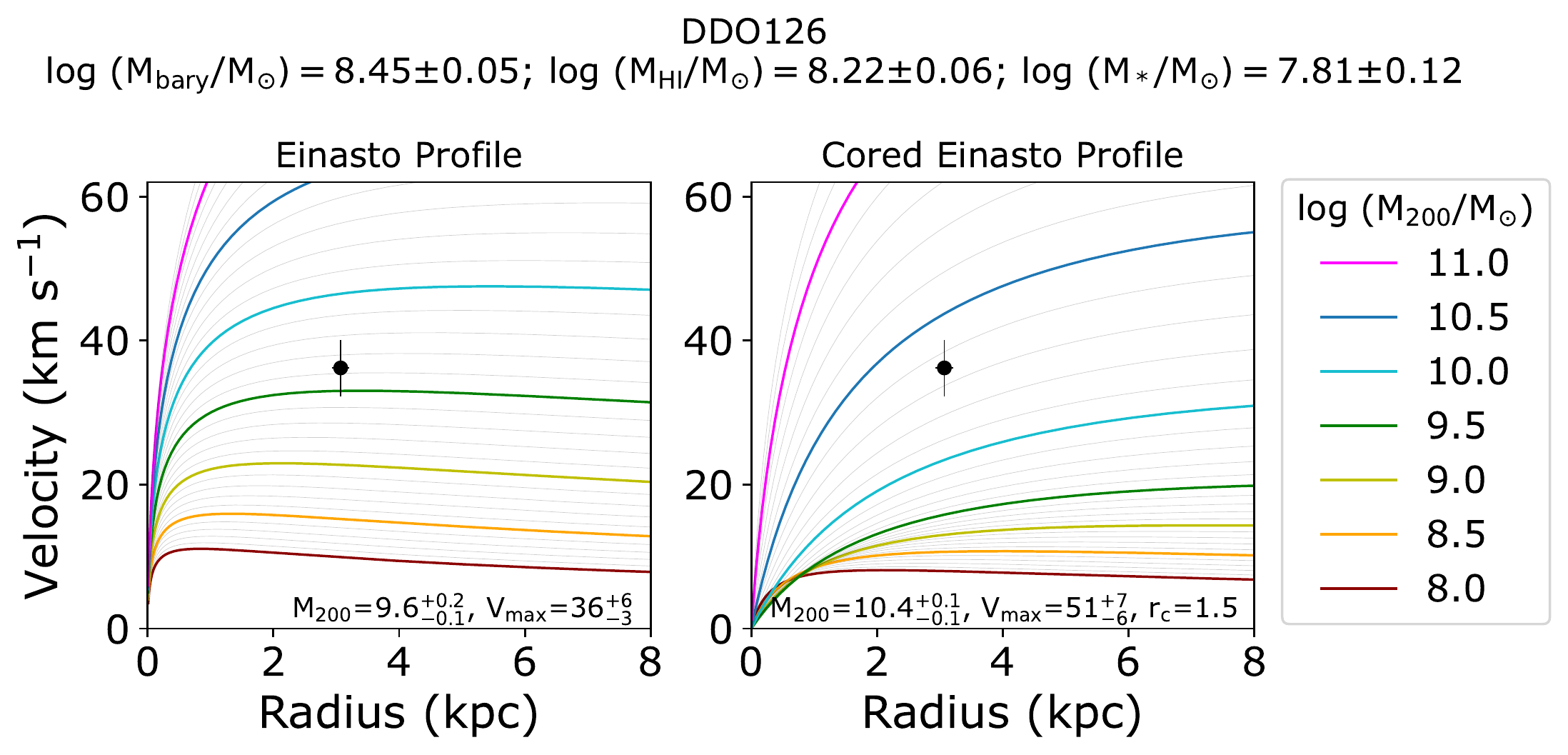}
\includegraphics[width=0.8\textwidth]{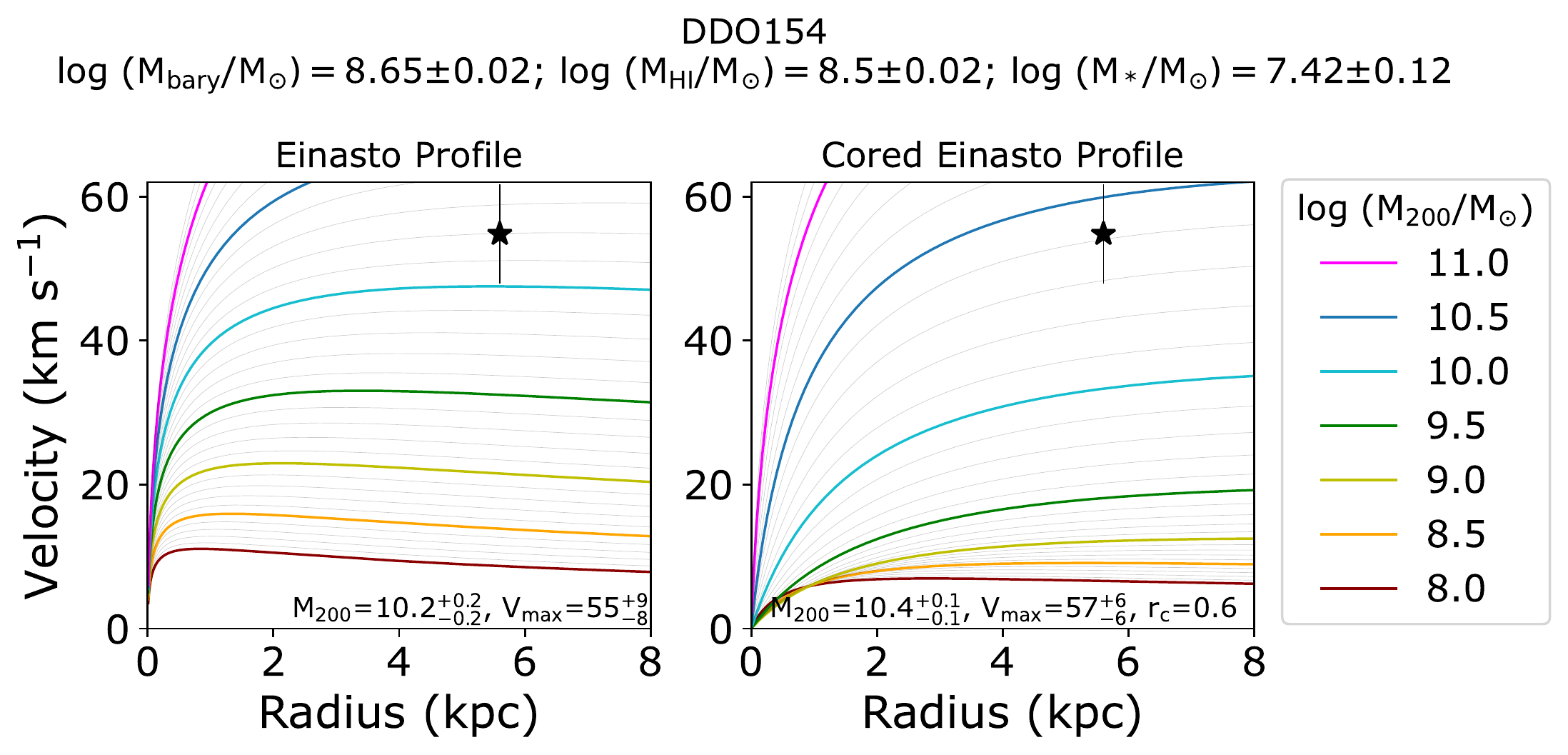}
\end{center}
\caption{Theoretical rotation curves for the LITTLE THINGS galaxies DDO53, DDO126, and DDO154 with the measured velocities corrected for asymmetric drift overplotted at the radii at which the velocity was measured. See Figure~\ref{fig:rot_curve_example} caption and Section~\ref{sec:rot_curves}.}
\label{fig:rot_curve_8}
\end{figure*}

\begin{figure*}
\begin{center}
\includegraphics[width=0.8\textwidth]{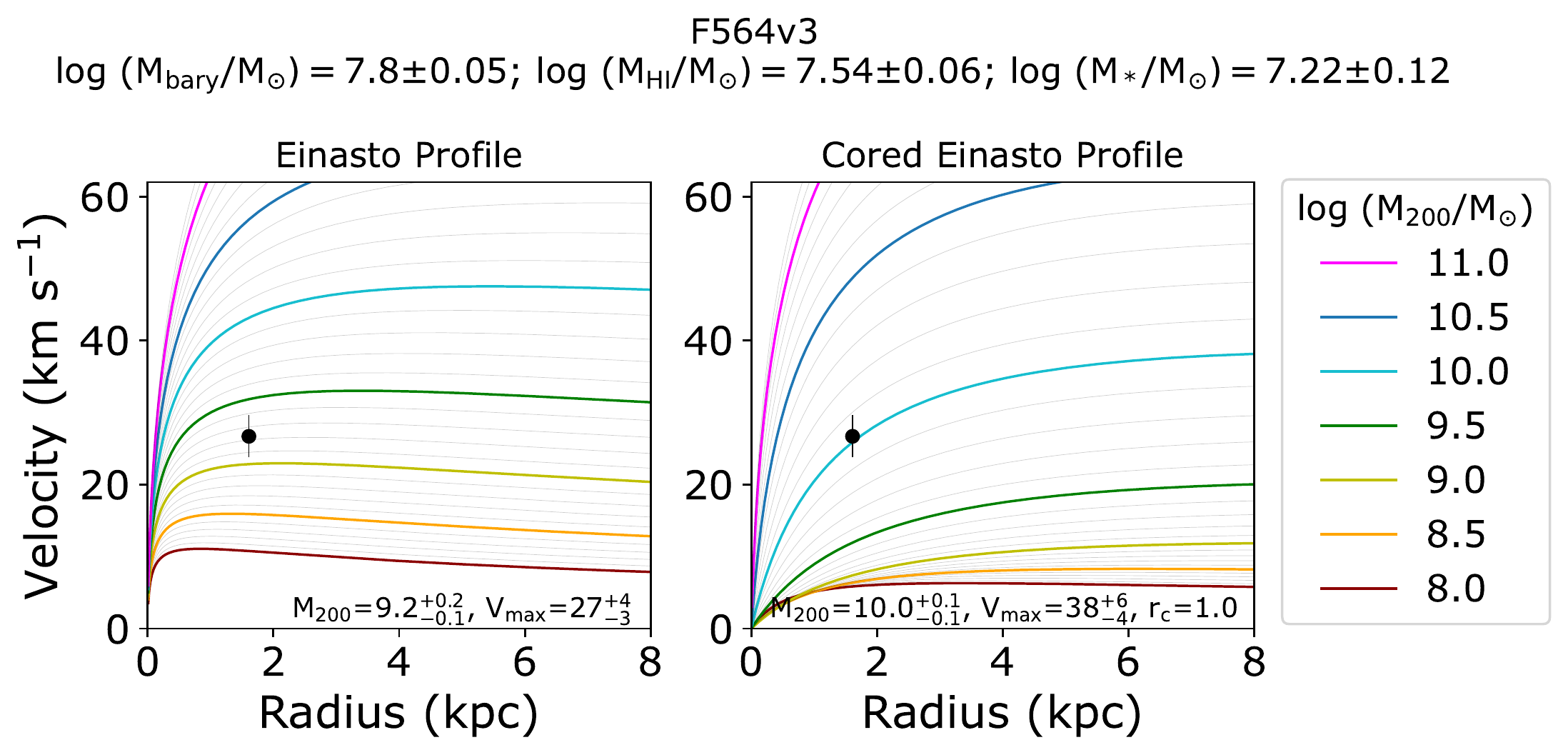}
\includegraphics[width=0.8\textwidth]{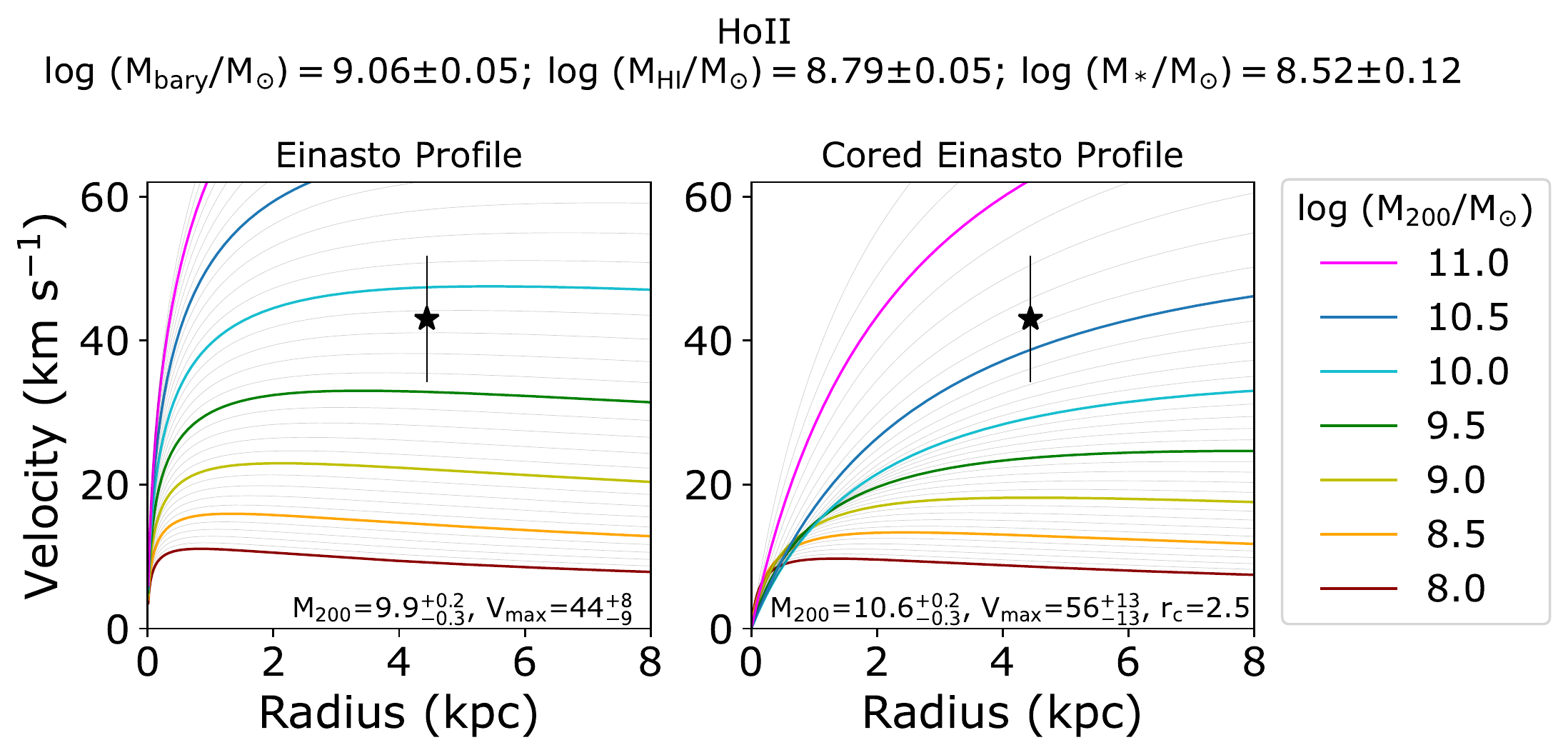}
\includegraphics[width=0.8\textwidth]{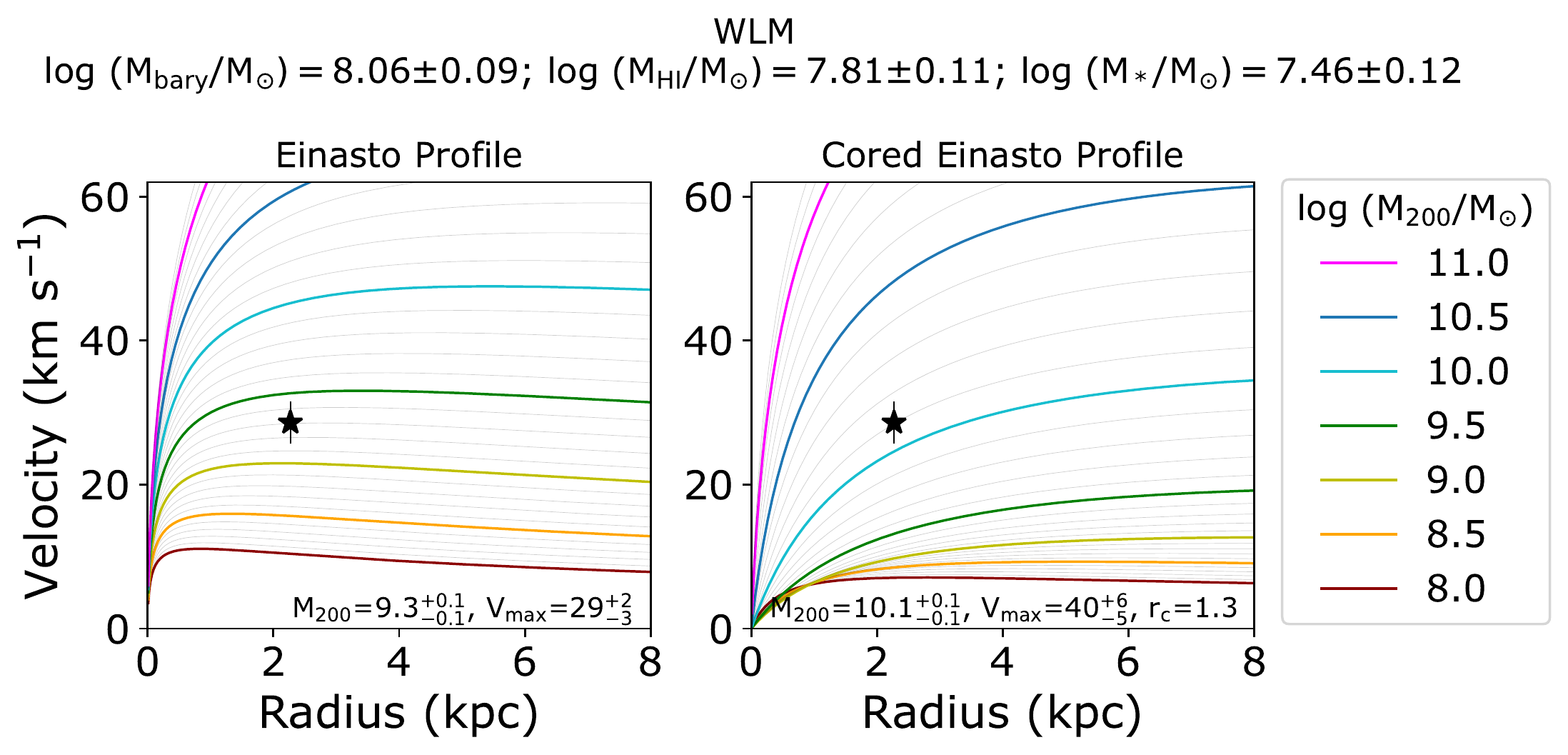}
\end{center}
\caption{Theoretical rotation curves for the LITTLE THINGS galaxies F564v3, HoII, and WLM with the measured velocities corrected for asymmetric drift overplotted at the radii at which the velocity was measured. See Figure~\ref{fig:rot_curve_example} caption and Section~\ref{sec:rot_curves}.}
\label{fig:rot_curve_9}
\end{figure*}

Figures~\ref{fig:rot_curve_2}-\ref{fig:rot_curve_9} present theoretical rotation curves generated using Einasto and cored Einasto density profiles for the remaining galaxies in our sample with our values of \vcor\ and \rmax\ overlaid. For galaxies where the measured velocities have clearly reached the flat part of the rotation curves, we use a star symbol; for all other galaxies we use a filled circle symbol. The methodology for generating the rotation curves is presented in Section~\ref{sec:dm}. Figures~\ref{fig:rot_curve_2}-\ref{fig:rot_curve_9} follow the same format and symbols used in Figure~\ref{fig:rot_curve_example}; see Section~\ref{sec:rot_curves} for additional details. Galaxies are ordered alpha-numerically within their original program of study, following the order listed in Table~\ref{tab:properties}. 

\end{document}